\documentclass[aps,prx,twocolumn,superscriptaddress,nofootinbib,notitlepage,longbibliography]{revtex4-1}

\usepackage{amsbsy}
\usepackage{amsfonts}
\usepackage{amsmath}
\usepackage{amstext}
\usepackage{amsthm}
\usepackage{amssymb}
\usepackage{bm}
\usepackage{color}
\usepackage{comment}
\usepackage{booktabs}
\usepackage{datetime}
\usepackage{dsfont}
\usepackage{dcolumn}
\usepackage{esint}
\usepackage{extarrows}
\usepackage{graphicx}
\usepackage[colorlinks,citecolor=blue]{hyperref} \usepackage{cleveref}
\usepackage{longtable,booktabs}
\usepackage{mathrsfs}
\usepackage{multirow}
\usepackage[super]{nth}
\usepackage{subfigure}
\usepackage{tensor}
\usepackage{url}
\usepackage{xfrac}

\hypersetup{colorlinks=true, linkcolor=blue, filecolor=magenta, urlcolor=blue,}

\begin{document}
\graphicspath{{figures/}}


\title{Two-Hole Ground State: Dichotomy in Pairing Symmetry}

\author{Jing-Yu Zhao} \affiliation{Institute for Advanced Study, Tsinghua University, Beijing 100084, China}
\author{Shuai A. Chen} \affiliation{Institute for Advanced Study, Tsinghua University, Beijing 100084, China}
\author{Hao-Kai Zhang} \affiliation{Institute for Advanced Study, Tsinghua University, Beijing 100084, China}
\author{Zheng-Yu Weng} \affiliation{Institute for Advanced Study, Tsinghua University, Beijing 100084, China}




\date{\today}

\begin{abstract}
    A single-hole ground state \emph{Ansatz} for the two-dimensional $t$-$J$ model has been recently studied by the variational Monte Carlo (VMC) method. 
    Such a doped hole behaves like a ``twisted'' non-Landau quasiparticle characterized by an emergent quantum number in agreement with exact numerics. 
    In this work, we further investigate the ground state of two holes by VMC. 
    It is found that the two holes strongly attract each other to form a pairing state with a new quantum number 
    the same as obtained by the numerical exact diagonalization and density matrix renormalization group (DMRG) calculations. 
    A unique feature of this pairing state is a \emph{dichotomy} in the pairing symmetry, i.e., 
    a \emph{d wave} in terms of the electron c operators and an \emph{s wave} in terms of the new quasiparticles,
    as explicitly illustrated in the ground state wave function. 
    A similar VMC study of a two-hole wave function for the $t$-$J$ two-leg ladder also yields a good agreement with the DMRG result. 
    We demonstrate that the pairing mechanism responsible for the strong binding here is not due to the long-range antiferromagnetic order 
    nor the resonating-valence-bound pairing in the spin background but is the consequence of the quantum phase-strings created by the hopping of holes. 
    The resulting spin-current pattern mediating the pairing force is explicitly illustrated in the VMC calculation. 
    Physical implications to superconductivity at finite doping are also discussed.
\end{abstract}

\maketitle

\tableofcontents

\section{Introduction}

The study of a doped antiferromagnetic (AFM) Mott insulator has been conducted intensively 
in the past three decades as one of the greatest challenges in condensed matter physics \cite{Mott1949,Anderson1987,Imada1998,Lee2006}. 
The mechanism of high-$T_c$ cuprate superconductors has been widely believed to be closely related to such a strongly correlated problem 
\cite{Anderson1987,Andersonbook,Lee2006,Zaanen2015}. 
While the low-energy physics of the AFM-Mott insulator at half filling as described by the Heisenberg Hamiltonian has been well understood \cite{Chakravarty1988,*Chakravarty1989,Auerbach1988a,*Auerbach1988,Liang1988,Manousakis1991}, 
the theoretical issues remain unsettled in the presence of doped holes, 
where superconducting (SC) condensation is expected to set in beyond some critical small doping concentration in two-dimensional (2D) or layered materials \cite{Lee2006,Zaanen2015}.

Among various theoretical proposals, the resonating-valence-bond (RVB) state \cite{Anderson1973,Anderson1987,Anderson2004} for high-temperature superconductivity stands out as one of the most influential and simplest ground state \emph{Ans{\"a}tzs}, 
which takes advantage of the essential properties of a doped Mott insulator: The local spin moments form singlet pairs due to the superexchange coupling. 
A tremendous body of studies around the RVB idea \cite{Baskaran1987,Kivelson1987,Lee1989,Fradkin1990,Andersonbook,Lee1992,Wen1996,Senthil2000,Sorella2002,Anderson2004,Lee2006,Edegger2007} 
has produced some insightful understandings on the doped Mott insulator and high-$T_c$ cuprate. 
At the same time, a fundamental question remains to be answered: 
How can an RVB-type ground state emerge from a spin AFM long-range-ordered (AFLRO) state at half filling by a doping effect? 

Another important observation made by Anderson \cite{Anderson1990} on the Mott physics is that a doped hole (electron) in the lower (upper) Hubbard band 
should generally induce a many-body response or ``phase-shift'' from the background electrons to accommodate such a doped hole (electron). 
This ``unrenormalizable phase-shift'' was conjectured \cite{Anderson1990} to cause the ``orthogonality catastrophe''  \cite{Anderson1967a,Anderson1967b}, 
leading to a non-Fermi-liquid behavior of the doped Mott insulator. 
Based on the $t$-$J$ model and later the Hubbard model, 
this many-body phase-shift has been precisely identified as the phase-string effect \cite{Sheng1996,*Weng1997,Wu2008,Zhang2014}, 
which can be mathematically formulated as a many-body Berry phase acquired by doped holes completing closed-paths motion. 
Protected by the Mott gap, such an effect fundamentally changes the Fermi statistics of the electrons \cite{Zhang2014,Zaanen2011}. 
In general, the Berry phase or the phase-string sign structure depends on the parity of spin-hole exchanges, 
which characterizes the intrinsic long-range mutual (spin-charge) entanglement \cite{Zheng2018a} in the doped Mott insulator. 

Along this line of thinking, an RVB state at finite doping, if it exists, must be \emph{qualitatively} modified, 
since each doped hole has to introduce an unrenormalizable phase-shift or ``irreparable phase-string,'' 
which would generally lead to a two-component RVB structure \cite{Weng2011b}. 
A new SC ground state \emph{Ansatz} of this type has been previously proposed \cite{Weng2011a}:
\begin{equation} 
	|\Psi_{G}\rangle = e^{\hat{\cal D}}|\mathrm{RVB}\rangle,
	\label{eqn:psgs} 	
\end{equation}
in which the charge pairing and the spin RVB pairing are explicitly separated. 
Here, the \emph{vacuum state} $|\mathrm{RVB}\rangle$ is a spin background in a short-range RVB state at finite doping \cite{Weng1998,*Weng1999}. 
It reduces to a long-range RVB state  $ |\phi_0\rangle $ to recover the AFLRO in the zero-hole limit. 
In contrast, the doped holes are created in spin-singlet pairs (another ``RVB'') by 
\begin{equation}
	\hat{\cal D}=\sum_{ij}{g}(i,j) \tilde{c}^{}_{i\uparrow}\tilde{c}^{}_{j\downarrow}~,
	\label{eqn:D} 	
\end{equation}
where $\tilde{c}$ is distinct from the bare electron operator $c$ by
\begin{equation}
	c_{i\sigma}\rightarrow \tilde{c}_{i\sigma}\equiv c_{i\sigma}e^{-i \hat{\Omega}_i }~,
	\label{eqn:ctilde}
\end{equation}
with $\hat{\Omega}_i$ [cf. Eq.~\eqref{eqn:phasestringoperator}] representing the many-body phase-shift or phase-string effect induced by the doped hole. 
Note that self-consistently the phase-shift field $\hat{\Omega}_i$ is also responsible for $|\phi_0\rangle\rightarrow |\mathrm{RVB}\rangle$ by doping \cite{Weng1998,*Weng1999,Ma2014}. 

A basic example to show the fundamental importance of the unrenormalizable phase-shift or phase-string effect is the single-hole case \cite{ Wang2015,Zhu2016,Chen2019}. 
The single-hole ground state may be reduced from Eq.~\eqref{eqn:psgs} as follows
\begin{equation} 
	|\Psi_{G}\rangle_{1h} = \sum_{i}{\phi}_h(i)\tilde{c}^{}_{i\sigma}|\phi_0\rangle,
	\label{eqn:1hgs} 	
\end{equation}
where $|\phi_0\rangle$ denotes the half-filling ground state of the Heisenberg model and the doped hole is created \emph{not} by $c$ but $\tilde{c}$ given in Eq.~\eqref{eqn:ctilde}. 
The single-hole wave function ${\phi}_h(i)$ in Eq.~\eqref{eqn:1hgs} is generally no longer a Bloch-wave-like ($\propto e^{i{\bf k}\cdot {r}_i}$), 
which can be determined instead as a variational parameter in the variational Monte Carlo (VMC) calculation \cite{Wang2015,Chen2019}. 
As the consequence of the many-body phase-shift operator $\hat{\Omega}_i$ acting on $|\phi_0\rangle$, 
the ground state \eqref{eqn:1hgs} acquires nontrivial angular momenta $L_z=\pm 1$ and novel ground state degeneracy in precise agreement 
with the exact diagonalization (ED) and density matrix renormalization group (DMRG) studies \cite{Zheng2018b}. 

Such a VMC approach has been powerful in establishing the best variational wave function $|\phi_0\rangle$ at half filling \cite{Liang1988} 
as well as the physical properties of the one-hole ground state $|\Psi_{G}\rangle_{1h}$ \cite{Wang2015,Chen2019} in comparison with the exact numerics. 
The single-hole problem illustrates how crucial the phase-string effect is to turn the doped hole into a non-Landau-like ``twisted'' quasiparticle with a novel quantum number. 
But the phase-shift field also generates some severe ``sign problem,'' i.e. unrenormalizable phase-string sign structure in $\hat{\cal D}$,  
to prevent an efficient VMC study of the \emph{Ansatz} state of Eq.~\eqref{eqn:psgs} 
beyond the mean-field and gauge theory approaches \cite{Weng2011a} at a finite doping. 
Therefore, to bridge the gap between the one-hole and a finite doping, and to reveal the hidden pairing mechanism of the doped antiferromagnet, 
a necessary and useful step is to use the same VMC approach to investigate the ground state of two holes injected into the spin background.
Actually, a simplified two-hole doped ladder with the injected holes restricted to move only along the chain (leg) direction was already analytically analyzed \cite{Chen2018}, 
which reveals a new pairing mechanism by a direct comparison with DMRG \cite{Zhu2018}. 

In this paper, we examine the pairing structure of the following two-hole ground state \emph{Ansatz} according to Eq.~\eqref{eqn:psgs}:
\begin{equation} 
	|\Psi_{G}\rangle_{2h} = \hat{\cal D}|\phi_0\rangle,
	\label{eqn:2hgs} 	
\end{equation}
which is determined by optimizing the pair amplitude $g(i,j)$ in Eq.~\eqref{eqn:D} by the VMC method. 
This is a paired state of the single twisted holes described by $\tilde{c}$. 
Because of the internal quantum numbers associated with $\tilde{c}$, the single-hole ground states are fourfold degenerate under an open boundary condition (OBC). 
The two-hole ground state is found to be nondegenerate with a spin-singlet pairing of two distinct twisted holes 
with a specific total quantum number $L_z=2\mod 4$ under the $C_4$ rotational symmetry, 
which has been previously already identified in the ED and DMRG calculations \cite{Zheng2018b}. 

A tightly bound pair is found in this two-hole ground state wave function, in which one hole predominantly distributes over a square of sites tightly packed around the other hole. 
It looks like a slightly anisotropic $s$-wave pairing of a nodeless $|g(i,j)|$, 
with the strongest pairing for two holes at the distance of $\sqrt{2}$ (in the units of the lattice constant). 
Nevertheless, if measured in terms of the ordinary Cooper pairing operator $c_{{\bf k}\uparrow}c_{-{\bf k}\downarrow}$ in the singlet channel, a well-defined pure $d$-wave symmetry is found. 
Such a ``dichotomy'' in the pairing symmetry is directly contradictory to an ordinary BCS pairing wave function for two Landau quasiparticles.
Here, the $d$-wave symmetry measured by the Cooper pairing can be attributed to the consequence of the phase-shift factor $e^{-i\hat{\Omega}}$ in combination with the nodeless $g(i,j)$. 
In other words, the two-hole wave function can be regarded as composed of an $s$-wave-like pairing of the twisted quasiparticles $\tilde{c}$, 
which is more robust than a BCS-like $d$-wave pairing state of the bare holes as a projecting or ``collapsing'' state of the former. 

The pairing mechanism in the two-hole ground state can be straightforwardly examined. 
By projecting the ground state wave function onto different \emph{fixed} hole positions, a ``rotonlike'' configuration of spin supercurrrents can be explicitly revealed, 
which represents the physical effect of the phase-strings created by the motion of holes. 
As a matter of fact, by turning off the phase-shift factor $e^{-i\hat{\Omega}}$, the strong binding between the holes diminishes immediately as shown in the VMC calculation, 
even if the AFLRO state $|\phi_0\rangle$ can be artificially tuned into a short-range RVB state in the dimer limit. 
It implies that the phase-strings not only renormalize the single holes to become the non-Landau twisted holes, 
but also are responsible crucially for their pairing in the two-hole ground state, which is consistent with earlier DMRG calculation for the two-hole ladder \cite{Zhu2014}. 

The two-hole ground state \emph{Ansatz} is also applied to the case in the two-leg ladder. 
Previously the single-hole ground state has already produced \cite{Wang2015} an excellent agreement with the DMRG. 
The present VMC calculation shows an overall agreement of the two-hole case with the DMRG results, 
including the pair-pair correlation and $d$-wave-like sign change of the Cooper pairing in the chain and rung directions. 
Turning off the phase-shift field in the wave function indeed leads to diminishing pairing strength, 
again indicating the pairing mechanism due to the phase-string effect. 

Finally, it is noted that, for an exact ground state, one should further treat the spin background $|\phi_0\rangle$ in either 
Eq. (\ref{eqn:1hgs}) or (\ref{eqn:2hgs}) as a variational one such that an additional ``longitudinal'' spin polaron effect due to 
the suppression of the AFM correlations by the doped hole(s) can be properly accounted for. 
By treating $|\phi_0\rangle$ as a very precise variational ground state of the Heisenberg model (cf. Appendix ~\ref{app:2}) 
in the present variational scheme, such a longitudinal spin polaron effect is neglected (cf. Sec.~\ref{sec:5D} as well as Ref. \onlinecite{Chen2019} for a detailed discussion). 
In other words, in the present two-hole and previous one-hole \cite{Chen2019} variational studies, 
the main focus is on the \emph{transverse} spin polaron effect due to the phase-string as characterized by the phase-shift factor $e^{-i\hat{\Omega}}$. 
According to the above-mentioned DMRG study in the two-leg ladder case, the residual longitudinal spin polaron effect is negligible as compared to the phase-string effect in the study of the pairing mechanism. A discussion of a variational treatment of $|\phi_0\rangle$ to incorporate the feedback effect of the doped holes (i.e., the residual longitudinal spin polaron) to further optimize the kinetic energy is given in Appendix \ref{app:3}. 

The rest of the paper is organized as follows. 
In Sec.~\ref{sec:2}, we will introduce the model and construct the two-hole \emph{Ansatz} wave function 
with an emergent phase-string phase factor being built in concretely. A VMC calculation is then carried out.
For completeness, a brief review of the single-hole ground state by a VMC calculation is also given.
In Sec.~\ref{sec:3}, the properties of the two-hole ground state are systematically analyzed, especially the pairing structure. 
A dichotomy of pairing symmetry, i.e., a $d$-wave symmetry measured in terms of the original $c$ operator of the electron, 
and an $s$-wave-like symmetry in terms of  twisted quasiparticles $\tilde{c}$, 
unveils a prototypical non-BCS feature. In Sec.~\ref{sec:4}, the binding energy and the pairing force originated from the phase-string effect are examined, 
which shows how a kinetic-energy-driven pairing mechanism can be clearly visualized by the hidden spin-current pattern generated by the hole hopping. 
In Sec.~\ref{sec:5}, we further discuss some important implications like the incoherent single-particle propagation and possible self-localization of the hole pair in the AFLRO phase 
and how the superconducting ground state at finite doping may evolve from the present two-hole ground state.  
Finally, the conclusion and perspectives are given in Sec.~\ref{sec:6}.

\section{The ground state \emph{Ansatz}}\label{sec:2}

\subsection{The $t$-$J$ model}

The standard $t$-$J$ model reads $H = \mathcal{P}_s(H_t+H_J)\mathcal{P}_s$, where 
\begin{align} 
    H_t & =-t\sum_{\langle ij\rangle,\sigma}(c_{i\sigma}^\dagger c^{}_{j\sigma}+H.c.), \label{eqn:hoppingH} \\ 
    H_J & =J\sum_{\langle ij\rangle}\left(\mathbf{S}_i\cdot\mathbf{S}_j-\frac{1}{4}n_in_j\right), \label{eqn:superexchangeH}
\end{align}
with the summations over $\langle ij\rangle$ denoting the nearest-neighbor (NN) bonds. Here $\mathbf{S}_i$ and $n_i$ are spin and electron number operators, respectively. 
The strong correlation nature of the $t$-$J$ model comes from the no-double-occupancy constraint $\sum_{\sigma} c^\dagger_{i\sigma}c^{}_{i\sigma}\leq 1$ on each site, 
which is imposed via the projection operator $\mathcal{P}_s$. The superexchange coupling constant $J$ is taken as the units with $t/J=3$ throughout the paper. 
The size of a square lattice is specified by $ N \equiv N_x\times N_y$, with $N_x=N_y=2\times M$ chosen to be even for an isotropic 2D square lattice and $ N \equiv N_x\times 2$ for a two-leg ladder.

\subsection{The single-hole ground state}

\begin{figure}[tb]
    \centering
    \includegraphics[width=0.44\textwidth]{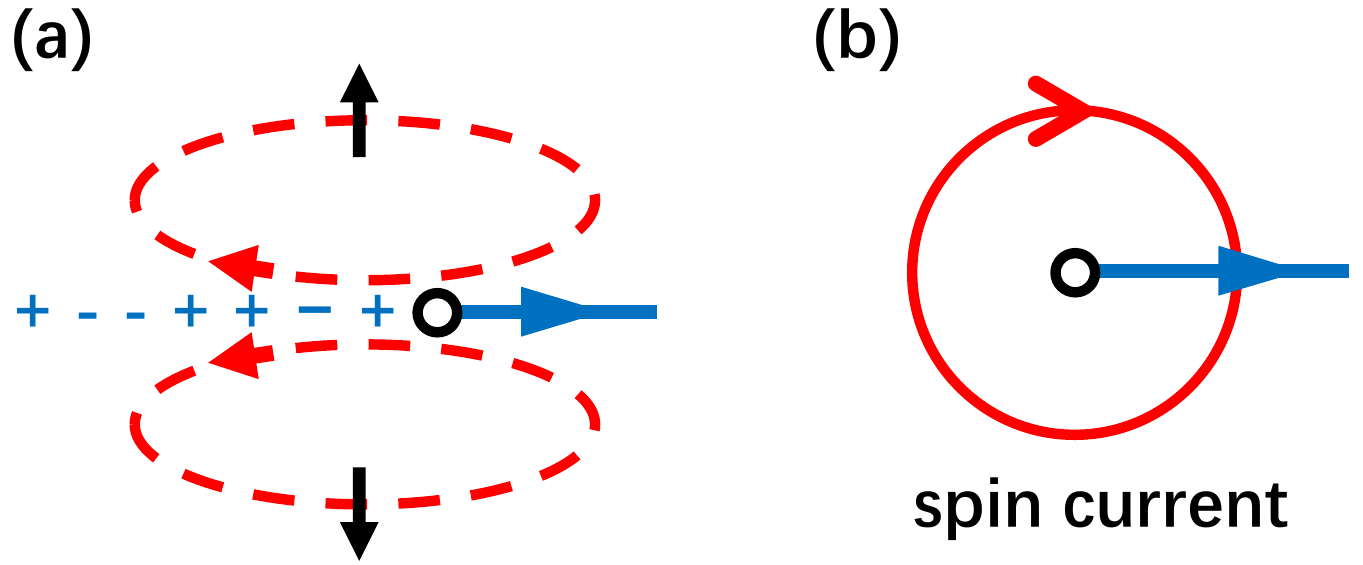} 
    \caption{The composite structure of a renormalized single-hole state discussed in Ref. \onlinecite{Chen2019}: 
        (a) The motion (blue arrow) of a doped hole (black circle) creates a sequence of $\pm$ signs (phase-string) depending on $\uparrow$ and $\downarrow$ of the backflow spins. 
        To avoid cancellation of the phase-string effect in the hopping integral, the spin backflow currents have to bend into a rotonlike pattern (red dashed line) around the hole; 
        (b) such a roton configuration can be explicitly manifested by a vortex of spin current around the hole as a \emph{net} effect for an extra $S_z=\pm 1/2$. 
        The chirality of the vortex leads to a double degeneracy of the states with an emergent angular momentum $L_z=\pm 1$.}    
    \label{fig:oneholeil}
\end{figure}

Before we proceed to investigate the two-hole ground state, it would be helpful for us to briefly outline some key results of the single-hole state in the previous study \cite{Chen2019,Wang2015}. 

The single-hole problem has been a focus of intensive study 
with a history almost as long as the high-$T_c$ problem. 
It had been believed for a long time that the single hole injected into the AFM spin background would behave like a Landau-like quasiparticle 
with translational symmetry and a finite spectral weight, known as a spin polaron \cite{Brinkman1970,Schmitt-Rink1988,Kane1989,Martinez1991,Bala1995,Sorella1992,Brunner2000,Imada2020}. 
Namely, the spin distortion of the AFM background in response to the motion of the hole was expected to merely dress the \emph{bare} hole's effective mass. 
It is particularly worth noting that, even though Shraiman and Siggia \cite{Shraiman1988a,Shraiman1989} have previously discovered an important transverse dipolar spin twist induced by the doped hole, 
their semiclassical approach fails to capture the \emph{correct} short-range singular structure of the spin current, 
reaching a conclusion of the dressed ``hole'' quasiparticle with a definite momentum in a translational fashion similar to the above obtained 
by considering only the longitudinal spin polaron effect. 
In other words, no Anderson's unrenormalizable phase-shift would be produced in the one-hole ground state to invalidate the Bloch wave behavior in the long wavelength. 

However, the recent ED and DMRG numerics \cite{Zheng2018b} reveal that something is fundamentally missing in this quasiparticle picture. 
It has been shown that actually there are \emph{hidden} spin currents always present in the spin background as generated by the motion of the doped hole in the degenerate ground states. 
Since it is always conserved in the $t$-$J$ model, the spin current itself can carry away momentum and change the translational symmetry of a renormalized hole 
even though the total (many-body) momentum remains conserved. 
It implies that the hole entity no longer obeys the Landau one-to-one correspondence assumption to behave like a Bloch wave.  

In fact, with a lattice rotational symmetry under the OBC, the spin current can generate a nontrivial angular momentum 
with a novel ground state degeneracy besides the normal degeneracy due to $S= 1/2$. 
As shown in Fig.~\ref{fig:oneholeil}(a), when a doped hole hops in an AFM background (always labeled by the quantum numbers $S$ and $S_z$), 
a string of spin-dependent $\pm$ signs (i.e., the phase-string) \cite{Sheng1996,Weng1997} is generally created, 
which leads to a backflow of \emph{spin current} [cf. Eqs.~\eqref{eqn:spincurrent} and \eqref{eqn:spincurrent1} for the definition under a conserved $S_z$] in response to the motion of the hole. 
However, due to the singlet nature of the spin background (in a finite but large sample size), a strong cancellation of the phase-string effect occurs in the hopping integral, 
unless the spin backflow bends into a rotonlike configuration for the backflow currents composed of two opposite spins [cf. Fig.~\ref{fig:oneholeil}(a)]. 
Note that there is also an opposite chirality of the roton configuration not shown in Fig.~\ref{fig:oneholeil}(a). 
Thus, due to the phase-string effect, the hidden spin currents must play an essential role to form a composite object to facilitate the hole's motion.

Furthermore, as introduced by the hole, there is an extra $S_z=\pm 1/2$ in the otherwise spin-singlet background, 
which can manifestly contribute to an explicit spin current as a \emph{vortex} illustrated in Fig.~\ref{fig:oneholeil} (b). 
It can be regarded as a \emph{net} effect of the underlying backflow pattern shown in Fig.~\ref{fig:oneholeil}(a). 
Under a $C_4$ symmetry such a hidden backflow configuration associated with the doped hole gives rise to a nontrivial angular momentum 
($L_z=\pm 1 $ for an $N=2M\times 2M$ lattice) as a new quantum number in addition to, say, $S_z=\pm 1/2$ associated with the hole composite.   

\begin{table}[tb]
    \centering
    \caption{
        Essential properties of the one-hole degenerate ground states
        from the single-hole variational wave function \emph{Ansatz} in Eq.~\eqref{eqn:singleholeansatz} on a 2D square lattice with $C_4$ symmetry (OBC). 
        $S_z$ denotes the total $z$-component spin, and $L_z$ is the orbital angular momentum.
        $J^{b,s}$ [defined in Eqs.~\eqref{eqn:spincurrent} and \eqref{eqn:spincurrent1}] 
        are the backflow and neutral spin currents accompanying the motion of a hole, respectively, and  
        $J^h_{ij} = it\sum _{\sigma} (c^\dagger_{i\sigma}c^{}_{j\sigma} - c^\dagger_{j\sigma}c^{}_{i\sigma})$ is the hole current.
        $\curvearrowright$ ($\curvearrowleft$) marks the current vortex circulation clockwise (anticlockwise).
        The phase-shift operator $\hat \Omega_i$ defined in Eq.~\eqref{eqn:phasestringoperator} originates from the phase-string effect and 
        captures the most essential feature of a single-hole wave function \emph{Ansatz} in Eq.~\eqref{eqn:singleholeansatz}. 
        }
    \begin{ruledtabular}
        \begin{tabular}{ccccc}
              & $|\Psi_{(+\uparrow)}\rangle$ & $|\Psi_{(-\uparrow)}\rangle$ & $|\Psi_{(+\downarrow)}\rangle$ & $|\Psi_{(-\downarrow)}\rangle$ \\
            \colrule
            Phase factor & $e^{-i\hat{\Omega}_i}$ & $e^{i\hat{\Omega}_i}$ & $e^{-i\hat{\Omega}_i}$ & $e^{i\hat{\Omega}_i}$ \\
            $S_z$ & $+1/2$ & $+1/2$ & $-1/2$ & $-1/2$\\
            $L_z$ & $+1$ & $-1$ & $-1$ & $+1$\\
            $J^{b,s}$  & $\curvearrowleft$ & $\curvearrowright$ & $\curvearrowleft$ & $\curvearrowright$\\
            $J^{h}$ & $\curvearrowright$ & $\curvearrowleft$ & $\curvearrowleft$ & $\curvearrowright$\\
        \end{tabular}	
    \end{ruledtabular}
    \label{tab:chirality}
\end{table}

Therefore, in the presence of the unrenormalizable phase-shift or the phase-string effect, 
the hole is renormalized to become a composite object as shown in Fig.~\ref{fig:oneholeil}. 
This picture is confirmed by the finite-size ED and DMRG calculations \cite{Zheng2018b}. 
Mathematically, the phase-string or the spin-current backflow pattern in Fig.~\ref{fig:oneholeil} can be characterized by a phase-shift factor 
$e^{- i\hat{\Omega}_i}$ already mentioned in the introduction. 
Here $\hat{\Omega}$ is explicitly defined by
\begin{equation}
    \hat{\Omega}_i = \sum_{l}\theta_i(l)n_{l\downarrow},
    \label{eqn:phasestringoperator}
\end{equation}
where $n_{l\downarrow}$ is the number operator of down spin at site $l$ and $\theta_i(l)=m \mathrm{Im}\ln(z_i-z_l)$ ($m=\pm$) 
denotes a statistical angle between site $i$ and site $l$ in a 2D plane with $z_i=x_i+iy_i$ being the complex coordinate of site $i$. 
Note that $\theta_i(l)$ satisfies the condition
\begin{equation} \label{theta}
\theta_i(l)-\theta_l(i)=\pm \pi ~,
\end{equation}
which results in a sign change, $e^{- i\hat{\Omega}_i}\rightarrow -e^{- i\hat{\Omega}_i}$, 
accompanying the exchange between a hole and a $\downarrow$ spin to precisely keep track of the phase-string effect. 

Consequently, a single-hole ground state \emph{Ansatz} is constructed as follows \cite{Chen2019}: 
\begin{equation}
    |\Psi_{(m\sigma)}\rangle_{{1h}} = \sum_i \phi_h^{(m\sigma)}(i)e^{-im\hat{\Omega}_i}c_{i\bar{\sigma}}|\phi_0\rangle,
    \label{eqn:singleholeansatz}
\end{equation}
where $|\phi_0\rangle$ is the ground state of the Heisenberg Hamiltonian at half filling. 
Here, $m=\pm$ represents opposite chiralities of the spin-current vortex, which is originally hidden in the definition of $\theta_i(l)$ above, 
but hereafter, redefinition on the statistical angle  $\theta_i(l)\equiv \mathrm{Im}\ln(z_i-z_l)$ in Eq.~\eqref{eqn:phasestringoperator} explicitly appears with 
$e^{- i\hat{\Omega}_i} \rightarrow e^{-im\hat{\Omega}_i}$ for the convenience of discussion below. 

The variational parameters $\phi_h^{(m\sigma)}(i)$ can be determined by optimizing the kinetic energy by VMC. 
Detailed analyses of this wave function \emph{Ansatz} for the 2D square lattice are given in Ref.~\onlinecite{Chen2019}. 
The agreement of the ground state properties between the \emph{Ansatz} states and the exact ED or DMRG results is shown there. 
For the 2D case under an OBC, the four degenerate ground states and their quantum numbers are listed in Table~\ref{tab:chirality} with $\sigma =\uparrow$, $\downarrow$ and $m=\pm$. 
Specifically, it shows that the hole acquires a nonzero angular momentum $L_z=\pm 1$ corresponding to the chiral spin current
(marked by $\curvearrowright$    and $\curvearrowleft$ in Table~\ref{tab:chirality})
in addition to $S_z=\pm 1/2$. 
Here the spin currents produced in the hopping (backflow spin current) and superexchange (neutral spin current) terms are defined by \cite{Chen2019}
\begin{align} \label{eqn:spincurrent}
    J_{ij}^b =& -i\frac{t}{2}\sum _{\sigma} \sigma (c^\dagger_{i\sigma}c^{}_{j\sigma} - c^\dagger_{j\sigma}c^{}_{i\sigma}),\\
    J_{ij}^s =& \,\frac{J}{2}i(S^+_iS^-_j-S^-_iS^+_j), \label{eqn:spincurrent1}
\end{align}
respectively. 

Finally, we mention that the single-hole ground state in Eq. (\ref{eqn:singleholeansatz}) was actually first successfully applied to the one-dimensional chain \cite{Zhu2016} 
and then two-leg ladder \cite{Wang2015}, respectively. 
Note that the statistical angle $\theta_i(l)$ in Eq.~\eqref{eqn:phasestringoperator} can be further optimized \cite{Wang2015} 
for the anisotropic $t$-$J$ two-leg ladder case under the constraint in Eq.~\eqref{theta}. 
But so long as Eq.~\eqref{theta} is satisfied, the phase-shift factor $e^{\mp i\hat{\Omega}_i}$ in Eq.~\eqref{eqn:singleholeansatz} 
is generally effective in keeping track of the singular phase-string effect. 

\begin{figure}[tb]
    \centering
    \includegraphics[width=0.38\textwidth]{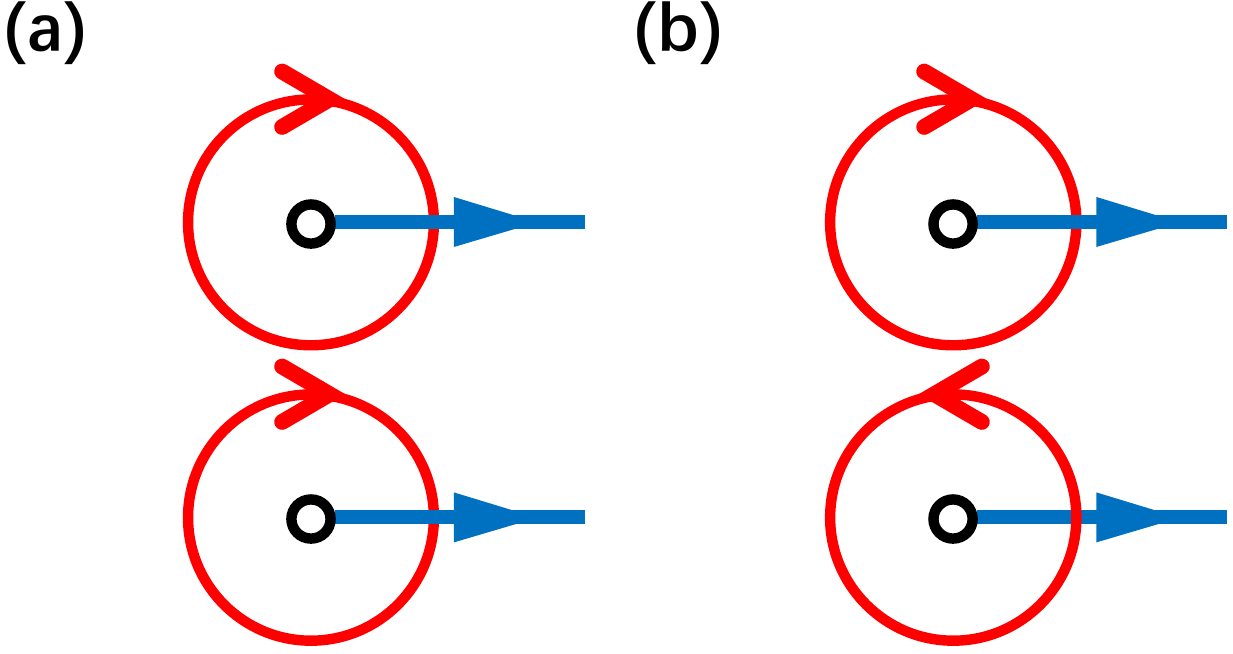}
    \caption{The two-hole ground state \emph{Ansatz}. 
        (a) Two holes are composed of the same chirality of spin current in $|\Psi_{+}\rangle_{{2h}} $; 
        (b) two holes are composed of the opposite chiralities of spin current in $|\Psi_{-}\rangle_{{2h}} $. 
        See the definitions in Eq.~\eqref{eqn:doubleholeansatz}. }
    \label{fig:twoholeil}
\end{figure}

\begin{table}[tb]
    \caption{The energies of the two-hole ground states $|\Psi_{0}\rangle_{2h}$ in Eq.~\eqref{eqn:bare2hole}
    and $|\Psi_{\pm}\rangle_{2h}$ in Eq.~\eqref{eqn:doubleholeansatz},
     on a $4\times 4$ lattice under OBC: 
        $E_{G}$ is the total energy; $E_t$ and $E_J$ are the kinetic and the superexchange energies, respectively; 
        $L_z = 0, \pm 1, 2$ denote the corresponding orbital angular momenta.}
    \label{tab:energy44}
    \begin{ruledtabular}
        \begin{tabular}{ccccc}
            & $E_{G}$ & $E_t$ & $E_J$ & $L_z$\\
            \colrule
            $|\Psi_{0}\rangle_{2h}$ & $-17.37$ & $-5.88$ & $-11.48$ & $0$ \\
            $|\Psi_{+}\rangle_{2h}$ & $-20.50$ & $-9.56$ & $-10.94$ & $\pm 1$\\
            $|\Psi_{-}\rangle_{2h}$ & $-22.51$ & $-11.20$ & $-11.31$ & $2$\\
            ED & $-24.98$  & $-14.57$ & $-10.42$ & $2$
        \end{tabular}
    \end{ruledtabular}
\end{table}

\subsection{Two-hole ground state \emph{Ansatz}}

In the above subsection, we briefly outline the ground state \emph{Ansatz} for the single-hole doped $t$-$J$ models in 2D and two-leg square lattice cases, respectively. 
It is straightforward to construct the corresponding two-hole ground state wave functions and study them based on a VMC approach. 

By noting that the single-hole ground state given in Eq.~\eqref{eqn:singleholeansatz} is generally fourfold degenerate 
(cf. Table~\ref{tab:chirality}) in an $N_x=N_y=2\times M $ lattice, with $S=1/2$ and angular momentum $L_z=\pm 1$ under the OBC, 
the two-hole ground state of $S^z=0$ may be constructed by 
\begin{equation}
    |\Psi_\eta\rangle_{2{h}} = \sum_{ij}g_\eta(i,j)
    c_{i\uparrow}^{}c_{j\downarrow}^{} e^{-i(\hat{\Omega}_i+\eta\hat{\Omega}_j)}
    |\phi_0\rangle+\cdots.
    \label{eqn:doubleholeansatz}
\end{equation}
Here, $|\phi_0\rangle$ describes a half-filling spin background. 
The subscript $\eta = \pm$ denotes two sectors of wave function \emph{Ansatz}, where the spin-current vortex pattern around each hole [Fig.~\ref{fig:oneholeil} (b)] 
has either the same chirality or the opposite one in forming a paired state as illustrated by Figs.~\ref{fig:twoholeil} (a) and \ref{fig:twoholeil}(b), respectively. 
The rest term $\cdots$ on the right-hand side of Eq.~\eqref{eqn:doubleholeansatz} involves the opposite chirality (the complex conjugate) of the phase-shift operator: 
$e^{i(\hat{\Omega}_i+\eta\hat{\Omega}_j)}$. 
For a comparison, we also consider a bare two-hole wave function without the phase-shift operator, which is denoted by $\eta=0$:
\begin{equation}
    |\Psi_{0}\rangle_{{2h}} = \sum_{ij}g_0(i,j)c_{i\uparrow}c_{j\downarrow}|\phi_0\rangle. 
    \label{eqn:bare2hole}
\end{equation}
A detailed analysis of the wave function symmetries is given in Appendix \ref{app:1}. 

Here, the pair amplitude $g_\eta (i,j)$ in $|\Psi_{\eta}\rangle_{{2h}} $ 
is treated as a variational parameter to be determined by VMC by optimizing the kinetic energy of the $t$-$J$ model for each $\eta=\pm, 0$. 
Note that the spin background $|\phi_0\rangle$ may be still chosen as the ground state of the half-filling Heisenberg model $H_J$. 
By doing so, one assumes that the other disturbance of the two doped holes to the spin background is negligible, except for the phase-string effect via $e^{\mp i(\hat{\Omega}_i+\eta\hat{\Omega}_j)}$. 
A further optimization of $|\phi_0\rangle$ in the VMC procedure is discussed below. 

The above three \emph{Ansatz} states of $|\Psi_{\eta}\rangle_{{2h}}$ are determined variationally in a VMC calculation similar to the one-hole case on a 2D square lattice. 
The variational ground state energies on a $4\times 4$ lattice are presented in Table~\ref{tab:energy44} in comparison with the ED result. 
According to Table~\ref{tab:energy44}, relative to the phase-string-free $|\Psi_{0}\rangle_{{2h}}$, 
both $|\Psi_{+}\rangle_{{2h}}$ and $|\Psi_{-}\rangle_{{2h}}$ have much improved variational energies as expected based on the single-hole case.


From Table~\ref{tab:energy44}, one sees that $|\Psi_{-}\rangle_{{2h}}$ has not only the lowest ground state energy, 
but also the same quantum number with the angular momentum $L_z=2 \mod 4$ as the ED result, 
which gives rises to a sign change of the total state under a $\pi/2$ rotational of the lattice. 
A combination of the degenerate ground states in the single-hole doped case leads to a nondegenerate $|\Psi_{-}\rangle_{{2h}}$ in the two-hole case, 
which is consistent with the ED and DMRG results in $4\times 4$ and larger sample sizes that the ground state 
of an even number of holes is generally nondegenerate. 

For comparison, the phase-string-free wave function $|\Psi_0\rangle_{{2h}}$ has an angular momentum $L_z=0$, 
while the same chirality wave function $|\Psi_{+}\rangle_{2h}$ corresponds to $L_z = \pm 1$ 
in which the two holes form a triplet pairing as indicated by the pair-pair correlator. 
Later, in Sec.~\ref{sec:4} and Appendix~\ref{app:3}, we further discuss the variational procedure to incorporate 
the additional feedback effect on $|\phi_0\rangle$ in Eq.~\eqref{eqn:doubleholeansatz}, 
which can then optimize $|\Psi_{+}\rangle_{{2h}}$ to result in an $L_z=2$ state with an energy comparable to that of $|\Psi_{-}\rangle_{2h}$. 
By contrast, $|\Psi_{-}\rangle_{2h}$ in Eq.~\eqref{eqn:doubleholeansatz}, 
with $|\phi_0\rangle$ as the half-filling ground state and $g_- (i,j)$ as the sole variational parameter, 
can naturally capture the nondegenerate two-hole ground state with $L_z=2$. 
Therefore, from now on, we mainly focus on such a variational $|\Psi_{-}\rangle_{{2h}}$ to further explore its internal structure and unconventional pairing properties. 

\begin{table}[tb]
    \caption{ Ground state energies of two-hole-doped $10\times 2$ ladder under the OBC obtained by VMC and ED. 
        $E_{{G}}$ is the total energy, $E_t$ and $E_J$ are the kinetic and the superexchange energies, respectively.}
    \label{tab:energy102}
    \begin{ruledtabular}
        \begin{tabular}{cccc}
              & $E_{{G}}$ & $E_t$ & $E_J$ \\
            \colrule
            $|\Psi_{0}\rangle_{{2h}}$ & $-21.03$ & $-6.29$ & $-14.73$ \\
            $|\Psi_{+}\rangle_{{2h}}$ & $-23.30$ & $-9.51$ & $-13.79$ \\
            $|\Psi_{-}\rangle_{{2h}}$ & $-25.70$ & $-11.07$ & $-14.63$ \\
            ED & $-27.74$  & $-14.00$ & $-13.74$ \\
        \end{tabular}	
    \end{ruledtabular}
\end{table}
Finally, we note that the ground state \emph{Ansatz} of Eq.~\eqref{eqn:doubleholeansatz} can be also directly applied for two holes injected into a two-leg Heisenberg ladder.  
The variational energy calculated on a $10\times 2$ ladder is shown in Table~\ref{tab:energy102}. 
Similar to the 2D square lattice, $|\Psi_{-}\rangle_{2h}$ for the two-leg ladder also gives rise to the best variational energy as compared with the ED result. 
Here, the phase-shift operator $\hat{\Omega}_i$ appearing in Eq.~\eqref{eqn:doubleholeansatz} may be also further \emph{optimized} for the two-leg ladder 
by tuning $\theta_i(l)$ under the constraint in Eq.~\eqref{theta}, 
which has been effectively done \cite{Wang2015} in the single-hole case as mentioned before. 
But, in the above VMC calculation, we still use the 2D version of $\theta_i(l)$, 
which is not qualitatively different from the optimized one for the isotropic $t$-$J$ model on the two-leg ladder \cite{Wang2015}. 
As already emphasized in the one-hole case, the crucial point is that the phase-shift factor $e^{\mp i \hat{\Omega}_i}$ under the constraint Eq.~\eqref{theta} 
can always explicitly keep track of the singular phase-string effect in Eq.~\eqref{eqn:doubleholeansatz}. 

\begin{figure}[tb]
    \centering
    \includegraphics[width=0.45\textwidth]{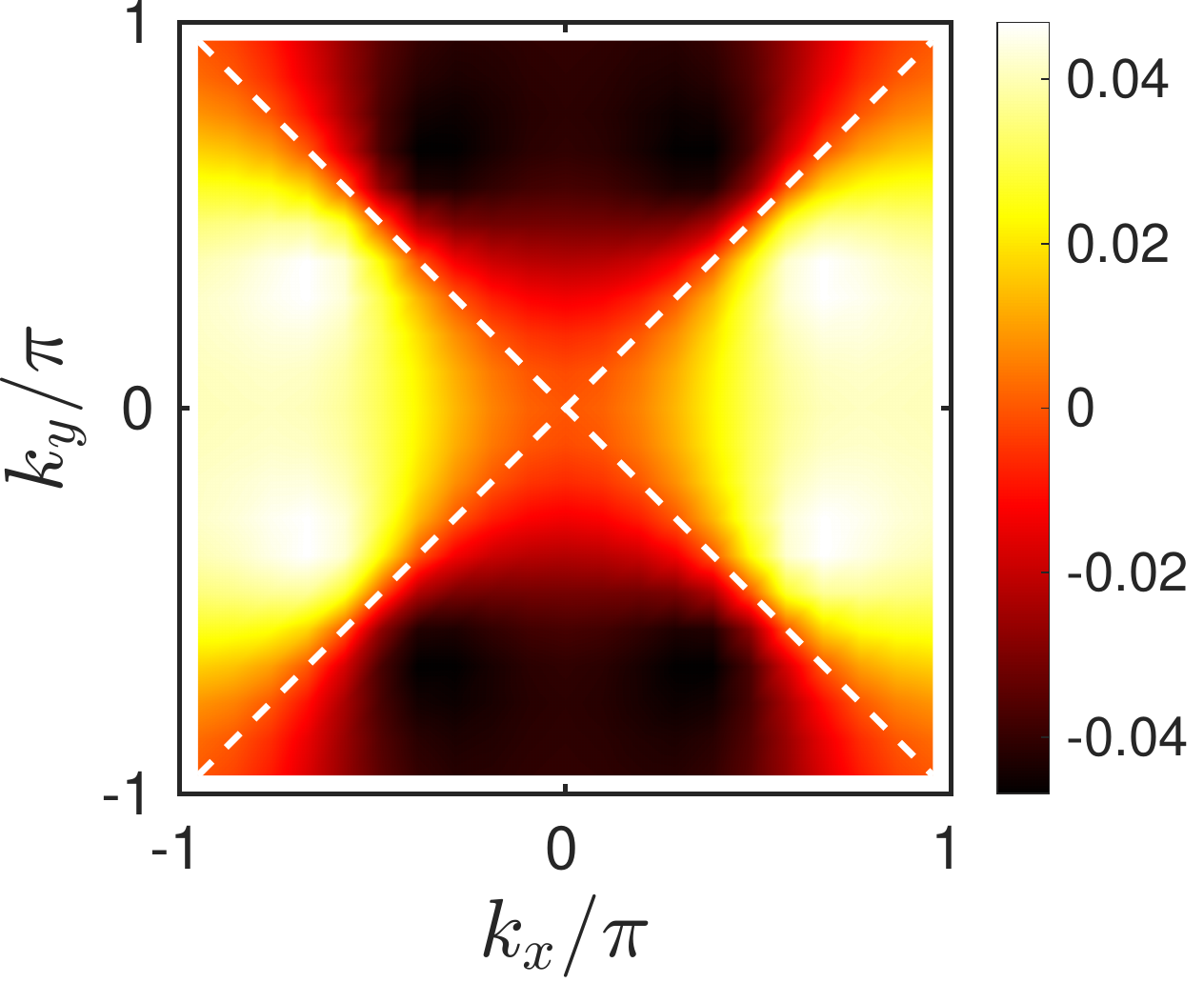}
    \caption{The $d$-wave symmetry of the Cooper pair component in the two-hole ground state $|\Psi_{-}\rangle_{{2h}}$, 
        as shown by the overlap $\Delta_{\mathbf{k}}^s$ defined in Eq.~\eqref{eqn:overlap} on a $20\times 20$ lattice. 
        The dashed (white) lines denote the nodal lines with the sign change across them.	}
    \label{fig:overlap2D}
\end{figure}

\begin{figure*}[thb]
    \centering
    \includegraphics[width=0.9\textwidth]{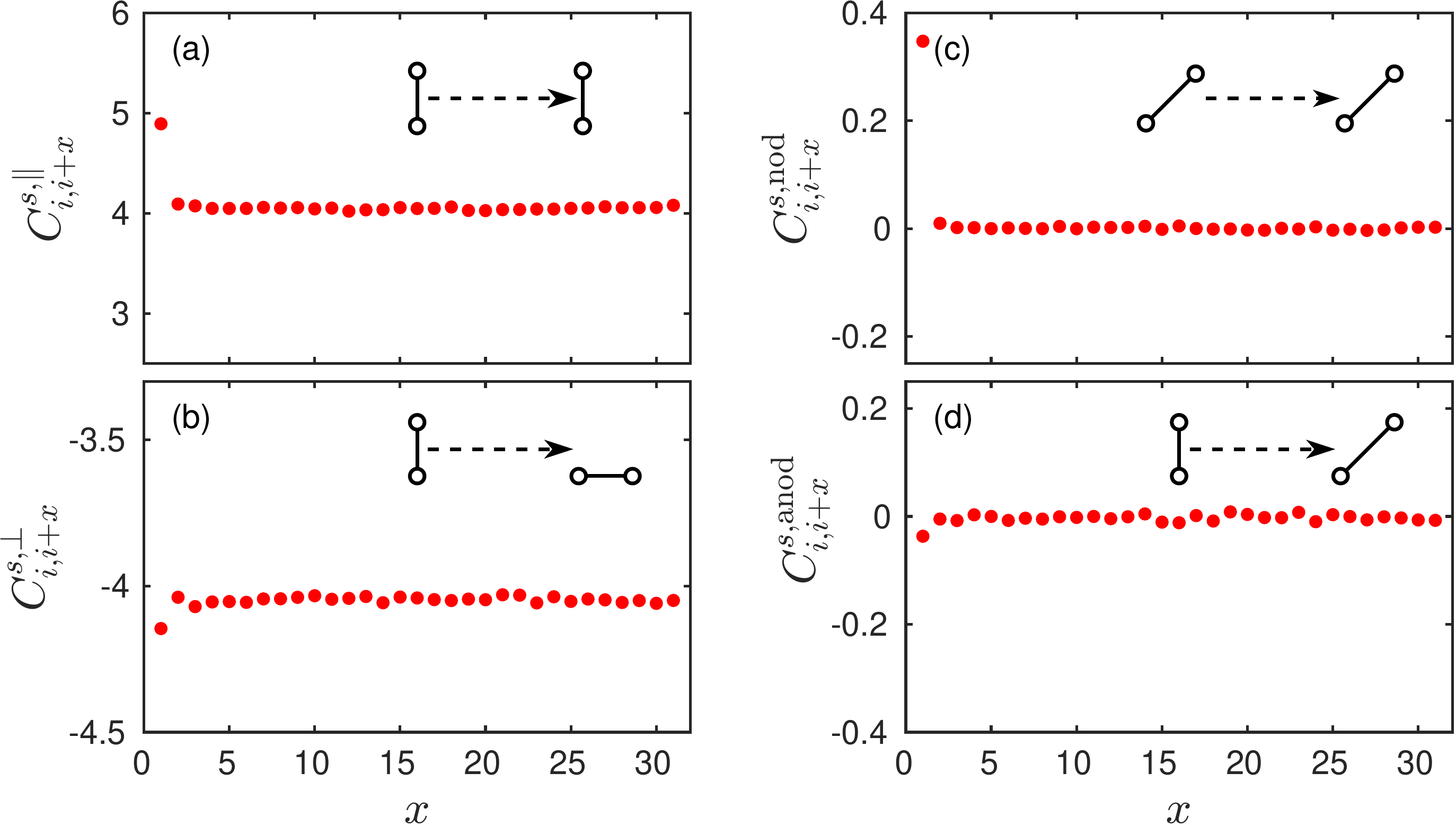}
    \caption{Pair-pair correlation in $|\Psi_{-}\rangle_{{2h}}$ with the pairing orientations indicated in the inset: 
        (a) and (b) show the correlations between the NN hole pairs at two perpendicular directions, 
        $C^{s,\parallel}_{i,j}$ and $C^{s,\perp}_{i,j}$, respectively; 
        (c) shows the NNN hole pair correlation $C^{s,\mathrm{nod}}_{i,j}$; 
        while (d) shows the correlation between the NN and NNN hole pairing, $C^{s,\mathrm{anod}}_{i,j}$. 
        The starting point $i$ is chosen at $i_x = N_x / 4 , i_y = N_y / 4$ with the ending point at $j_x = i_x + x , j_y = i_y$. 
        Note that the variational parameter $g_{-}(i,j)$ fitted in Eq.~\eqref{eqn:fitteddwave} with $|g_-(i,j)|$ set as 1 
        is used in the calculation of the correlators at a large lattice size $64\times 64$.}
    \label{fig:pairpair2D}
\end{figure*}

\section{Pairing structure}\label{sec:3}

In the previous section, we construct a two-hole variational ground state $|\Psi_{-}\rangle_{{2h}}$ in Eq.~\eqref{eqn:doubleholeansatz}, 
in which the pairing amplitude $g_- (i,j)$ is determined by VMC to result in a nondegenerate ground state consistent with the ED calculation. 
In the following, we explore the intrinsic pairing structure in such a ground state.  

\subsection{$d$-wave pairing symmetry}

Define the pair operator of the electrons as
\begin{equation} 
    \hat{\Delta}^s_{\mathbf{k}} = c_{\mathbf{k}\uparrow}c_{-\mathbf{k}\downarrow}-
    c_{\mathbf{k}\downarrow}c_{-\mathbf{k}\uparrow},
    \label{eqn:scoderparak}
\end{equation}
which involves two holes at momenta $\mathbf{k}$ and $-\mathbf{k}$ in a spin-singlet channel. 
Then, the pairing structure of the two-hole ground state $|\Psi_{-}\rangle_{{2h}}$ may be measured by 
\begin{equation}
    \Delta^s_{\mathbf{k}} \equiv \tensor[_{2h}]{\langle\Psi_-|\hat{\Delta}_{\mathbf{k}}^s|\phi_0\rangle}{}~,
    \label{eqn:overlap}
\end{equation}
which is just the usual definition of the Cooper pairing order parameter in the two-hole limit. 

The calculated $\Delta^s_{\mathbf{k}}$ on a $20\times 20$ lattice is shown in Fig.~\ref{fig:overlap2D}. 
The nodal lines of the pairing order parameter (white dashed lines) with a sign change after a $\pi/2$ rotation in the $\mathbf{k}$ space 
clearly illustrate a $d$-wave symmetry of $\Delta^s_{\mathbf{k}}$, 
which is also consistent with the angular momentum $L_z=2$ of $|\Psi_{-}\rangle_{{2h}}$. 

The $d$-wave pairing symmetry can be also directly probed in the pair-pair correlation function. Define the pair operator on lattice 
\begin{equation} 
    \hat{\Delta} ^s _{ij} = c^{}_{i\uparrow}c^{}_{j\downarrow} - c^{}_{i\downarrow}c^{}_{j\uparrow} 
    \label{eqn:scoderpara}
\end{equation}
and the corresponding pair-pair correlations 
\begin{equation}
    \begin{aligned} 
        C^{s,\parallel}_{i,j} =& \left\langle\hat{\Delta}^{s}_{i,i+\hat{\mathbf{e}}_y}\left(\hat{\Delta}^{s}_{j,j+\hat{\mathbf{e}}_y}\right)^\dagger\right\rangle,\\
        C^{s,\perp}_{i,j} =& \left\langle\hat{\Delta}^{s}_{i,i+\hat{\mathbf{e}}_y}\left(\hat{\Delta}^{s}_{j,j+\hat{\mathbf{e}}_x}\right)^\dagger\right\rangle,\\ 
        C^{s,\mathrm{anod}}_{i,j} = &\left\langle\hat{\Delta}^{s}_{i,i+\hat{\mathbf{e}}_y}\left(\hat{\Delta}^{s}_{j,j+\hat{\mathbf{e}}_x+\hat{\mathbf{e}}_y}\right)^\dagger\right\rangle,\\
        C^{s,\mathrm{nod}}_{i,j} = &\left\langle\hat{\Delta}^{s}_{i,i+\hat{\mathbf{e}}_x+\hat{\mathbf{e}}_y}\left(\hat{\Delta}^{s}_{j,j+\hat{\mathbf{e}}_x+\hat{\mathbf{e}}_y}\right)^\dagger\right\rangle,
    \end{aligned}
    \label{eqn:paircorrelator}
\end{equation}
for the NN bonds and the next-nearest-neighboring (NNN) bonds, respectively (cf. the insets in Fig.~\ref{fig:pairpair2D}). 
Here $\mathbf e_{x}$($\mathbf e_{y}$) denotes the unit vector along the $x$($y$) direction. 
As illustrated in Figs.~\ref{fig:pairpair2D}(a) and \ref{fig:pairpair2D}(b), the sign change between $C^{s,\parallel}_{i,i+x}$ and $C^{s,\perp}_{i,i+x}$ clearly confirms the $d$-wave symmetry. 

It is noted that in Fig.~\ref{fig:pairpair2D} a large 2D square lattice of $64\times 64$ is used to show the $d$-wave behavior in real space. 
Here, in carrying out the numerical calculation, we use a fitted parameter $g_-(i,j)$ for the NN and NNN sites, with the phases given by
\begin{equation}
    g_{-}(i,j) = e^{i\phi_{ij}^0}e^{-i\theta_i(j)}|g_-(i,j)| 
    \label{eqn:fitteddwave}
\end{equation}
where the first phase factor on the right-hand side of Eq.~\eqref{eqn:fitteddwave} is given by
\begin{equation}
    \phi^0_{ij}=\frac{1}{2}\sum_{l(\neq i,j)}[\theta_i(l)-\theta_j(l)].
    \label{eqn:phiij0}
\end{equation}
Namely, based on the variational results of the sample sizes up to $N=20\times 20$, 
the quantity $g_{-}(i,j) \times e^{-i\phi_{ij}^0}e^{i\theta_i(j)}$ is found to be a pure $s$-wave-like constant amplitude $|g_-(i,j)|$ (see below), 
up to a global phase, for both the NN and NNN bonds. 

As shown in Fig.~\ref{fig:pairpair2D}(c), the nodal direction pair-pair correlation function is actually finite at very short distance before decaying quickly to zero. 
It suggests that, in contrast to a pure $d$-wave symmetry of Cooper pairing, 
the two doped holes actually do form pairing along the diagonal direction with a finite $|g_-(i,j)|$, 
but its phase coherence is \emph{not} maintained over a couple of lattice constants because of the phase-shift factor 
$e^{\mp i\hat{\Omega}_i}$  in $|\Psi_{-}\rangle_{2h}$. 
In the following, we show a dichotomy in the pairing structure of $|\Psi_{-}\rangle_{2h}$, 
which significantly goes beyond a simple $d$-wave pairing of the bare holes.


\begin{figure}[thb]
    \centering
    \includegraphics[width=0.35\textwidth]{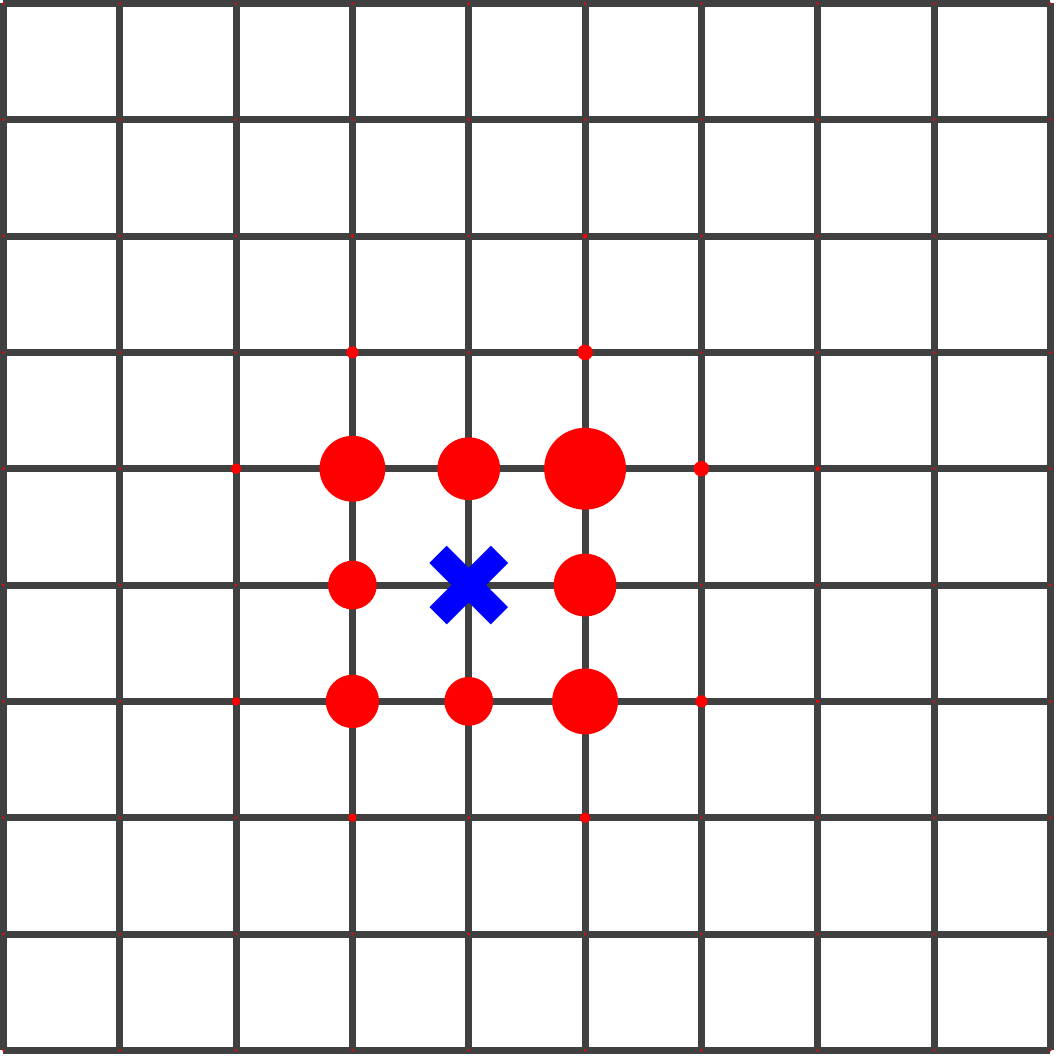} 
    \caption{ The hole-hole density correlator $\langle n^h_in^h_j\rangle$ of $|\Psi_{-}\rangle_{2h}$. 
        Here a hole is fixed at site $i$ (labeled by the blue cross), and $j$ runs over the other sites of the lattice. 
        The size of red bullets represents the strength of  the correlator, which indicates an $s$-wave-like tightly pairing of the two holes in a $10\times 10$ lattice.}
    \label{fig:nihnjh}
\end{figure}

\subsection{Dichotomy: $s$-wave-like pairing between the holes}

The tight binding between the two doped holes can be also directly measured by the hole-hole density correlation
\begin{equation}
    N_{ij}^h \equiv \langle n_i^h n_j^h\rangle,
    \label{eqn:nihnjh}
\end{equation}
where $n_i^h $ is the hole number operator at site $i$.  As shown in Fig.~\ref{fig:nihnjh}, two holes indeed form a tightly bound pair. 
For a hole sitting at a given (blue cross) site, 
the second hole distributes predominantly at the neighboring sites (red full dots), forming a square shape around it, 
which is much smaller than the whole lattice. 
Note that the weight on the square shape slightly deviates from a $C_4$ symmetry, because its center (blue cross) 
is off the center of the $10\times 10$ lattice in Fig.~\ref{fig:nihnjh}.

Figure \ref{fig:nihnjh} reveals a surprising fact that the hole pairing configuration actually resembles an $s$-wave-like pairing with 
the largest weight even at the four diagonal NNN sites of distance $\sqrt{2}$, 
where the weight is supposed to be zero according to a conventional $d$-wave pairing symmetry. 
Such an enhanced NNN diagonal pairing between the holes is already observed by earlier numerical calculations 
\cite{White1997,Chernyshev1998,Poilblanc1994,Sorella2002,Mezzacapo2016}.

By noting 
\begin{equation}
    N_{ij}^h \propto |g_{-}(i,j)|^2,
    \label{eqn:g2}
\end{equation}
such a generalized $s$-wave-like paring is, thus, related to $|g_{-}(i,j)|$, 
i.e., the pair amplitude for the twisted quasiholes created by $\tilde{c}_i=c_ie^{\mp i\hat{\Omega}_i}$ 
in the ground state $|\Psi_{-}\rangle_{2h}$ of Eq.~\eqref{eqn:doubleholeansatz}, as mentioned in the above subsection. 
Even though $|g_{-}(i,j)|$ is the largest at $j=i+\hat{\mathbf{e}}_x+\hat{\mathbf{e}}_y$, 
Fig.~\ref{fig:pairpair2D} (c) shows an exponential decay of the Cooper pair correlation along the diagonal direction beyond 
a couple of lattice constants due to $e^{\mp i\hat{\Omega}_i}$. 
Clearly, the phase-shift operator $e^{\mp i\hat{\Omega}_i}$ plays an essential role in the dichotomy of the pairing symmetry. 
In this sense, the $d$-wave symmetry of Cooper pair may be regarded as ``emergent'', 
which is in sharp contrast to the intrinsic $s$-wave-like pairing of the non-Landau quasiparticles described by $\tilde{c}$.

Therefore, the ground state $|\Psi_{-}\rangle_{2h}$ manifests a dichotomy in pairing symmetry, 
which indicates that the hole pairing in the doped $t$-$J$ model is non-BCS-like. 
Indeed, such a dichotomy has already been found in a two-leg $t$-$J$ model \cite{Chen2018} in the limit that the interchain hopping 
$t_{\perp}\rightarrow 0$ such that the phase-shift operator $\hat{\Omega}_i$ can be analytically derived in a one-dimensional form. 
In particular, the hole NNN pairing along the diagonal bond also gets substantially suppressed in sharp contrast to a much enhanced coherent NNN pairing 
in terms of the twisted quasiparticles explicitly identified there \cite{Chen2018}. 
In the following, we further examine the two-hole pairing state in the two-leg ladder in the isotropic limit of $t_{\perp}=t$ based on the present VMC scheme.

One may notice two weak maxima and minima points in the nodal regions in Fig.~\ref{fig:overlap2D},  which is in contrast to a simple BCS $d$-wave pairing $\Delta_k\sim \cos k_x - \cos k_y$. 
As emphasized above, the fundamental pairing in the two-hole ground state is actually an $s$-wave pairing of two twisted quasi-particles, 
which is exhibited as a strong real-space pairing with even a slight enhancement along the diagonal direction in Fig.~\ref{fig:nihnjh}. 
On the other hand, the overlap of the ground state with a conventional Cooper pair wave function indeed shows a $d$-wave symmetry, 
which is realized via the phase factor $e^{-i\hat{\Omega}}$ to suppress the diagonal amplitude. 
The maxima at, say, $k_x\sim 3\pi/4$ in Fig.~\ref{fig:overlap2D} should, thus, be a trade-off effect of such a $d$-wave phase suppression against the original strong $s$-wave pairing along the nodal direction.

\subsection{Two-hole ground state in the two-leg $t$-$J$ ladder}

\begin{figure}[tb]
    \centering
    \includegraphics[width=0.45\textwidth]{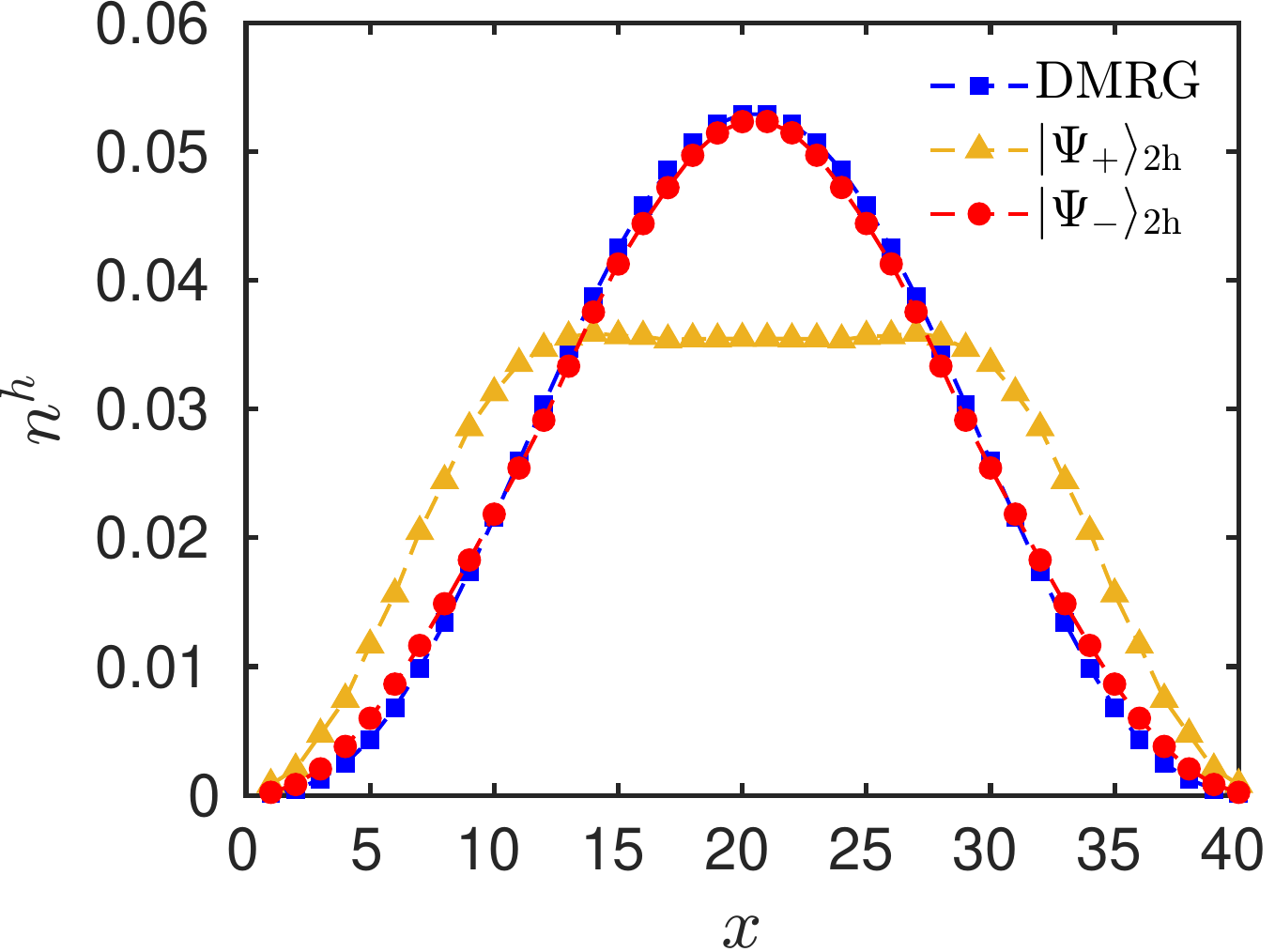} 
    \caption{ The density distribution of two holes in a $40\times 2$ ladder obtained by VMC and DMRG methods.}
    \label{fig:nh2legladder}
\end{figure}

Previously, we have briefly discussed applying the two-hole wave function \emph{Ansatz} in Eq.~\eqref{eqn:doubleholeansatz} 
to the case of an isotropic $t$-$J$ two-leg ladder. 
The VMC calculation of the variational energies in comparison with ED is presented in Table~\ref{tab:energy102}, 
in which it shows that $|\Psi_{-}\rangle_{2h}$ is the most competitive in the variational energy.

In Fig.~\ref{fig:nh2legladder}, the hole density distributions in a $40\times 2$ ladder obtained by both variational methods and DMRG are shown. 
It can be seen that $|\Psi_{-}\rangle_{2h}$ gives rise to almost the same hole distribution as the DMRG result. 

\begin{figure}[tb]
    \centering
    \includegraphics[width=0.45\textwidth]{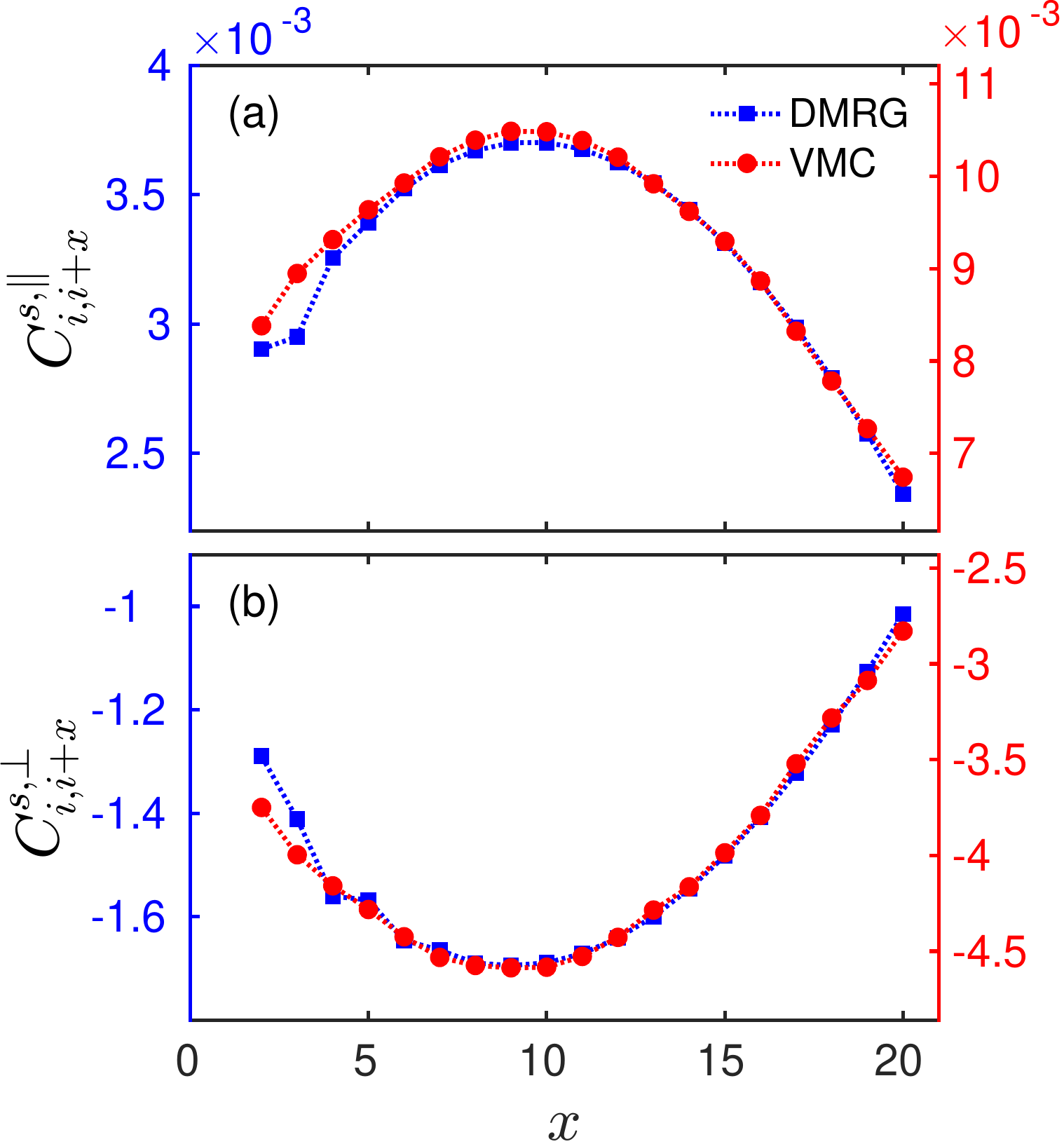} 
    \caption{ Pair-pair correlations in $|\Psi_{-}\rangle_{2h}$ on a $N=40\times 2$ ladder (red circle). 
        Here, $C^{s,\parallel}_{i,j}$ and $C^{s,\perp}_{i,j}$ shown in the upper and lower, respectively, indicate the $d$-wave-like pairing symmetry. 
        The DMRG result (blue square) is also shown for comparison. The starting point $i$ is chosen as $i_x = N_x / 4 , i_y = 1$, and the ending point is at $j_x = i_x + x , j_y = 1$.  }
    \label{fig:pair2legladder}
\end{figure}

The pair-pair correlation functions $C^{s,\parallel}_{i,j}$ and $C^{s,\perp}_{i,j}$ are also calculated 
for the two-leg ladder as shown in Fig.~\ref{fig:pair2legladder} for the ground state $|\Psi_-\rangle_{2h}$. 
The $d$-wave sign changes are clearly seen there. 
Here, the apophysis of the line shape in Fig.~\ref{fig:pair2legladder} may be related to the hole density distribution as indicated in Fig.~\ref{fig:nh2legladder}. 
The agreement of the line shape with the DMRG result in Fig.~\ref{fig:pair2legladder} implies that $|\Psi_-\rangle_{2h}$ 
indeed captures the long-wavelength physics of the two-hole ground state in an isotropic two-leg ladder. 
Note that the magnitude of the pair-pair correlation for $|\Psi_-\rangle_{2h}$ calculated by VMC is about a couple of times stronger than DMRG, 
which may be ascribed to a local longitudinal polaron effect in $|\phi_0\rangle$ not considered in the variational wave function. 
In particular, it includes the aforementioned phase-shift factor in the isotropic form, 
which should be adjusted for the two-leg case as shown before in the one-hole case \cite{Wang2015}. 

\begin{figure}[t]
	\centering
	\includegraphics[width=0.45\textwidth]{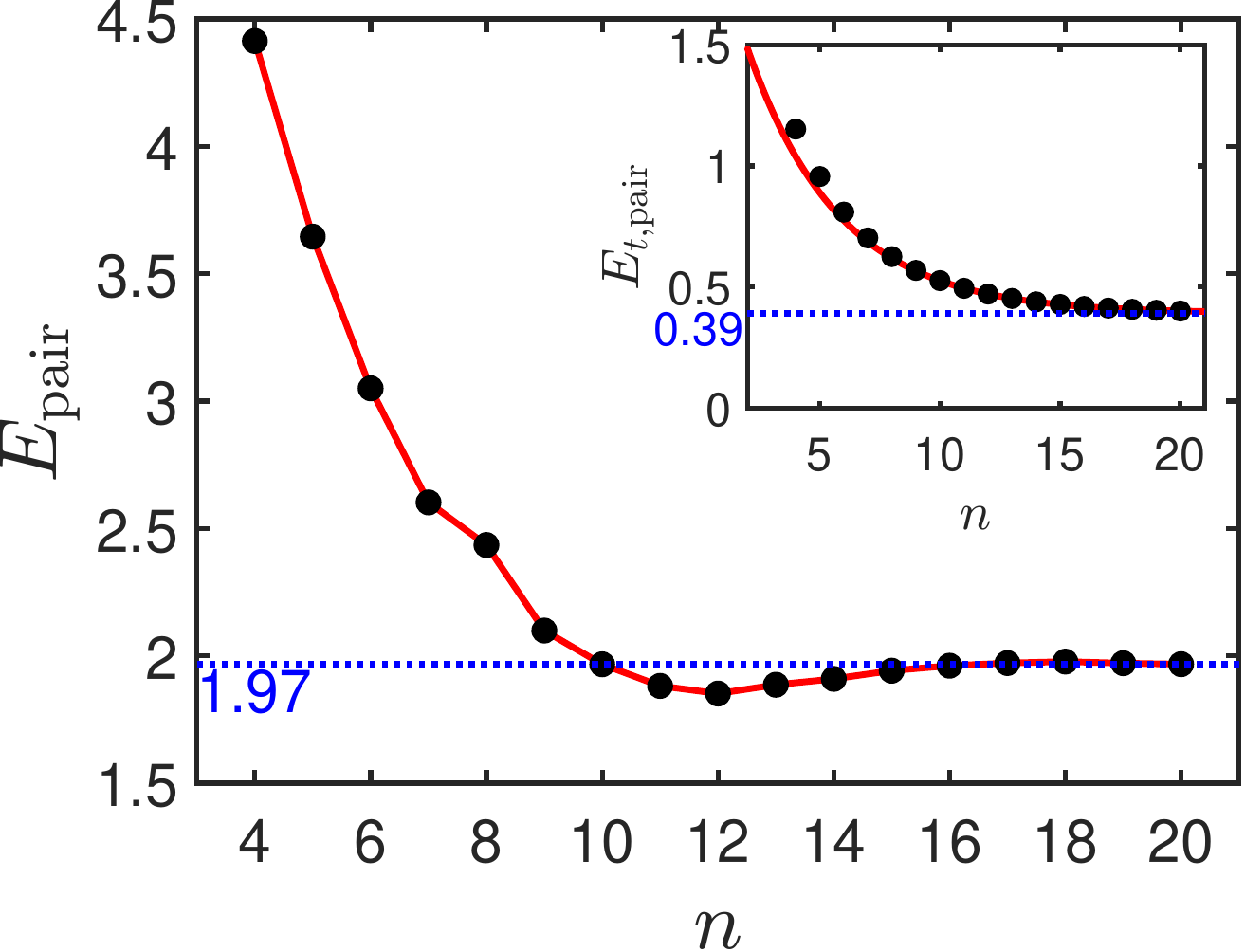} 
	\caption{The binding energy $E_{\mathrm{pair}}$ as a function of the side length $n$ of a square subregion with the two holes 
    confined inside which is embedded within a larger $24\times 24$ lattice of the spin background $|\phi_0\rangle$ as the half-filling ground state. 
    Inset: the kinetic energy part of the binding energy, $E_{t,\mathrm{pair}}$, versus $n$ with an exponential decay, 
    saturating at a typical area of approximately $4 \times 4$ as indicated by the fitting curve (red) given by Eq.~\eqref{Epair}. }
\label{fig:paire}
\end{figure}

\section{Pairing Mechanism}\label{sec:4}

In the last section, the pairing structure in the wave function \emph{Ansatz} $|\Psi_{-}\rangle_{2h}$ is investigated. 
A strong binding of the two doped holes is found with a novel feature of dichotomy in the pairing symmetry, which indicates a non-BCS nature of the hole pairing. 
In the following, we further explore the detailed pairing mechanism, 
which points to a new kinetic-energy-driven pairing force beyond the RVB pairing mechanism. 

\begin{figure*}[tb]
    \centering
    \includegraphics[width = 1\textwidth]{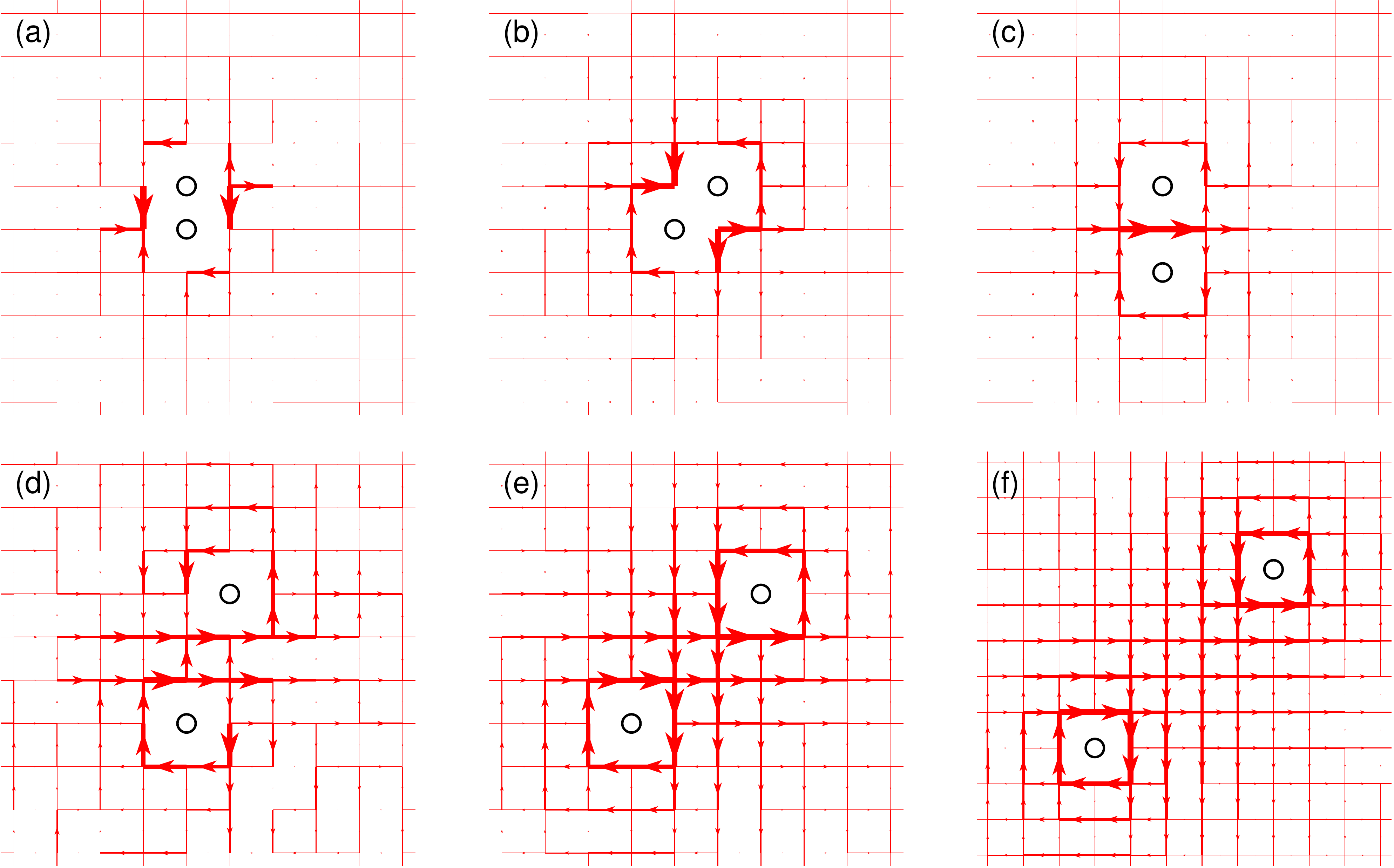}
    \caption{The spin current patterns in $|\Psi_{-}\rangle_{2h}$ with two holes projected to different locations on a $16\times 16$ lattice 
        (only the central part is shown). 
        The dipolar configuration of the spin currents is clearly indicated with the increase of the hole-hole distance from (a) to (f), 
        which illustrates the crucial role of the phase-string effect in the hole pairing as opposed to the absence of both in $|\Psi_{0}\rangle_{2h}$. 
        Note that an opposite chirality of the spin-current patterns is also present in $|\Psi_{-}\rangle_{2h}$, which is not shown here for simplicity. }
    \label{fig:spincurrent}
\end{figure*}

\subsection{Binding energy}
\label{sec:4A}

So far, we mainly focus on the two-hole ground state in the VMC calculation. 
Given the half-filling spin background $|\phi_0\rangle$, the variational procedure to determine $g_\eta(i,j)$ 
can be reduced to a quadratic eigenvalue problem as discussed in Appendix~\ref{app:2}. 
Consequently, a series of excited states in the form of two-hole \emph{Ansatz} wave function Eq.~\eqref{eqn:doubleholeansatz} 
may also be available if one restricts the Hilbert space to ignore spin excitations of the vacuum state $|\phi_0\rangle$ in Eq.~\eqref{eqn:doubleholeansatz}. 

For such low-lying excited states obtained, two holes remain in a tightly bound pair as indicated by the hole-hole density correlator similar to that shown 
in Fig.~\ref{fig:nihnjh} for the ground state. 
On the other hand, the hole pair becomes broken up in higher excited states, where no real space binding is seen in the hole-hole density correlators. 
For a finite-size sample, we can identify the energy difference between the first unpaired state and the ground state 
as an estimation of the binding energy $E_{\mathrm{pair}}$ for the hole pairing. 

In practice, to reduce the finite-size effect, we fix the half-filling ground state $|\phi_0\rangle$ obtained in a larger lattice size $N_x \times N_y$, say, at $24\times 24$. 
Then we calculate $E_{\mathrm{pair}}$ with the two holes restricted within a central subregion $n\times n$: $n\leq 24$. 
Figure~\ref{fig:paire} shows the scaling behavior of $E_{\mathrm{pair}}$ as a function of $n$, 
which saturates to $1.97$ in units of $J$. In the inset in Fig.~\ref{fig:paire}, 
a scaling behavior of the hopping energy $E_{t,\mathrm{pair}}$ is further shown. 
Note that, in the simplest VMC procedure, it is $E_{t,\mathrm{pair}}$ that is minimized since the vacuum state $|\phi_0\rangle$ is fixed. 
Here, $E_{t,\mathrm{pair}}$ saturates to a constant value with a length scale $\lambda_0$, following the scaling behavior of 
\begin{equation}\label{Epair}
    E_{t,\mathrm{pair}}=1.84e^{-n/{\lambda_0}}+0.39 
\end{equation}
in units of $J$ with $\lambda_0=3.82$ in the units of lattice constant, which corresponds to the typical pair size of approximately $4\times 4$.

Finally, we note that a self-consistent method is used to determine the binding energy within the wave function \emph{Ansatz} Eq. \eqref{eqn:doubleholeansatz}, 
by comparing the energy difference between the ground state and the first unpaired excited state. 
In this way, the binding energy is contributed solely by the phase-string effect via the phase-shift factor. 
A possible error may occur if the so-called longitudinal spin polaron effect as induced by the doped hole(s) in the spin background $|\phi_0\rangle$ also contributes to the pairing. 
In the present approach, such an effect is assumed negligible (cf. the discussion in Sec. \ref{sec:5} D).

\subsection{Pairing force and spin currents}



A rather tight spatial pairing in Fig.~\ref{fig:nihnjh} indicates a strong pairing force between two doped holes in an AFM spin background. 
Here, the phase-shift factor $e^{\mp i\hat{\Omega}_i}$ not only is essential in favoring the kinetic energy in a single-hole case, but also plays a critical role in the pairing of two holes. 
As a matter of fact, without incorporating it, the variational wave function $|\Psi_{0}\rangle_{2h}$ in Eq.~\eqref{eqn:bare2hole} 
cannot produce a sensible pairing even if the spin background $|\phi_0\rangle$ is artificially tuned from long-range to short-range spin-spin correlations. 
In other words, the pairing mechanism here cannot be attributed to the purely RVB pairing in $|\phi_0\rangle$. 
Rather it is due to the phase-shift $e^{\mp i\hat{\Omega}_i}$ in $|\Psi_{-}\rangle_{2h}$ [Eq.~\eqref{eqn:doubleholeansatz}] that represents the phase-string effect associated with the hopping term.

It has been previously shown that physically $e^{\mp i\hat{\Omega}_i}$ can be visualized by a spin-current vortex around a hole in the single-hole ground state 
(cf. Fig.~\ref{fig:oneholeil} and Ref.~\cite{Chen2019}). 
Similarly, we may measure the spin-current pattern for a \emph {fixed} configuration of two holes in $|\Psi_{-}\rangle_{2h}$. 
For each term with a given chirality on the right-hand side of Eq.~\eqref{eqn:doubleholeansatz}, 
one may compute the corresponding spin current based on $J_{ij}^s$ defined in Eq.~\eqref{eqn:spincurrent1}.

Figures~\ref{fig:spincurrent}(a)-\ref{fig:spincurrent}(f) show the patterns of the neutral spin current $\langle J_{ij}^s \rangle$ at different hole locations, 
contributed only by one of the terms with a given chirality of spin currents in Eq.~\eqref{eqn:doubleholeansatz}. 
The patterns with exactly an opposite chirality of spin currents in Eq.~\eqref{eqn:doubleholeansatz} are not shown for simplicity. 
It can be seen that the spin currents surrounding the two doped holes form rotonlike configurations 
with two holes sitting at the cores of the roton. 
The spin currents get strongly canceled as the two holes approach each other, 
while the spin-current vortices (antivortices) become stronger as the two holes are separated in a farther distance, approaching to the single-hole limit.

Such a non-RVB mechanism of the hole pairing has been first found \cite{Zhu2016} in the two-hole ground state of the two-leg $t$-$J$ ladder by the DMRG calculation, 
in which the strong binding between the holes is shown to disappear once the phase-string effect is precisely turned off 
without altering the short-range antiferromagnetic correlation of the spin background in the so-called $\sigma\cdot t$-$J$ model. 
Furthermore, in a simplified two-leg $t$-$J$ ladder with $t_{\perp}=0$, a stringlike pairing potential due to the phase-string effect 
can be explicitly identified \cite{Chen2018} in an analytical form, which results in the strong binding consistent with the DMRG result. 
In the present 2D square lattice case, the ED and DMRG calculations also show \cite{Zheng2018b} that, 
once the phase-string is removed in the $\sigma\cdot t$-$J$ model, the angular momentum of the two-hole ground state changes from $L_z=2$ 
to a trivial $L_z=0$ consistently with the description of $|\Psi_{0}\rangle_{2h}$ with diminished hole pairing. 

Therefore, we self-consistently establish a novel pairing mechanism for two holes injected into the Mott insulator as described by the $t$-$J$ model. 
The starting point is that a single doped hole does not propagate like a translational invariant Bloch wave 
and its wave function in Eq. (\ref{eqn:singleholeansatz}) must involve a spin-current vortex produced by the irreparable phase-string effect 
via the phase-shift factor $e^{\mp i\hat{\Omega}_i}$, which is essential to facilitate the hopping in an AFM spin background. 
Although the kinetic energy can be significantly lowered by such a phase-shift field, 
the single hole also becomes highly incoherent in sharp contrast to a coherent Landau quasiparticle. 
As such, it is shown by the VMC calculation that two doped holes show a strong incentive to form a tightly bound pair, 
which can further reduce the kinetic energy cost by eliminating the residual spin currents due to 
$e^{\mp i\hat{\Omega}_i}$ via the above roton pattern in the wave function $|\Psi_{-}\rangle_{2h}$.

\section{Discussion}\label{sec:5}

Given the above understanding of the hole pairing mechanism in the two-hole limit of a doped Mott insulator, 
in the following, we discuss several remaining issues that are beyond the scope of the present VMC approach.

\subsection{Single-particle coherence versus incoherence: A dual picture of AFM correlations}

So far, a good agreement between the VMC results and the exact numerics of ED and DMRG has been reached in a finite-size calculation. 
An important remaining issue is about the long-distance behavior of a hole or a pair of holes in the AFLRO phase. 
We point out below that the coherence or incoherence of the hole propagation 
is determined by the AFM correlations in the spin background in the long-wavelength limit. 

\begin{figure}[tb]
    \centering
    \includegraphics[width = 0.44\textwidth]{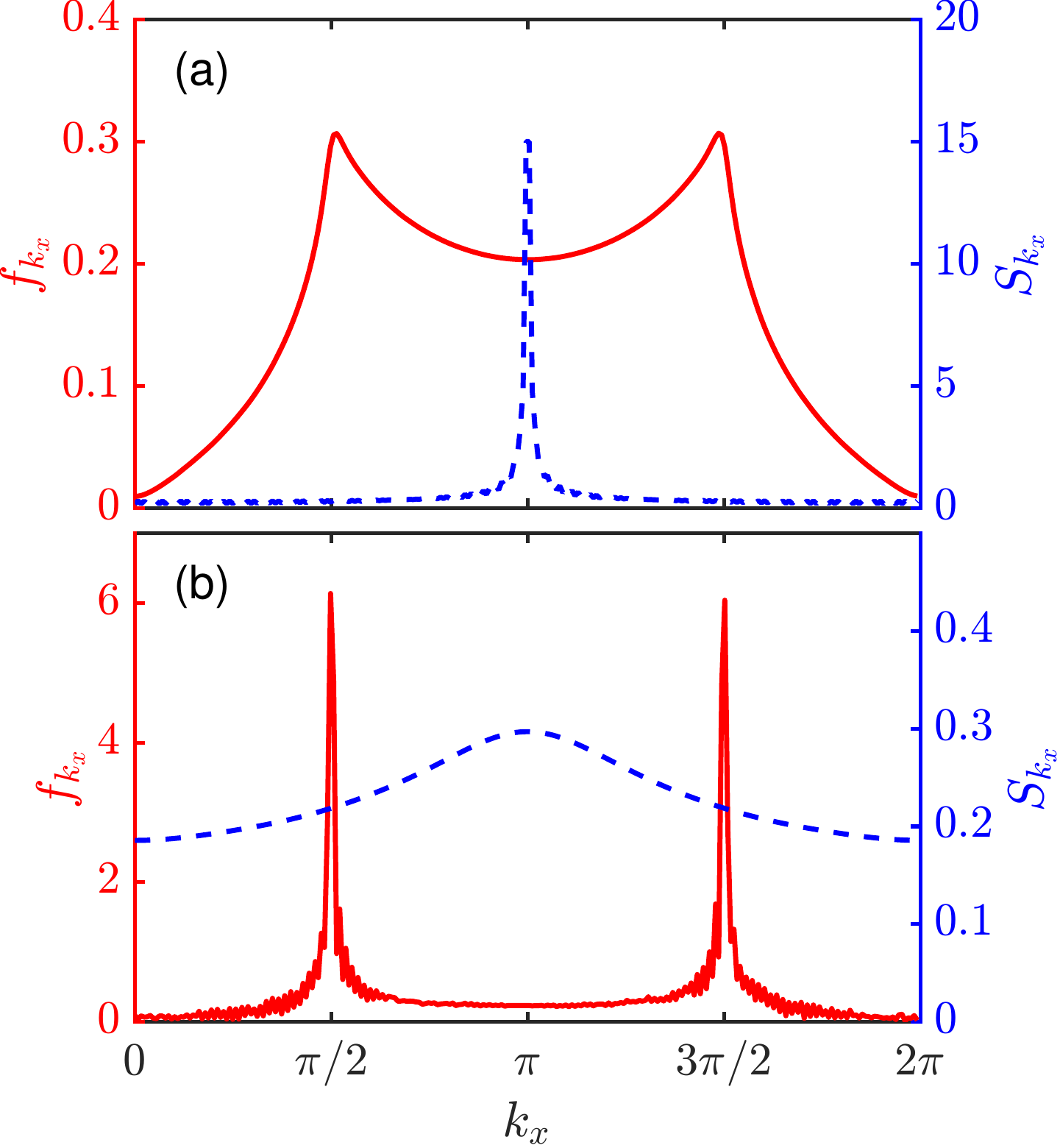}
    \caption{The mutual duality shown by the Fourier transformations, $f_{k_x}$ and $S_{k_x}$, of the gauge invariant vortex correlations $f_{jj'}$ in Eq.~\eqref{fij} (red) and the spin-spin correlation 
        $\langle\phi_0|\mathbf{S}_i\cdot\mathbf{S}_{j=i+x}|\phi_0\rangle$ (blue) along the $x$ direction, which are measured on (a) the AFLRO state and (b) an RVB state in the dimer limit.}
    \label{fig:phasek}
\end{figure}

The phase-shift operator $e^{\mp i\hat{\Omega}_i}$  plays an essential role in the construction of the ground states 
for both the single-hole and two-hole problems [cf. Eqs.~\eqref{eqn:1hgs} and ~\eqref{eqn:doubleholeansatz}]. 
For example, the single-hole propagator in the ground state of Eq.~\eqref{eqn:1hgs} 
\begin{equation}
    G_{jj'} \equiv \tensor[_{1h}]{ \langle\Psi_{\mathrm{G}}|c^{}_{j\sigma}c^\dagger_{j'\sigma} |\Psi_{\mathrm{G}}\rangle}{_{1h}}
\end{equation}
is proportional to
\begin{equation}
    f_{jj'}\equiv e^{\mp i\sum_{j\rightarrow j'}\phi_{kl}^0}
    \langle \phi_0|n_{j\sigma}e^{\pm i(\hat{\Omega}_j-\hat{\Omega}_{j'})}n_{j'\sigma}|\phi_0\rangle ~,
    \label{fij}
\end{equation}
which falls off exponentially as a function of $|i-j|$ if $|\phi_0\rangle$ is an AFLRO state 
but becomes a power-law decay if $|\phi_0\rangle$ is artificially tuned into a short-range RVB state, as seen in the VMC simulation below. 
Here the phase factor $\sum_{j\rightarrow j'}\phi_{kl}^0$ is added to replace the variational parameters $\phi_h^*(j)\phi^{}_h(j')$ to make $f_{jj'}$ gauge invariant, 
which is a summation over the link variables $\phi_{kl}^0$ [cf. Eq.~\eqref{eqn:phiij0}] connecting $i$ and $j$. 
For simplicity, one may consider $j=i+x$ along the $x$ axis and calculate the following  
Fourier transformation of Eq.~\eqref{fij} along the $k_{x}$ direction 
\begin{equation}
    f_{k_x} \equiv \sum_x e^{-ik_xx}f_{j,j+x}~,
\end{equation}
as shown in Fig.~\ref{fig:phasek}, where $f_{k_x}$ peaks at $\pm \pi/2$ are substantially broadened in the AFLRO state 
but become very sharp in a short-range RVB state of $|\phi_0\rangle$ in a dimer limit. 

Thus, the single-hole behavior as controlled by the long-range behavior of $ f_{jj'}$, 
which is, in turn, decided by the spin-spin correlation in $|\phi_0\rangle$ via the phase-shift factor  $e^{\mp i\hat{\Omega}_i}$, 
should be rather incoherent in an AFLRO background, while it becomes coherent once the AFLRO disappears in a short-range RVB state of $|\phi_0\rangle$. 

\subsection{Possible localization in the AFLRO state}

Two holes are shown to form a tightly bound pair in Fig.~\ref{fig:nihnjh}. 
It means that one may reasonably treat such a pairing entity as a building block with more holes doped into the antiferromagnet in the dilute limit 
so long as the spin-spin correlation length is much larger than the pair size. 
As emphasized before, $|\Psi_{-}\rangle_{2h}$ involves two \emph{opposite} vortices of spin currents 
that are canceled out in a distance far away from the hole pair (cf. Fig.~\ref{fig:spincurrent}). 
Thus, the long-range AFM correlation should not be affected by the presence of such a tightly bound hole pair. 
In other words, the AFLRO is expected to persist in the presence of a finite concentration of such hole pairs 
each with a scale of approximately $4\times 4$. 

Naively, if a hole pair can propagate coherently, a finite density of them should be superconducting. 
However, in the AFLRO state, the incoherence of a single hole discussed above based on the long-distance behavior of $f_{ij}$ 
indicates the self-localization of the doped hole. 
Similarly, the long-wavelength coherence or incoherence of the hole pair should be also sensitive to the spin-spin correlation in the background. 

Nevertheless, the very mechanism of pairing is to remove the leading frustration effect of $e^{\mp i\hat{\Omega}_i}$ 
in the ground state of Eq.~\eqref{eqn:doubleholeansatz}. 
The question is if there would be still a residual effect of the phase-string effect not canceled out by the pairing 
to further localize the hole pair in the AFLRO phase. 
Here, we point out that the conserved spin currents associated with the holes become dissipationless ``supercurrents'' in the AFLRO phase. 
On a general ground, the vortex-antivortex configuration of spin current or the roton in Fig.~\ref{fig:spincurrent} 
is expected to be an immobile object \cite{Feynman1956}, in contrast to a ``polaron'' of spin amplitude distortion 
\emph{tightly} bound to the hole pair to form a rigid translational invariant object. 
In particular, $|\Psi_{-}\rangle_{2h}$ also involves an opposite chirality of spin currents (not shown in Fig.~\ref{fig:spincurrent}) 
as the superposition state in Eq.~\eqref{eqn:doubleholeansatz}. 
Thus, as a whole, such a hole pair state is not translationally invariant, which indirectly implies a tendency for the self-trapping 
of the hole pair once the spin background becomes AFM ordered. 

Furthermore, locally the NN and NNN pairings in $|\Psi_{-}\rangle_{2h}$ are closely connected by the hopping term, 
which even makes the NNN pairing stronger (cf. Fig.~\ref{fig:nihnjh}). 
But only the NN Cooper pair shows a longer-range coherence in Fig.~\ref{fig:pairpair2D}(a), 
whereas a quick exponential decay for the NNN Cooper pair in Fig.~\ref{fig:pairpair2D}(c) clearly indicates the incoherence or localization 
of the NNN component caused by the phase-shift field. 
It explains and reconciles the dichotomy in the pairing symmetries of the $s$-wave-like local pairing versus the $d$-wave-like Cooper paring in large distance. 
Then, as the superposition of both delocalized and localized components, the hole pair as a whole quantum entity is generally expected to be localized too 
in a long distance, although a more careful study beyond the present VMC approach is further needed to resolve this important issue. 


\subsection{The ground state with more doped holes}

How can the present approach to the two-hole problem provide an understanding of the SC state \emph{Ansatz} 
in Eq.~\eqref{eqn:psgs} at finite doping?

Note that, in the hole pair creation operator $\hat{\cal D}$ defined in Eq.~\eqref{eqn:D} for the SC state, 
$\tilde{c}_{i\sigma}$ is defined in Eq.~\eqref{eqn:ctilde} in terms of the \emph{same} phase-shift factor $e^{\mp i\hat{\Omega}_i}$ such that the translational symmetry is formally retained in the ground state \eqref{eqn:psgs} 
if the short-range AFM vacuum state $|\mathrm{RVB}\rangle$ is translationally invariant. 
However, such a state in the two-hole limit reduces to $\hat{\cal D}  |\mathrm{RVB}\rangle $, which looks more like $|\Psi_{+}\rangle_{2h}$ 
[cf. Eq.~\eqref{eqn:doubleholeansatz} with $\eta=+1$] than the true ground state $|\Psi_{-}\rangle_{2h}$ studied in the present work. 
But we point out below that the distinction between $|\mathrm{RVB}\rangle$ and $|\phi_0\rangle$ can reconcile this difference, 
which also means that there must be a phase transition to separate these two phases.

By a reorganization in Eq.~\eqref{eqn:fitteddwave}, the ground state $|\Psi_-\rangle_{2h}$
may be rewritten as 
\begin{equation}
  \begin{aligned}
       |\Psi_-\rangle_{2h}=& \sum_{ij}g_-(i,j) \tilde{c}_{i\uparrow}^{}\tilde{c}_{j\downarrow}^{}
        \left[e^{\pm 2i\hat{\Omega}_i}|\phi_0\rangle\right]+\cdots \\
          \end{aligned}
    \label{eqn:diff2same}
\end{equation}
by introducing $\tilde{c}_{i\alpha}^{}={c}_{i\alpha}^{}e^{\mp i\hat{\Omega}_i}$, which may be further expressed as 
\begin{equation} 
    \begin{aligned}
        &|\Psi_-\rangle_{2h} \rightarrow& \sum_{ij}g_-(i,j)\tilde{c}_{i\uparrow}^{}\tilde{c}_{j\downarrow}^{}|\mathrm{RVB}\rangle +\cdots, 
    \end{aligned}
    \label{psai+} 
\end{equation}
by denoting $e^{\pm 2i\hat{\Omega}_i}|\phi_0\rangle \rightarrow |\mathrm{RVB}\rangle $.
Note that Eq.~\eqref{psai+} would also resemble $|\Psi_{+}\rangle_{2h}$ defined in Eq.~\eqref{eqn:doubleholeansatz} ($\eta=+1$) if
$|\phi_0\rangle$ is replaced by $|\mathrm{RVB} \rangle \sim e^{\pm 2i\hat{\Omega}_i}|\phi_0\rangle $ there. 
Although the original $|\Psi_{+}\rangle_{2h}$ has a higher VMC energy and a distinct quantum number $L_z=\pm 1$ than that of $|\Psi_{-}\rangle_{2h}$ with $|\phi_0\rangle$ as an AFLRO state, 
now with $|\phi_0\rangle\rightarrow |\mathrm{RVB}\rangle$, one may repeat a VMC procedure for such a new state given in Eq.~\eqref{psai+}. 
Here, both $|\mathrm{RVB}\rangle$ and parameter $g_{-}(i,j)$ are treated variationally, and, in particular, 
$|\mathrm{RVB}\rangle$ satisfies an eigenequation in the variational procedure that can be diagonalized by ED with each given $g_{-}(i,j)$ in an iteration procedure of the VMC 
(details are given in Appendix \ref{app:3}). 

As a result, a nondegenerate $|\Psi_+\rangle_{2h}$ can be found with the correct angular momentum $L_z=2 \mod 4$, 
consistent with $|\mathrm{RVB} \rangle \sim e^{\pm 2i\hat{\Omega}_i}|\phi_0\rangle $ in Eq.~\eqref{psai+}. 
The variational kinetic energy of approximately $-11.94 J$, is even slightly lower than that of the ground state $|\Psi_{-}\rangle_{2h}$ shown in Table~\ref{tab:energy44}, 
with a slightly higher superexchange energy of approximately $-10.35 J$. 
Furthermore, it is easy to find the overlap $\langle \mathrm{RVB} |\phi_0\rangle \sim 10^{-13}$, i.e., almost orthogonal, and, 
on the other hand, $\langle \mathrm{RVB}|e^{\pm 2i\hat{\Omega}_j}|\phi_0\rangle \sim 0.2$, which is of order of one to validate Eq.~\eqref{psai+}.

With the pairing creation operator in Eq.~\eqref{eqn:D},  
a generalization of Eq.~\eqref{psai+} implies a ground \emph{Ansatz} for the $N_h$ hole case: 
\begin{equation}
    |\Psi_{{G}}\rangle\propto \hat{\cal D}^{N_h/2}|\mathrm{RVB}\rangle,
    \label{eqn:gs} 	
\end{equation}
which is essentially the same as Eq.~\eqref{eqn:psgs} with the total hole number fixed at a finite doping concentration, where
\begin{equation}
    |\mathrm{RVB}\rangle {\sim} e^{i\sum_{ij}n^h_i\hat{\Omega}_j} |\phi_0\rangle.
    \label{eqn:modifiedRVB}
\end{equation}
At finite dopings, $ |\mathrm{RVB}\rangle$ can be shown to become short ranged by a mean-field theory \cite{Weng1998,*Weng1999}. 
This is the reason why the vacuum state in Eq.~\eqref{eqn:gs} is denoted by $ |\mathrm{RVB}\rangle$ to distinguish it from $|\phi_0\rangle$ 
of the AFLRO state of the Heisenberg model at no doping. 

An off diagonal long range order of the pairing amplitude can be achieved by $\langle \hat{\cal D}\rangle \neq 0$ in Eq.~\eqref{eqn:gs} at finite doping \cite{Weng2011a}. 
According to Eq.~\eqref{psai+}, the local pairing symmetry should resemble that of $|\Psi_{-}\rangle_{2h} $ for two holes 
discussed in the present work so long as the short-range AFM correlations persist in $|\mathrm{RVB}\rangle$. 
Indeed, a dichotomy of an $s$-wave symmetry of the pairing amplitude ${g}(i,j)$ for the twisted quasiholes created by $\hat{\cal D}$ in Eq.~\eqref{eqn:modifiedRVB}
and a $d$-wave symmetry of the Cooper pairing order parameter have been previously found in a generalized mean-field theory \cite{Ma2014} leading to Eq.~\eqref{eqn:gs}. 

Therefore, by properly incorporating the hole-induced phase-shift field $e^{\mp i\hat{\Omega}_i}$, 
an evolution of the ground state from the AFLRO to the SC as a function of doping may be mathematically realizable, 
although a phase transition should occur from long-range AFM ordered $|\phi_0\rangle$ to an emergent RVB state $|\mathrm{RVB}\rangle$. 
It leads to a non-BCS-like SC wave function \eqref{eqn:gs} at finite doping, in which the local pairing structure should be similar to that of 
$|\Psi_{-}\rangle_{2h}$ studied in the AFLRO phase in the present work, but it is in a translational invariant form in contrast to the latter. 
In particular, the AFLRO must be replaced by an \emph{emergent} RVB state self-consistently to establish the transition of the charge pairing from localized state to superconducting. 

A possible non-Landau-type phase transition has been studied before in a topological gauge theory formulation \cite{Kou2003a, Ye2011}. 
The hole pairing in the AFLRO phase with a phase transition to the SC phase is also discussed in the semiclassical form of vortex-antivortex pairing 
\cite{Kou2003b,Capati2015}. 
Such long-wavelength physics in the dilute concentration of holes is beyond the scope of the present two-hole VMC work. 
Nevertheless, the short-distance pairing structure 
is expected to remain similar as characterized by Eq.~\eqref{eqn:doubleholeansatz}. 
A further exploration of the phases and the phase transition in the dilute doping regime should first incorporate this short-range physics correctly.

\subsection{Comparison with other approaches and experiments}
\label{sec:5D}

As the building block of superconductivity, the pairing mechanism of doped holes has been under intensive study. 
Numerical methods, including ED \cite{Dagotto1990,Poilblanc1993,Poilblanc1994,Chernyshev1998}, 
DMRG \cite{White1997,Zhu2014,Jiang2018,Jiang2019}, and VMC \cite{Shih1998,Sorella2002,Misawa2014,Mezzacapo2016}, 
are employed to analyze the pairing structure of the two-hole ground state and excitations. 
Consistent results, including a strong pairing of $d$-wave symmetry and a maximal hole density correlator at distance $\sqrt{2}$, have been obtained. 
In comparison with these numerical results, the variational wave function Eq.~\eqref{eqn:2hgs} proposed in this work 
captures essentially all the important features with using only a few number of variational parameters 
(approximately $N^2$ where $N = N_x\times N_y$ is the lattice site number). 

The structure of the two-hole ground state in Eq. (\ref{eqn:doubleholeansatz}) ($\eta=-1$) is basically fixed with the fitted form of the 
variation parameter $g_-(i,j) $ given in Eq.~\eqref{eqn:fitteddwave}. 
Introducing more variational parameters including considering the ``spin polaron'' effect in $|\phi_0\rangle$ may lead to a better variational ground state. 
However, the most essential features, i.e., the dichotomy in pairing structure, 
with a $d$-wave-like pairing symmetry in terms of the bare electron operators and the $s$-wave-like real space pairing of the holes, 
is already well captured in the simplest variational \emph{Ansatz}. 
The phase-shift field $\hat{\Omega}_i$ in Eq. (\ref{eqn:doubleholeansatz}) plays a fundamental role here, 
which keeps track of the singular phase-string effect.

It has been long believed that the superexchange interaction plays the role of the driven force for the pairing in the cuprate 
in place of the electron-phonon interaction in conventional BCS theory. 
The pairing mechanisms due to the RVB spin pairing \cite{Anderson1987,Baskaran1987,Anderson2004}, 
the semiclassical long-range spin texture \cite{Shraiman1989b,Shraiman1990}, or the ``spin bag'' \cite{Schrieffer1988,Weng1990}, 
were proposed and extensively explored. 
However, the recent DMRG calculation for the two-leg ladder reveals \cite{Zhu2014} that the superexchange interaction is only a necessary condition. 
By comparing the $t$-$J$ model and the so-called $\sigma\cdot t$-$J$ model, in which the superexchange term is the same, 
it shows that the phase-string sign structure induced by the hopping term in the former is sufficient to lead to the formation of strong hole pairing, 
whereas both are absent in the latter. 
The present results clearly support this opinion, in which the phase-string as encoded in the phase-shift factor 
$e^{\mp i\hat{\Omega}_i}$ is crucial in the variational two-hole wave function $|\Psi_{-}\rangle_{2h}$ for the $t$-$J$ model, 
including the strong binding of two holes. 
But, due to the absence of $e^{\mp i\hat{\Omega}_i}$ in $|\Psi_{0}\rangle_{2h}$, which works for the $\sigma\cdot t$-$J$ model, 
the pairing is diminished even if the superexchange term may be optimized via the vacuum $|\phi_0\rangle$. 

Finally, we briefly discuss some direct experimental implications based on the understanding reached in the present two-hole study. 
It has been a long-standing puzzle why the nodal $d$-wave superconductivity in the cuprates is so stable against nonmagnetic impurities and disorders. 
Here, the dichotomy in the pairing symmetry provides a natural explanation. 
Namely, it may be the intrinsic $s$-wave pairing of the doped holes, 
which is expected to remain well present with the reduction of the spin correlation length to a finite larger size at finite doping, 
that is robust against the strong potential disorders. 
On the other hand, generally, a $d$-wave pairing symmetry should be detected via a single-particle probe like tunneling and photoemission experiments; 
the $s$-wave-like pairing structure of the twisted holes may also be observed via a scanning tunneling microscopy (STM) 
if \emph{directly} probing into the $\mathrm{CuO_2}$ layer where the strong correlation takes place. 
There have been already some recent STM experiments reporting a possible $s$ wave or the so-called U-shape $dI/dV$ signal 
coexistent with the $d$ wave or V-shape signal in some specially designed setups of the cuprates \cite{Zhong2016,Ren2016,Zhu2021}. 
More careful theoretical examinations on these issues are needed in the future.

\subsection{Effect of the NNN hopping integral $t'$}
\begin{figure}[tb]
    \centering
    \includegraphics[width = 0.5\textwidth]{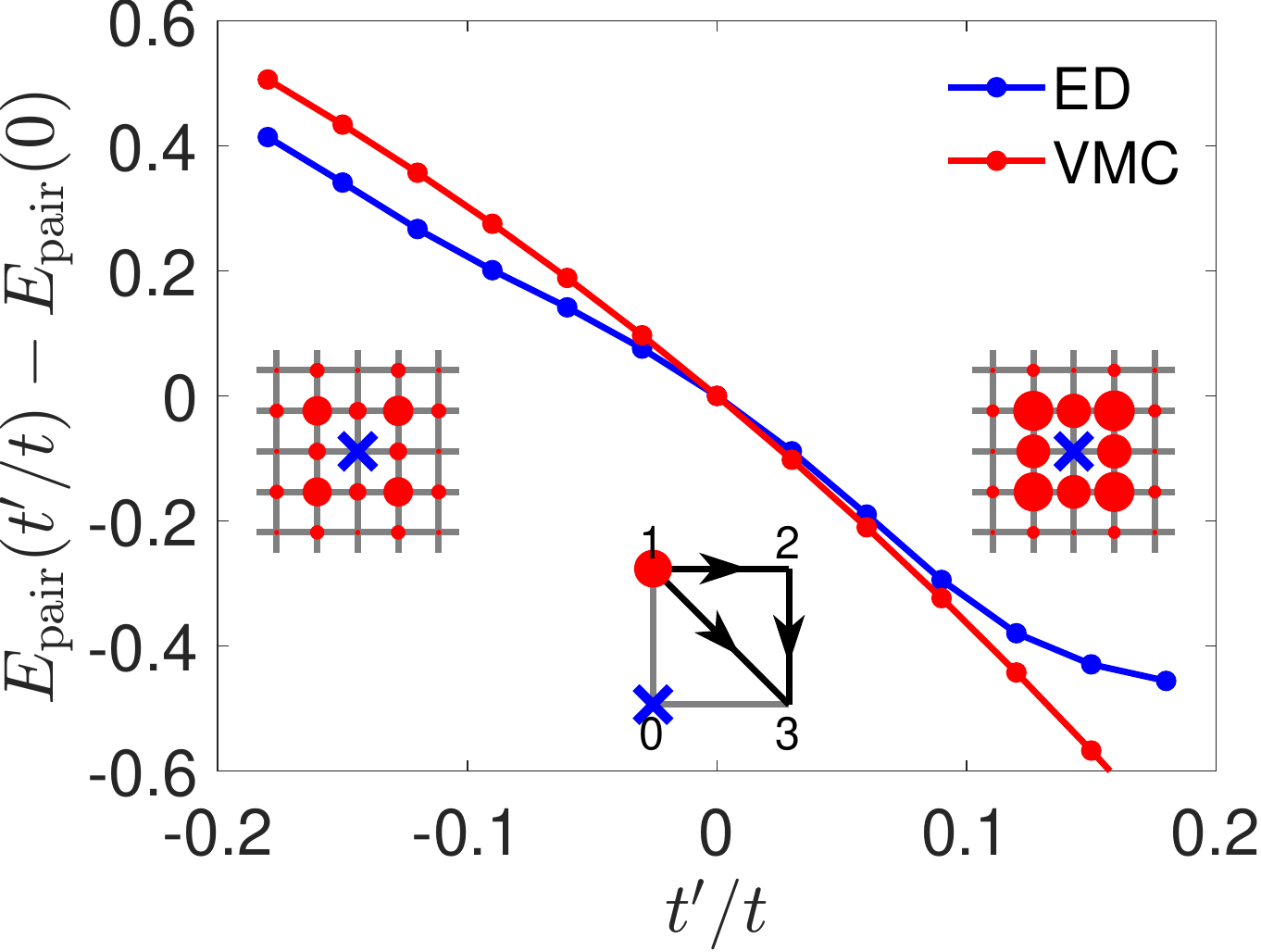}
    \caption{The two-hole pairing energy shift in $E_{\mathrm{pair}}(t'/t)$ as a function of $t'/t$ due to the NNN hopping $t'$ term added to the $t$-$J$ model. 
        The VMC result (red dot) is determined by the same method used in Sec.~\ref{sec:4} with the largest subregion $n = 20$ 
        which is almost saturated to the extrapolation value. 
        Insets: Similar to the plot in Fig.~\ref{fig:nihnjh}, the blue cross denotes one hole position and the red dots the distributions of the other hole 
        as given by the hole-hole density correlator $\langle n_i^hn_j^h\rangle $ at $t'/t=-0.2$ (left top) and $t'/t=0.2$ (right top), respectively. 
        The middle bottom inset further indicates the additional hopping of the hole along the diagonal direction (see the text). 
        The ED result (blue dot) is also shown in the main panel as calculated by the definition in Eq.~\eqref{eqn:ebound} on a $4\times 4$ lattice with the OBC 
        (which is comparable to the intrinsic size of the bound pair determined by VMC). }
    \label{fig:deltatp}
\end{figure}

Besides the competition between the AFM background and the NN hopping of the holes in the $t$-$J$ model, 
in a realistic case like the cuprate compound, longer-range hopping such as the NNN hopping $t'$ is also present as given by
\begin{equation}
    H_{t'} = -t'\sum_{\langle\langle ij\rangle\rangle,\sigma} (c^\dagger_{i\sigma}c^{}_{j\sigma}+H.c.),
    \label{eqn:htp}
\end{equation}
where $\langle\langle ij\rangle\rangle$ denotes the NNN sites. 

At small $t'/t$, one may consider $H_{t'}$ as a perturbation to the $t$-$J$ model, and the novel pairing mechanism found in the latter is not expected to be drastically changed. To the leading order of approximation, one may still treat the effect of NNN hopping term within the VMC scheme based on the two-hole variational \emph{Ansatz} studied in the $t$-$J$ case above. 

The resulting shift in the pairing energy $E_{\mathrm{pair}}$ due to $H_{t'}$ is shown in Fig.~\ref{fig:deltatp} as a function of $t'/t$. The VMC method to determine $E_{\mathrm{pair}}(t'/t)$ is the same as that stated in Sec.~\ref{sec:4A}, i.e. by calculating the energy difference between the first unpaired state and the ground state. 
Here only the largest subregion $n = 20$ in Fig.~\ref{fig:paire} is shown which is already quite close to the saturation value. 
In Fig.~\ref{fig:deltatp}, the pairing energy shift as calculated by ED is also shown for comparison, which is
based on a more conventional definition 
\begin{equation}
    \label{eqn:ebound}
    E_{\mathrm{pair}} = E_{2h}+E_{1h}-2E_{1h}~,
\end{equation}
where $E_{2h}$, $E_{1h}$, and $E_{1h}$ denote the two-hole, one-hole, and half-filling ground state energies, respectively. 
Although two methods are different, the pairing energy difference 
$E_{\mathrm{pair}}(t'/t) - E_{\mathrm{pair}}(0)$ shown in Fig.~\ref{fig:deltatp} agrees with each other remarkably at small $|t'/t|$. 
Qualitatively the binding between the two doped holes becomes stronger as $t'/t$ increases from $t'/t<0$ to $t'/t>0$. 
Here, the ED is carried out on a $4\times4$ lattice under the OBC, which is comparable to the intrinsic size of the two-hole bound state found in the VMC calculation. 
In the latter, the real space hole-hole density correlator $\langle n_i^hn_j^h\rangle$ shows a tighter binding with the increase of $t'/t$, 
as indicated in the two insets in Fig.~\ref{fig:deltatp} from top left to top right
with one hole projected at one site (blue cross). 

Physically, a positive $t'/t$ term favors the $1\rightarrow 3$ hopping process shown in the middle bottom inset in Fig.~\ref{fig:deltatp}, which corresponds to a matrix element $-t'<0$. On the other hand, the $1\rightarrow 2 \rightarrow 3$ process corresponds to a matrix element $-t^2<0$, where the minus sign comes from the phase-string sign,$(\pm 1)\times (\mp 1)=-1$, 
on a (short-range) AFM background where site $2$ and site $3$ tend to have opposite spins. 
Then, the $1\rightarrow 3$ and the $1\rightarrow 2\rightarrow 3$ processes strengthen each other for a positive $t'$, 
making the NN pairing stronger to further lower the pairing energy. 

Finally, we remark that with the increase of strength of $|t'/t|$, eventually the phase-string sign structure is expected to be scrambled by the NNN hopping such that the present VMC wave function breaks down. In this limit, a more conventional Landau quasiparticle picture, similar to that in the $\sigma\cdot t$-$J$ model, is recovered. The novel pairing mechanism discussed in the present work should disappear too. As a matter of fact, a DMRG study in a two-leg ladder system clearly shows \cite{Sun2019} such a crossover as a function of $t'/t$.  


\section{Conclusion}\label{sec:6}

In this paper, we have studied a two-hole ground state wave function in both the 2D lattice and two-leg ladder $t$-$J$ models. 
The VMC results are in good agreement with the ED and DMRG numerics. 
Here, the key feature in the wave function is the presence of a many-body nonlocal phase-shift associated with each doped hole, 
which keeps track of the precise phase-string sign structure of the $t$-$J$ model. 
For a one-hole ground state in a 2D square lattice with $C_4$ symmetry (OBC), an emergent novel quantum number, 
i.e., an angular momentum $L_z=\pm 1 $, is a manifestation of such an effect. 
In the present two-hole case, the same effect leads to $L_z=2 \mod 4$, 
which characterizes a nondegenerate ground state as precisely predicted by the exact numerical calculations. 
Therefore, the spin transverse distortion or spin current generated by the motion of a doped hole is essential in the construction of the ground state 
with the correct new quantum number. 
This is in sharp contrast to the previous approaches in the literature, in which the spin longitudinal or magnetization distortion 
or the ``spin bag'' effect have been emphasized instead.

Here, we have found that two doped holes in the ground state do form a tightly bound pair in both 2D and two-leg ladder, 
indicating the existence of a strong binding force in such doped Mott insulators. 
The hole binding diminishes once the phase-shift field in the wave function is turned off, 
even if the spin background remains in the AFLRO or RVB state. 
This strongly suggests that the pairing glue originates from the phase-string effect, 
which is explicitly illustrated by a rotonlike pattern of spin currents surrounding two holes. 
In particular, the pairing is kinetic energy driven to facilitate the hole hopping, which is kind of counterintuitive. 
It is distinct from the usual pairing mechanism of potential energy driven like exchanging AFM spin fluctuations or the RVB pairing. 

Such an unconventional pairing structure results in a dichotomy in the pairing symmetry. 
Namely, although a $d$-wave symmetry is clearly exhibited in the Cooper channel of bare electronic holes, 
a \emph{stronger} $s$-wave-like pairing will manifest if probed in terms of the ``twisted holes.'' 
Here the twisted hole has been shown to replace the usual Landau quasiparticle as a new quasiparticle in the single-hole doped case, 
with an emergent novel angular momentum $L_z=\pm 1$. 
In this sense, the $s$-wave pairing of twisted holes reflects a more intrinsic aspect of the two-hole ground state, 
which is, thus, more ``robust'' as compared to the $d$-wave symmetry in the Cooper pair channel. 

The important lesson that we have learned in the present approach is that doping a quantum spin antiferromagnet is very singular and nonperturbative at short distance. 
In order to correctly tackle the long-wavelength physics at dilute but finite doping, 
one has to first handle the short-range physics carefully, since a conventional perturbative renormalization group approach 
may well fail as clearly demonstrated in this work. 
In the last section, we have briefly discussed the possible long-wavelength physics including the self-localization of the hole pair in the AFLRO phase 
as well as the phase transition to a true superconducting phase beyond a critical doping. 
A future theoretical effort beyond the present VMC method will be needed to further deal with theses important issues of long distance including the phase transition. 


\begin{acknowledgments}
    Stimulating discussions and earlier collaborations with Qingrui Wang, Yang Qi, Long Zhang, and Donna Sheng are acknowledged. 
    This work is partially supported by MOST of China (Grant No. 2017YFA0302902) and Natural Science Foundation of China (Grant No. 11534007). \\
\end{acknowledgments}

\appendix
\section{Symmetries of the \emph{Ansatz} wave function}\label{app:1} For a system with a nondegenerate ground state, 
a good \emph{Ansatz} wave function is expected to conserve the symmetries of its Hamiltonian. 
In the following, we show that, for the symmetries to be maintained in the ground state wave function,
the opposite chirality counterpart (the $\cdots$ term) in Eq.~\eqref{eqn:doubleholeansatz}, i.e., 
\begin{equation}
    \begin{aligned}
        |\Psi_\eta\rangle_{2h} = & \sum_{ij}g_{\eta}(i,j)c_{i\uparrow}^{}c_{j\downarrow}^{} e^{-i(\hat{\Omega}_i+\eta\hat{\Omega}_j)}|\phi_0\rangle + \cdots \\ 
        =&\sum_{ij,m=\pm}g_{\eta,m}(i,j)c_{i\uparrow}^{}c_{j\downarrow}^{} e^{-im(\hat{\Omega}_i+\eta\hat{\Omega}_j)}|\phi_0\rangle,
    \end{aligned}
    \label{eqn:complex2real}
\end{equation}
has to be present. Here, we introduce $m = \pm$ as an explicit index of the chirality and 
$g_{\eta,\pm}(i,j)$ as variational parameters corresponding to $m=\pm$, respectively.

First let us examine the time reversal symmetry. Under the time reversal transformation $\hat{T}$, one has
\begin{align}
    \hat{T}c_{i\sigma}\hat{T}^{-1} &= \sigma c_{i\bar{\sigma}}^{}, \\
    \hat{T}e^{-im\hat{\Omega}_i}\hat{T}^{-1} &= e^{+im \hat{T}\hat{\Omega}_i\hat{T}^{-1}} \nonumber \\
        &= e^{im\sum_{l(\neq i)} \theta_i(l)}e^{-im\hat{\Omega}_i},
\end{align}
with using the constraint $n_{l\uparrow}=1-n_{l\downarrow}$ in $|\phi_0\rangle$. 
A time reversal invariant wave function is expected to transform as 
\begin{equation}
    \begin{aligned}
        \hat{T} |\Psi_{\eta}\rangle =& \sum_{i,j,m}
        e^{im\left[\sum_{l(\neq i)}\theta_i(l) +\eta\sum_{l(\neq j)}\theta_j(l)\right]}\\
        &g^{*}_{\eta,m}(i,j)c_{j\uparrow}c_{i\downarrow}e^{-im(\hat{\Omega}_i+\eta\hat{\Omega}_j)}|\phi_0\rangle\\
        =&\sum_{i,j,m} e^{im\left[\sum_{l(\neq i)}\theta_i(l) +\eta\sum_{l(\neq j)}\theta_j(l)\right]}\\
        &g^*_{\eta,\eta m}(j,i)c_{i\uparrow}c_{j\downarrow}e^{-im(\hat{\Omega}_i+\eta\hat{\Omega}_j)}|\phi_0\rangle\\
        =& \alpha_{T} |\Psi_\eta\rangle,
    \end{aligned}
\end{equation}
which requires  
\begin{equation}
    \label{eqn:TRsym}
    g^*_{\eta,\eta m}(j,i) = \alpha_T g_{\eta,m}(i,j)
    e^{-im\left[\sum_{l(\neq i)} \theta_{i}(l)+\eta\sum_{l(\neq j)} \theta_{j}(l)\right]} ~,
\end{equation}
where $\alpha_T$ is an arbitrary global phase factor with $|\alpha_T|=1$. 

Next consider a discrete $Z_2$ symmetry of spin flip $\hat{F}$, which is defined as 
\begin{align}
    \hat{F}c_{i\sigma}\hat{F}^{-1} =& c_{i\bar{\sigma}},\\
    \hat{F}e^{-im\hat{\Omega}_i}\hat{F}^{-1} =& e^{-im\sum_{l(\neq i)} \theta_i(l)}e^{im\hat{\Omega}_i} . 
\end{align}
It is expected that the ground state would remain the same under this transformation up to a global phase 
\begin{equation}
    \begin{aligned}
        \hat{F}|\Psi_\eta\rangle = & -\sum_{i,j,m}e^{-im\left[\sum_{l(\neq i)}\theta_i(l)+\eta\sum_{l(\neq i)}\theta_i(l)\right]}\\
         &g_{\eta,m}(i,j)c_{j\uparrow}c_{i\downarrow}e^{im(\hat{\Omega}_i+\eta\hat{\Omega}_j)}|\phi_0\rangle\\
        = & -\sum_{i,j,m}e^{im\left[\sum_{l(\neq i)}\theta_i(l)+\eta\sum_{l(\neq i)}\theta_i(l)\right]}\\
        &g_{\eta,-\eta m}(j,i)c_{j\uparrow}c_{i\downarrow}e^{im(\hat{\Omega}_i+\eta\hat{\Omega}_j)}|\phi_0\rangle\\
        =& \alpha_F|\Psi_\eta\rangle.
    \end{aligned}
\end{equation}
We swap the summation indexes $i,j$ and replace $m$ by $-\eta m$ in deriving the second equality above. 
Again, $\alpha_F$ is an arbitrary global phase factor with $|\alpha_F|=1$. 
Actually, by doing the spin flip transformation twice, we further get $\alpha_F^2=1$, which leads to $\alpha_F = \pm 1$. 
The last equality gives rise to the following constraint on a spin-flip invariant wave function 
\begin{equation}
    \label{eqn:Z2sym}
    g_{\eta,-\eta m}(j,i) = -\alpha_Fg_{\eta,m}(i,j) 
    e^{-im\left[\sum_{l\neq i} \theta_{i}(l)+\eta\sum_{l\neq j}\theta_{j}(l)\right]}.
\end{equation}

For a wave function satisfying both of the above symmetries, 
we can fix the arbitrary phase factor $\alpha_T=-\alpha_F$ and combine the two constraints into one: 
\begin{equation}
    g^*_{\eta,m}(i,j) = g_{\eta,-m}(i,j),
\end{equation}
which is equivalent to saying that a nondegenerate ground state should be a real one, and no \emph{net} currents are expected to exist. 
This constraint is verified numerically for the variational ground state $|\Psi_-\rangle_{2h}$ studied in the main text, 
which is nondegenerate with $L_z=2 \mod 4$ under the $C_4$ symmetry. 

\section{Variational Monte Carlo procedure}\label{app:2}

The Monte Carlo procedure used to optimize the variational energy and to measure other observables of the two-hole doped wave function \emph{Ansatz} \eqref{eqn:doubleholeansatz} is outlined in this appendix. 
Similar procedures have been previously utilized first in the single-hole doped $t$-$J$ model for both two-leg ladder and 2D lattice cases  \cite{Wang2015,Chen2019}, and then in the two-hole doped problem in the $t_\perp = 0$ two-leg ladder \cite{Chen2018}. 

At half filling, the $t$-$J$ model reduces to the Heisenberg model, 
whose ground state $ |\phi_0\rangle$ for a bipartite lattice can be well simulated variationally by the Liang-Doucot-Anderson-type 
(bosonic RVB) wave function \cite{Liang1988}. Namely,
\begin{equation} 
    |\phi_0\rangle = \sum_v\omega_v|v\rangle,
    \label{eqn:Liang}
\end{equation}
where the valence bond (VB) state 
\begin{equation}
    |v\rangle = |(a_1,b_1)\cdots(a_n,b_n)\rangle 
\end{equation}
consists of singlet pairs $|(a,b)\rangle = |\uparrow_a\downarrow_b\rangle-|\downarrow_a\uparrow_b\rangle$, 
with $a$ and $b$ from different sublattices, $A$ and $B$, respectively. 
Here, the VB states are not orthogonal to each other, and the overlap between two different VB states is given by
\begin{equation}
    \langle v'|v\rangle = 2^{N^{\mathrm{loop}}_{v',v}},
\end{equation}
where $N^{\mathrm{loop}}_{v',v}$ is the number of loops in the transposition-graph covers $(v',v)$ \cite{Liang1988}. 

The variational parameter $w_v$ of each valence bond state $|v\rangle$, which is always positive according to the Marshall sign rule, 
can be further fractionalized as $w_v = \prod _{(a_i,b_i)\in v}h(a_i-b_i)$, 
where $h(a_i-b_i) > 0$ is a positive function dependent on the distance $a_i-b_i$ between sites $a_i$ and $b_i$. 
By tuning the long-range behavior of $h(a-b)$, different spin background can be obtained. 
For a power law $h(l)\sim l^{-p}$ and $p<5$, $|\phi_0\rangle$ is an AFLRO state; 
while for $h(l)\sim l^{-p}$ and $p\geq 5$ or when the maximum bond length is finite, a short-ranged state can be obtained. 
Specifically, a dimer limit state is obtained when the maximum bond length is chosen as 1. 

Here, we follow Ref.~\cite{Liang1988} to express all the observables we need in the form of expectation values on the spin background $|\phi_0\rangle$. 
As a simple example, the spin-spin correlation $\mathbf{S}_i\cdot \mathbf{S}_j$ of the spin background $|\phi_0\rangle$ can be calculated as  
\begin{equation}
    \frac{\langle\phi_0|\mathbf{S}_i\cdot \mathbf{S}_j|\phi_0\rangle}{\langle\phi_0|\phi_0\rangle} = 
    \sum_{v',v}P(v',v)\frac{\langle v'|\mathbf{S}_i\cdot \mathbf{S}_j|v\rangle}{\langle v'|v\rangle},
    \label{eqn:sisjMC_eg}
\end{equation}
where 
\begin{equation}
    P(v',v) = \frac{w_{v'}w_v\langle v'|v\rangle}{\langle\phi_0|\phi_0\rangle}
    \label{eqn:probability}
\end{equation}
is a probability distribution satisfying $\sum_{v',v}P(v',v) = 1$. 
Equation \eqref{eqn:sisjMC_eg} can then be evaluated using the standard Monte Carlo method as long as one knows how to calculate 
$\frac{\langle v'|\mathbf{S}_i\cdot \mathbf{S}_j|v\rangle}{\langle v'|v\rangle}$, which is 
$$\frac{\langle v'|\mathbf{S}_i\cdot \mathbf{S}_j|v\rangle}{\langle v'|v\rangle} = \left\{
\begin{aligned} &(-1)^{i+j} \frac{3}{4}, &&i,j \in \mathrm{sl}\\ 
&0, && i,j \notin \mathrm{sl}\end{aligned}\right.~,$$
where $i,j \in \mathrm{sl}$ means that $i$ and $j$ belong to the same loop in the transposition-graph covers $(v',v)$.

A similar Monte Carlo procedure can also be done for the two-hole wave function $|\Psi_\eta\rangle_{2h}$ with minor adjustment. 
For an arbitrary observable $\hat{O}$, its expectation value over $|\Psi_\eta\rangle_{2h}$ is
\begin{equation}
    \langle\hat{O}\rangle \equiv 
    \frac{\tensor[_{2h}]{\langle \Psi_\eta|\hat{O}|\Psi_\eta\rangle}{_{2h}}}
    {\tensor[_{2h}]{\langle \Psi_\eta|\Psi_\eta\rangle}{_{2h}}}
    =\frac{\tensor[_{2h}]{\langle \Psi_\eta|\hat{O}|\Psi_\eta\rangle}{_{2h}}}{\langle \phi_0|\phi_0\rangle}
    \frac{\langle \phi_0|\phi_0\rangle}{\tensor[_{2h}]{\langle \Psi_\eta|\Psi_\eta\rangle}{_{2h}}}.
    \label{eqn:opeO0}
\end{equation}
By fixing the normalization condition of $|\Psi_\eta\rangle_{2h}$ as 
\begin{equation}
    \frac{\tensor[_{2h}]{\langle \Psi_\eta|\Psi_\eta\rangle}{_{2h}}}{\langle \phi_0|\phi_0\rangle} = 1~,
    \label{eqn:norm}
\end{equation}
we further get
\begin{widetext}
\begin{equation}
    \begin{aligned}
        \langle\hat{O}\rangle \equiv
        \frac{\tensor[_{2h}]{\langle \Psi_\eta|\hat{O}|\Psi_\eta\rangle}{_{2h}}}{\langle \phi_0|\phi_0\rangle}
        =&\sum_{i'j'm',ijm} g^*_{\eta,m'}(i',j')g_{\eta,m}(i,j)
        \frac{\langle\phi_0|\hat{\Lambda}_{i'j'}^{-m'} c^{\dagger}_{j'\downarrow}c^{\dagger}_{i'\uparrow} \hat{O} 
        c^{}_{i\uparrow}c^{}_{j\downarrow}\hat{\Lambda}_{ij}^{m}|\phi_0\rangle} {\langle \phi_0|\phi_0\rangle}\\
        =&\sum_{i'j'm',ijm} g^*_{\eta,m'}(i',j')g_{\eta,m}(i,j)
        \left[\sum_{v',v}P(v',v)\frac{\langle v'|\hat{\Lambda}_{i'j'}^{-m'} c^{\dagger}_{j'\downarrow}c^{\dagger}_{i'\uparrow} \hat{O} 
        c^{}_{i\uparrow}c^{}_{j\downarrow}\hat{\Lambda}_{ij}^{m}|v\rangle} {\langle v'|v\rangle}\right]\\
        =&\mathbf{g}^\dagger \mathbf{O}\mathbf{g},
    \end{aligned}
    \label{eqn:opeO}
\end{equation}
where for simplicity the phase operator is written as $\hat{\Lambda}_{ij}^m = e^{-im[\sum_{l\neq i}\theta_i(l)n_{l\downarrow}+\eta\sum_{l\neq j}\theta_j(l)n_{l\downarrow}]}$. 
Equation \eqref{eqn:opeO} is a quadratic form of the variational parameters $g_{\eta,m}(i,j)$, which are written as a vector $\mathbf{g}$ in the last line. 
$\mathbf{O}$ is a Hermitian matrix, whose matrix elements are 
\begin{align}
    \mathbf{O}_{ijm}^{i'j'm'} = \sum_{v',v}P(v',v)\frac{\langle v'|\hat{\Lambda}_{i'j'}^{-m'} c^{\dagger}_{j'\downarrow}c^{\dagger}_{i'\uparrow} \hat{O} 
        c^{}_{i\uparrow}c^{}_{j\downarrow}\hat{\Lambda}_{ij}^{m}|v\rangle} {\langle v'|v\rangle}~,\label{eqn:matrixO}
\end{align}
which have the same form as that of Eqs.~\eqref{eqn:sisjMC_eg} and \eqref{eqn:probability}. 
Therefore, the same Monte Carlo procedure in Ref.~\cite{Liang1988} can be directly used here as long as one gets the formulas of 
$\frac{\langle v'|\hat{\Lambda}_{i'j'}^{-m'} c^{\dagger}_{j'\downarrow}c^{\dagger}_{i'\uparrow} \hat{O} 
c^{}_{i\uparrow}c^{}_{j\downarrow}\hat{\Lambda}_{ij}^{m}|v\rangle} {\langle v'|v\rangle}$. 
The loop update procedure introduced in Ref.~\cite{Sandvik2010} is used in the Monte Carlo sampling procedure to accelerate calculation. 

Specifically, the normalization condition \eqref{eqn:norm} can also be written as the form of Eq.~\eqref{eqn:opeO}, 
with $\hat{O}$ taken as the identity operator $\hat{\mathds{1}}$: 
\begin{equation}
    1 = \langle \hat{\mathds{1}}\rangle =\mathbf{g}^\dagger \mathbf{A}\mathbf{g}~,
    \label{eqn:ope1}
\end{equation}
which is also a quadratic form serving as the normalization condition for the variational parameters $g_{\eta,m}(i,j)$. 
The corresponding matrix of the identity operator $\hat{\mathds{1}}$ is denoted as $\mathbf{A}$ here [cf. Eq.~\eqref{eqn:measurenorm} below], 
whose matrix elements can also be obtained by Eq.~\eqref{eqn:matrixO}. 
Note that $\mathbf{A}$ is not an identity matrix as a result of indistinguishability of identical holes 
created by $c_{i\uparrow}c_{j\downarrow}$ and $c_{i\downarrow}c_{j\uparrow}$. 

\subsection{Explicit expressions for some observables}

In calculating the matrix elements in Eq.~\eqref{eqn:matrixO}, one may find that all the calculations can be transformed 
to evaluating expressions of the following form: 
\begin{equation}
    \frac{\langle v'|\hat{\Lambda}^{-m'}_{i'j'} n_{k_1\sigma_1}n_{k_2\sigma_2}\cdots n_{k_n\sigma_n}
    S_{l_1}^{\sigma_1}S_{l_2}^{\sigma_2}\cdots S_{l_s}^{\sigma_s}
    \hat{\Lambda}^{m}_{ij}|v\rangle }{\langle v'|v\rangle }~.
    \label{eqn:nnnsss}
\end{equation}

Making use of this expression, we can again take advantage of the loop configurations in the transposition-graph covers $(v',v)$, 
by dividing the loops into relevant loops which contain sites $k_1\cdots k_n$ or $l_1\cdots l_s$ and irrelevant loops which do not. 
When the phase factor $\hat{\Lambda}^m_{ij}$ is not considered, the irrelevant loops are just unity  
while the relevant loops determine whether $n_{k_1\sigma_1}n_{k_2\sigma_2}\cdots n_{k_n\sigma_n} S_{l_1}^{\sigma_1}S_{l_2}^{\sigma_2}\cdots S_{l_s}^{\sigma_s}$
operators are compatible with  the loop configuration $(v',v)$, or, in other words, whether the term is nonzero or zero. 
After the phase factor $\hat{\Lambda}^m_{ij}$ is added back, the irrelevant loops contribute a factor 
$\frac{1}{2}(e^{i\theta^A_{\alpha}}+e^{i\theta^B_{\alpha}})$ now, 
while the relevant loops give a phase factor $\frac{1}{2}e^{i\theta_\alpha^{\mathrm{rel}}}$ besides the duty of checking compatibility, where
\begin{equation}
    \theta_\alpha^{A,B} = \sum_{\substack{l\in\mathrm{loop}\ \alpha \\ l\in A,B}} m'[\theta_{i'}(l)+\eta\theta_{j'}(l)] - m[\theta_{i}(l)+\eta\theta_{j}(l)]
    \label{eqn:looptheta}
\end{equation}
is the summation of all the phases along a loop $\alpha$ and $\theta_\alpha^{\mathrm{rel}}$ depends on the spin configuration on the relevant loops. 

Therefore, as long as we expand the matrix elements of Eq.~\eqref{eqn:matrixO} in the form of Eq.~\eqref{eqn:nnnsss}, the programming progress is straightforward. 
In the following we give this expansion explicitly for some important operators. 
\begin{enumerate}

\item 
First for the normalization condition $\hat{O}=\hat{\mathds{1}}$, 
the corresponding matrix $\mathbf{A}$ has matrix elements
\begin{equation}
    \mathbf{A}^{i'j'm'}_{ijm}= \sum_{v',v}P(v',v)\frac{1}{\langle v'|v\rangle}
    \left[\delta_{ii'}\delta_{jj'}\langle v'|\hat{\Lambda}^{-m'}_{ij}n_{i\uparrow}n_{j\downarrow}\hat{\Lambda}^{m}_{ij}|v\rangle
    -\delta_{ij'}\delta_{ji'}\langle v'|\hat{\Lambda}^{-m'}_{ji}S_i^-S_j^+\hat{\Lambda}^{m}_{ij}|v\rangle \right].
    \label{eqn:measurenorm}
\end{equation}

\item
The hopping Hamiltonian $\hat{H}_t$ defined in Eq.~\eqref{eqn:hoppingH} connects pairs of sites with only one site different, 
resulting in the corresponding matrix $\mathbf{H}_t$ with elements
\begin{equation}
    \begin{aligned}
        (\mathbf{H}_{t})_{ijm}^{i'j'm'} = t \sum_{v',v}P(v',v)\frac{1}{\langle v'|v\rangle}\sum_{\alpha = x,y}
        &\left[\delta_{ii'}\delta_{jj'\pm\mathbf{e}_\alpha}\langle v'|\hat{\Lambda}^{-m'}_{ij'}n_{i\uparrow}(n_{j\downarrow}n_{j'\downarrow}
        +S_j^+S_{j'}^-)\hat{\Lambda}^{m}_{ij}|v\rangle\right.\\
        &\left. +\delta_{jj'}\delta_{ii'\pm\mathbf{e}_\alpha}\langle v'|\hat{\Lambda}^{-m'}_{i'j} n_{j\downarrow} (n_{i\uparrow}n_{i'\uparrow}
        +S^-_iS^+_{i'})\hat{\Lambda}^{m}_{ij}|v\rangle \right.\\
        &\left. -\delta_{ji'}\delta_{ij'\pm\mathbf{e}_\alpha}\langle v'|\hat{\Lambda}^{-m'}_{j'i}S^-_{i}(S^+_{j}n_{j'\uparrow}+n_{j\downarrow}S^
        +_{j'})\hat{\Lambda}^{m}_{ij} |v\rangle \right.\\
        &\left. -\delta_{ij'}\delta_{ji'\pm\mathbf{e}_\alpha}\langle v'|\hat{\Lambda}^{-m'}_{ji'}S^-_{j}(S^+_{i}n_{i'\uparrow}
        +n_{i\downarrow}S^+_{i'})\hat{\Lambda}^{m}_{ij} |v\rangle \right],\\
    \end{aligned}
    \label{eqn:measurehopping}
\end{equation}
where $\mathbf{e}_{x,y}$ are the $x$ and $y$ direction unit vectors. 

\item
For the superexchange Hamiltonian $\hat{H}_J$ defined in Eq.~\eqref{eqn:superexchangeH}, the corresponding matrix $\mathbf{H}_J$ is 
\begin{equation}
    \begin{aligned}
        (\mathbf{H}_{J})_{ijm}^{i'j'm'} = \frac{J}{2} \sum_{v',v}P(v',v)\frac{1}{\langle v'|v\rangle}\sum_{\langle kl\rangle(\neq i,j)}&
        \left[\delta_{ii'}\delta_{jj'}\langle v'|\hat{\Lambda}^{-m'}_{ij}n_{i\uparrow}n_{j\downarrow}
        (S_k^+S_l^-+S_k^-S_l^+-n_{k\uparrow}n_{l\downarrow}-n_{k\downarrow}n_{l\uparrow})\hat{\Lambda}^{m}_{ij}|v\rangle \right.\\ 
        &\left.+ \delta_{ij'}\delta_{ji'}\langle v'|\hat{\Lambda}^{-m'}_{ji}S_i^-S_j^+
        (n_{k\uparrow}n_{l\downarrow}+n_{k\downarrow}n_{l\uparrow}-S_k^+S_l^--S_k^-S_l^+)\hat{\Lambda}^{m}_{ij}|v\rangle \right].
    \end{aligned}
    \label{eqn:measureAF}
\end{equation}

\item
Calculation of the overlap Eq.~\eqref{eqn:overlap} is a bit different with other variables. 
We first normalize Eq.~\eqref{eqn:overlap} as 
\begin{equation}
    \begin{aligned}
        \Delta_{\mathbf{k}}^s =& \frac{\tensor[_{2h}]{\langle\Psi_\eta|\hat{\Delta}_{\mathbf{k}}^s|\phi_0\rangle}{}}
        {\sqrt{\tensor[_{2h}]{\langle\Psi_\eta|\Psi_\eta\rangle}{_{2h}}\langle\phi_0|\phi_0\rangle}} 
        =\sum_{i'j'm'}g^*_{\eta,m'}(i',j') \frac{ \langle\phi_0|\hat{\Lambda}^{-m'}_{i'j'}c^\dagger_{j'\downarrow}c^{\dagger}_{i'\uparrow}\hat{\Delta}_{\mathbf{k}}^s|\phi_0\rangle} {\langle\phi_0|\phi_0\rangle}  \\
        =&\sum_{i'j'm',ij}g^*_{\eta,m'}(i',j')\frac{1}{N}e^{i\mathbf{k}\cdot(\mathbf{r}_i - \mathbf{r}_j)}\sum_{v',v}P(v',v)
         \frac{ \langle v'|\hat{\Lambda}^{-m'}_{i'j'}c^\dagger_{j'\downarrow}c^{\dagger}_{i'\uparrow}\hat{\Delta}_{ij}^s\hat{\Lambda}_{ij}^0|v\rangle} {\langle v'|v\rangle},\\
    \end{aligned}
\end{equation}
which is a similar quadratic formula with Eq.~\eqref{eqn:opeO} if we treat $\frac{1}{N}e^{i\mathbf{k}\cdot(\mathbf{r}_i - \mathbf{r}_j)}$ as $g_{\eta,0}(i,j)$. 
The corresponding matrix elements are then 
\begin{equation}
    \label{eqn:measureoverlap}
    \begin{aligned}
        (\mathbf{\Delta}_{\mathbf{k}}^s)_{ij0}^{i'j'm'} \equiv &\sum_{v',v}P(v',v)
        \frac{\langle v'|\hat{\Lambda}^{-m'}_{i'j'}c^\dagger_{j'\downarrow}c^{\dagger}_{i'\uparrow}\hat{\Delta}_{ij}^s\hat{\Lambda}_{ij}^0|v\rangle}{\langle v'|v\rangle}\\
        =&\sum_{v',v}P(v',v)\frac{1}{\langle v'|v\rangle}
        \left[\delta_{ii'}\delta_{jj'}\langle v'|\hat{\Lambda}^{-m'}_{ij}(n_{i\uparrow}n_{j\downarrow}-S_{i}^+S_{j}^-)|v\rangle 
        +\delta_{ij'}\delta_{ji'}\langle v'|\hat{\Lambda}^{-m'}_{ji}(n_{i\downarrow}n_{j\uparrow}-S_{i}^-S_{j}^+)|v\rangle \right].
    \end{aligned}
\end{equation}

\item
For the singlet pair-pair correlators defined in Eq.~\eqref{eqn:paircorrelator}, written together as $\hat{C}_{k'l',kl}^s = \hat{\Delta}_{k'l'}^s(\Delta_{kl}^s)^\dagger$ here with $k'\neq k,l$ and $l'\neq k,l$, 
the matrix elements are as follows: 
\begin{equation}
    \begin{aligned}
        (\mathbf{C}_{k'l',kl}^s)_{ijm}^{i'j'm'} = \sum_{v',v}P(v',v)\frac{1}{\langle v'|v\rangle}
        &\left[\delta_{i'k'}\delta_{j'l'}\delta_{ik}\delta_{jl} 
        \langle v'|\hat{\Lambda}^{-m'}_{k'l'}(n_{k'\uparrow}n_{l'\downarrow}-S_{k'}^+S_{l'}^-)(n_{k\uparrow}n_{l\downarrow}-S_k^-S_l^+)\hat{\Lambda}^{m}_{kl}|v\rangle \right.\\ 
        &\left.\delta_{i'k'}\delta_{j'l'}\delta_{il}\delta_{jk} 
        \langle v'|\hat{\Lambda}^{-m'}_{k'l'}(n_{k'\uparrow}n_{l'\downarrow}-S_{k'}^+S_{l'}^-)(n_{k\downarrow}n_{l\uparrow}-S_k^+S_l^-)\hat{\Lambda}^{m}_{lk}|v\rangle \right.\\ 
        &\left.\delta_{i'l'}\delta_{j'k'}\delta_{ik}\delta_{jl} 
        \langle v'|\hat{\Lambda}^{-m'}_{l'k'}(n_{k'\downarrow}n_{l'\uparrow}-S_{k'}^-S_{l'}^+)(n_{k\uparrow}n_{l\downarrow}-S_k^-S_l^+)\hat{\Lambda}^{m}_{kl}|v\rangle \right.\\ 
        &\left.\delta_{i'l'}\delta_{j'k'}\delta_{il}\delta_{jk} 
        \langle v'|\hat{\Lambda}^{-m'}_{l'k'}(n_{k'\downarrow}n_{l'\uparrow}-S_{k'}^-S_{l'}^+)(n_{k\downarrow}n_{l\uparrow}-S_k^+S_l^-)\hat{\Lambda}^{m}_{lk}|v\rangle \right].\\ 
    \end{aligned}
    \label{eqn:measurepairpair}
\end{equation}

\item
The hole-hole density correlator $\hat{N}_{kl}^h$ defined in Eq.~\eqref{eqn:nihnjh} is essentially the same with the unit operator 
\begin{equation}
    (\mathbf{N}_{kl}^h)_{ijm}^{i'j'm'} = (\delta_{ik}\delta_{jl}+\delta_{ik}\delta_{jl})\sum_{v',v}P(v',v)\frac{1}{\langle v'|v\rangle}
    \left[\delta_{ii'}\delta_{jj'}\langle v'|\hat{\Lambda}^{-m'}_{ij}n_{i\uparrow}n_{j\downarrow}\hat{\Lambda}^{m}_{ij}|v\rangle
    -\delta_{ij'}\delta_{ji'}\langle v'|\hat{\Lambda}^{-m'}_{ji}S_i^-S_j^+\hat{\Lambda}^{m}_{ij}|v\rangle \right]~.
    \label{eqn:measurenihnjh}
\end{equation}

\item
The hole density can be obtained by a simple summation of $N_{ij}^h$: 
\begin{equation}
    \langle \hat{n}_i^h\rangle = \sum _{j(\neq i)} N_{ij}^h.
\end{equation}

\item
The neutral spin current $\hat{J}_{kl}^s$ defined in Eq.~\eqref{eqn:spincurrent1} has similar matrix elements with the superexchange term: 
\begin{equation}
    \begin{aligned}
        (\mathbf{J}_{kl}^s)^{i'j'm'}_{ijm} = -i\frac{J}{2} (1-\delta_{\langle ij\rangle,\langle kl\rangle})\sum_{v',v}P(v',v)\frac{1}{\langle v'|v\rangle}&
        \left[\delta_{ii'}\delta_{jj'}\langle v'|\hat{\Lambda}^{-m'}_{ij}n_{i\uparrow}n_{j\downarrow}
        (S_k^+S_l^--S_k^-S_l^+)\hat{\Lambda}^{m}_{ij}|v\rangle\right.\\ 
        &\left.- \delta_{ij'}\delta_{ji'}\langle v'|\hat{\Lambda}^{-m'}_{ji}S_i^-S_j^+ (S_k^+S_l^--S_k^-S_l^+)\hat{\Lambda}^{m}_{ij}|v\rangle \right] ~,
    \end{aligned}
    \label{eqn:measureJs}
\end{equation}
where $\delta_{\langle ij\rangle,\langle kl\rangle} = 0$ when $i\neq k,l$ and $j\neq k,l$ and $\delta_{\langle ij\rangle,\langle kl\rangle} = 1$ otherwise. 
The above expressions can also be used to calculate $J^s_{kl}$ with the two doped holes projected at $i$ and $j$. 
\end{enumerate}

\subsection{Variational procedure for optimizing $g_\eta(i,j)$}

In the variational procedure, we take the operator $\hat{O}$ in Eq.~\eqref{eqn:opeO} as the kinetic Hamiltonian $\hat{H}_{t}$ 
\begin{equation} 
    E_{t} \equiv \langle \hat{H}_t\rangle = \mathbf{g}^\dagger \mathbf{H}_{t} \mathbf{g} \label{eqn:generalizeddiag_E}.
\end{equation}
Combined with the normalization condition Eq.~\eqref{eqn:ope1}, 
the procedure to optimize $E_t$ turns out to be a generalized eigenvalue problem 
\begin{equation}\label{eqn:generalizeddiag}
    \mathbf{H}_t\mathbf{g} = E_t \mathbf{A}\mathbf{g}. 
\end{equation}

Besides the ground state energy, Eq.~\eqref{eqn:generalizeddiag} can also give excited energies and corresponding variational parameters $\mathbf{g}$, 
which are used in Sec.~\ref{sec:4} in obtaining the binding energy. 
Here, only the kinetic energy $E_t$ is taken into account in the variational procedure following Ref.~\cite{Chen2019}. 
The superexchange energy $E_J$ is omitted, as inclusion of $E_J$ causes severe boundary effect and makes the calculation time unacceptable for a large lattice. 
However, calculation on $10\times 10$ lattice with the holes constrained on the central $8\times 8$ sites shows that the $E_J$ term seldom changes the physics, 
which validates our omission of $E_J$ on a larger lattice. 

\end{widetext}

\section{Feedback effect on the spin background}\label{app:3}

In the basic variational procedure in the main text, the spin background $|\phi_0\rangle$ in Eq.~\eqref{eqn:doubleholeansatz} 
is taken to be the half-filling ground state of $H_J$ by ignoring the feedback effect from motion of holes. 
The main doping effect is explicitly incorporated as the phase-string sign structure in the two-hole ground state \emph{Ansatz} $|\Psi_-\rangle_{2h}$ in the VMC simulation. 
However, this assumption may no longer hold true at finite doping as the spin background evolves into a short-ranged AFM state. 
Indeed, the \emph{Ansatz} state $|\Psi_+\rangle$ or Eq.~\eqref{eqn:psgs} is expected to work 
to capture the essential physics where the feedback effect on the spin background from hole motion comes into play at finite doping.

\begin{table}[tb]
    \centering
    \caption{The energies of the generalized two-hole ground states [Eqs.~\eqref{eqn:doubleholeansatz_ge} and \eqref{eqn:doubleholeansztz_mix}] on a $4\times 4$ lattice: 
        $E_{{G}}$ is the total energy; $E_t$ and $E_J$ are the kinetic and the superexchange energies, respectively; 
        $L_z$ denotes the corresponding orbital angular momentum.}
    \begin{ruledtabular}
        \begin{tabular}{lcccc}
            ~& $E_{{G}}$ & $E_t$ & $E_J$ & $L_z$\\
            \colrule
            $|\Psi_{+}\rangle_{2h}$ & $-20.50$ & $-9.56$ & $-10.94$ & $\pm 1$\\
            $|\Psi_{-}\rangle_{2h}$ & $-22.51$ & $-11.20$ & $-11.31$ & $2$\\
            $|\widetilde{\Psi}_+\rangle_{2h}$ & $-22.29$ & $-11.94$ & $-10.35$ & $2$ \\
            $|\widetilde{\Psi}_-\rangle_{2h}$ & $-22.80$ & $-11.92$ & $-10.88$ & $2$ \\
            $|\widetilde{\Psi}_{\mathrm{mix}}\rangle_{2h}$ & $-22.94$ & $-12.16$ & $-10.78$ & $2$ \\
            ED & $-24.98$  & $-14.57$ & $-10.42$ & $2$
        \end{tabular}	
    \end{ruledtabular}
    \label{tab:relaxRVB}
\end{table}

In order to analyze this feedback effect, in this appendix, we iteratively optimize the spin background as well as variational parameters $g_\eta(i,j)$ 
for both kinds of ground state \emph{Ansatz} $|\Psi_\pm\rangle_{2h}$. 
Generally, a wave function \emph{Ansatz} can be written as 
\begin{equation}
    |\widetilde{\Psi}_\eta\rangle_{2h} = \sum_{ij}g_\eta(i,j)c_{i\uparrow}^{}c_{j\downarrow}^{} 
    e^{-i(\hat{\Omega}_i+\eta\hat{\Omega}_j)}|\mathrm{RVB}\rangle + \cdots.
    \label{eqn:doubleholeansatz_ge}
\end{equation}
Here $|\mathrm{RVB}\rangle$, when expanded in terms of an Ising basis $|\{\sigma\}\rangle$, 
\begin{equation}
    |\mathrm{RVB}\rangle = \sum_{\{\sigma\}}w[\{\sigma\}]|\{\sigma\}\rangle,
\end{equation}
represents a spin background with parameters $w[\{\sigma\}]$ to be determined by 
optimizing the ground state energy with fixed variational parameters $g_\eta(i,j)$.

Iterate calculations are done to get the best variational energy; i.e., 
first choose $w^0[\{\sigma\}]$ as the ground state of $H_J$ and find variational parameters $g^0_\eta(i,j)$ to optimize the kinetic energy, 
then fix the variational parameters $g^0_\eta(i,j)$ and find new $w^1[\{\sigma\}]$ to optimize the ground state energy, 
and then fix the spin background as $w^1[\{\sigma\}]$ to find new variational parameters $g^1_\eta(i,j)$ to optimize the kinetic energy and so on, 
until a convergent ground state energy is obtained. 

Variational energies and the quantum numbers calculated via the generalized wave function \emph{Ansatz} Eq.~\eqref{eqn:doubleholeansatz_ge} on a $4\times 4$ lattice 
are shown in Table~\ref{tab:relaxRVB}, with the results of original Eq.~\eqref{eqn:doubleholeansatz} also shown here for comparison. 
It can be seen that the energy of $|\widetilde{\Psi}_+\rangle_{2h}$ is improved significantly, while the energy of $|\widetilde{\Psi}_-\rangle_{2h}$ is improved less. 
These results confirm that the feedback effect indeed is insignificant for $|\Psi_-\rangle_{2h}$ but plays an essential role for the same chirality state $|\Psi_+\rangle_{2h}$, with its ground state angular momentum changed from $L_z=\pm 1$ to $L_z=2$ the same as that of $|\Psi_-\rangle_{2h}$. 

In other words, now both $|\Psi_{\pm}\rangle_{2h}$ converge to the same nondegenerate ground state in the VMC procedure. 
The overlap $|\tensor[_{2h}]{\langle \widetilde{\Psi}_-|\widetilde{\Psi}_+\rangle}{_{2h}}| = 0.50$ also confirms this. 
Therefore, at least for the short-range physics, the above two kinds of wave function \emph{Ansatz} can lead to the same ground state. 
Actually, as stated in Eqs.~\eqref{psai+} and \eqref{eqn:modifiedRVB} in the main text, 
a vortex $e^{2i\hat{\Omega}_i}$ is expected to be generated automatically in the $|\mathrm{RVB}\rangle$ state of $|\widetilde{\Psi}_+\rangle_{2h}$ due to the feedback effect, 
which reconciles with the antiphase factor in $|\widetilde{\Psi}_-\rangle_{2h}$. 
Inspired by the similarity of $|\widetilde{\Psi}_+\rangle_{2h}$ and $|\widetilde{\Psi}_+\rangle_{2h}$, we can further mix these two states 
\begin{equation}
    |\widetilde{\Psi}_{\mathrm{mix}}\rangle_{2h} =c_1 |\widetilde{\Psi}_+\rangle_{2h} + c_2|\widetilde{\Psi}_-\rangle_{2h} 
    \label{eqn:doubleholeansztz_mix}
\end{equation}
and optimize the variational parameters $g_\pm(i,j)$ to get an improved ground state energy, which is also shown in Table~\ref{tab:relaxRVB}. 





\bibliography{ref/nameAbrv.bib,ref/refs.bib}

\begin{thebibliography}{78}%
\makeatletter
\providecommand \@ifxundefined [1]{%
 \@ifx{#1\undefined}
}%
\providecommand \@ifnum [1]{%
 \ifnum #1\expandafter \@firstoftwo
 \else \expandafter \@secondoftwo
 \fi
}%
\providecommand \@ifx [1]{%
 \ifx #1\expandafter \@firstoftwo
 \else \expandafter \@secondoftwo
 \fi
}%
\providecommand \natexlab [1]{#1}%
\providecommand \enquote  [1]{``#1''}%
\providecommand \bibnamefont  [1]{#1}%
\providecommand \bibfnamefont [1]{#1}%
\providecommand \citenamefont [1]{#1}%
\providecommand \href@noop [0]{\@secondoftwo}%
\providecommand \href [0]{\begingroup \@sanitize@url \@href}%
\providecommand \@href[1]{\@@startlink{#1}\@@href}%
\providecommand \@@href[1]{\endgroup#1\@@endlink}%
\providecommand \@sanitize@url [0]{\catcode `\\12\catcode `\$12\catcode
  `\&12\catcode `\#12\catcode `\^12\catcode `\_12\catcode `\%12\relax}%
\providecommand \@@startlink[1]{}%
\providecommand \@@endlink[0]{}%
\providecommand \url  [0]{\begingroup\@sanitize@url \@url }%
\providecommand \@url [1]{\endgroup\@href {#1}{\urlprefix }}%
\providecommand \urlprefix  [0]{URL }%
\providecommand \Eprint [0]{\href }%
\providecommand \doibase [0]{http://dx.doi.org/}%
\providecommand \selectlanguage [0]{\@gobble}%
\providecommand \bibinfo  [0]{\@secondoftwo}%
\providecommand \bibfield  [0]{\@secondoftwo}%
\providecommand \translation [1]{[#1]}%
\providecommand \BibitemOpen [0]{}%
\providecommand \bibitemStop [0]{}%
\providecommand \bibitemNoStop [0]{.\EOS\space}%
\providecommand \EOS [0]{\spacefactor3000\relax}%
\providecommand \BibitemShut  [1]{\csname bibitem#1\endcsname}%
\let\auto@bib@innerbib\@empty
\bibitem [{\citenamefont {Mott}(1949)}]{Mott1949}%
  \BibitemOpen
  \bibfield  {author} {\bibinfo {author} {\bibfnamefont {N.~F.}\ \bibnamefont
  {Mott}},\ }\bibfield  {title} {\enquote {\bibinfo {title} {{The Basis of the
  Electron Theory of Metals, with Special Reference to the Transition
  Metals}},}\ }\href {\doibase 10.1088/0370-1298/62/7/303} {\bibfield
  {journal} {\bibinfo  {journal} {Proc. Phys. Soc. A}\ }\textbf {\bibinfo
  {volume} {62}},\ \bibinfo {pages} {416--422} (\bibinfo {year}
  {1949})}\BibitemShut {NoStop}%
\bibitem [{\citenamefont {Anderson}(1987)}]{Anderson1987}%
  \BibitemOpen
  \bibfield  {author} {\bibinfo {author} {\bibfnamefont {P.~W.}\ \bibnamefont
  {Anderson}},\ }\bibfield  {title} {\enquote {\bibinfo {title} {{The
  Resonating Valence Bond State in $\mathrm{La}_2\mathrm{CuO}_4$ and
  Superconductivity}},}\ }\href {\doibase 10.1126/science.235.4793.1196}
  {\bibfield  {journal} {\bibinfo  {journal} {Science}\ }\textbf {\bibinfo
  {volume} {235}},\ \bibinfo {pages} {1196--1198} (\bibinfo {year}
  {1987})}\BibitemShut {NoStop}%
\bibitem [{\citenamefont {Imada}\ \emph {et~al.}(1998)\citenamefont {Imada},
  \citenamefont {Fujimori},\ and\ \citenamefont {Tokura}}]{Imada1998}%
  \BibitemOpen
  \bibfield  {author} {\bibinfo {author} {\bibfnamefont {Masatoshi}\
  \bibnamefont {Imada}}, \bibinfo {author} {\bibfnamefont {Atsushi}\
  \bibnamefont {Fujimori}}, \ and\ \bibinfo {author} {\bibfnamefont
  {Yoshinori}\ \bibnamefont {Tokura}},\ }\bibfield  {title} {\enquote {\bibinfo
  {title} {{Metal-Insulator Transitions}},}\ }\href {\doibase
  10.1103/RevModPhys.70.1039} {\bibfield  {journal} {\bibinfo  {journal} {Rev.
  Mod. Phys.}\ }\textbf {\bibinfo {volume} {70}},\ \bibinfo {pages}
  {1039--1263} (\bibinfo {year} {1998})}\BibitemShut {NoStop}%
\bibitem [{\citenamefont {Lee}\ \emph {et~al.}(2006)\citenamefont {Lee},
  \citenamefont {Nagaosa},\ and\ \citenamefont {Wen}}]{Lee2006}%
  \BibitemOpen
  \bibfield  {author} {\bibinfo {author} {\bibfnamefont {Patrick~A.}\
  \bibnamefont {Lee}}, \bibinfo {author} {\bibfnamefont {Naoto}\ \bibnamefont
  {Nagaosa}}, \ and\ \bibinfo {author} {\bibfnamefont {Xiao~Gang}\ \bibnamefont
  {Wen}},\ }\bibfield  {title} {\enquote {\bibinfo {title} {{Doping a Mott
  Insulator: Physics of High-Temperature Superconductivity}},}\ }\href
  {\doibase 10.1103/RevModPhys.78.17} {\bibfield  {journal} {\bibinfo
  {journal} {Rev. Mod. Phys.}\ }\textbf {\bibinfo {volume} {78}},\ \bibinfo
  {pages} {17--85} (\bibinfo {year} {2006})},\ \Eprint
  {http://arxiv.org/abs/cond-mat/0410445} {arXiv:cond-mat/0410445} \BibitemShut
  {NoStop}%
\bibitem [{\citenamefont {Anderson}(1997)}]{Andersonbook}%
  \BibitemOpen
  \bibfield  {author} {\bibinfo {author} {\bibfnamefont {P.W.}\ \bibnamefont
  {Anderson}},\ }\href@noop {} {\emph {\bibinfo {title} {The Theory of
  Superconductivity in the High-$T_c$ Cuprates}}},\ Princeton series in
  physics\ (\bibinfo  {publisher} {Princeton University Press},\ \bibinfo
  {year} {1997})\BibitemShut {NoStop}%
\bibitem [{\citenamefont {{Keimer}}\ \emph {et~al.}(2015)\citenamefont
  {{Keimer}}, \citenamefont {{Kivelson}}, \citenamefont {{Norman}},
  \citenamefont {{Uchida}},\ and\ \citenamefont {{Zaanen}}}]{Zaanen2015}%
  \BibitemOpen
  \bibfield  {author} {\bibinfo {author} {\bibfnamefont {B.}~\bibnamefont
  {{Keimer}}}, \bibinfo {author} {\bibfnamefont {S.~A.}\ \bibnamefont
  {{Kivelson}}}, \bibinfo {author} {\bibfnamefont {M.~R.}\ \bibnamefont
  {{Norman}}}, \bibinfo {author} {\bibfnamefont {S.}~\bibnamefont {{Uchida}}},
  \ and\ \bibinfo {author} {\bibfnamefont {J.}~\bibnamefont {{Zaanen}}},\
  }\bibfield  {title} {\enquote {\bibinfo {title} {{From Quantum Matter to
  High-Temperature Superconductivity in Copper Oxides}},}\ }\href {\doibase
  10.1038/nature14165} {\bibfield  {journal} {\bibinfo  {journal} {Nature}\
  }\textbf {\bibinfo {volume} {518}},\ \bibinfo {pages} {179--186} (\bibinfo
  {year} {2015})}\BibitemShut {NoStop}%
\bibitem [{\citenamefont {Chakravarty}\ \emph {et~al.}(1988)\citenamefont
  {Chakravarty}, \citenamefont {Halperin},\ and\ \citenamefont
  {Nelson}}]{Chakravarty1988}%
  \BibitemOpen
  \bibfield  {author} {\bibinfo {author} {\bibfnamefont {Sudip}\ \bibnamefont
  {Chakravarty}}, \bibinfo {author} {\bibfnamefont {Bertrand~I.}\ \bibnamefont
  {Halperin}}, \ and\ \bibinfo {author} {\bibfnamefont {David~R.}\ \bibnamefont
  {Nelson}},\ }\bibfield  {title} {\enquote {\bibinfo {title} {{Low-Temperature
  Behavior of Two-Dimensional Quantum Antiferromagnets}},}\ }\href {\doibase
  10.1103/PhysRevLett.60.1057} {\bibfield  {journal} {\bibinfo  {journal}
  {Phys. Rev. Lett.}\ }\textbf {\bibinfo {volume} {60}},\ \bibinfo {pages}
  {1057--1060} (\bibinfo {year} {1988})}\BibitemShut {NoStop}%
\bibitem [{\citenamefont {Chakravarty}\ \emph {et~al.}(1989)\citenamefont
  {Chakravarty}, \citenamefont {Halperin},\ and\ \citenamefont
  {Nelson}}]{Chakravarty1989}%
  \BibitemOpen
  \bibfield  {author} {\bibinfo {author} {\bibfnamefont {Sudip}\ \bibnamefont
  {Chakravarty}}, \bibinfo {author} {\bibfnamefont {Bertrand~I.}\ \bibnamefont
  {Halperin}}, \ and\ \bibinfo {author} {\bibfnamefont {David~R.}\ \bibnamefont
  {Nelson}},\ }\bibfield  {title} {\enquote {\bibinfo {title} {{Two-Dimensional
  Quantum Heisenberg Antiferromagnet at Low Temperatures}},}\ }\href {\doibase
  10.1103/PhysRevB.39.2344} {\bibfield  {journal} {\bibinfo  {journal} {Phys.
  Rev. B}\ }\textbf {\bibinfo {volume} {39}},\ \bibinfo {pages} {2344--2371}
  (\bibinfo {year} {1989})}\BibitemShut {NoStop}%
\bibitem [{\citenamefont {Auerbach}\ and\ \citenamefont
  {Arovas}(1988)}]{Auerbach1988a}%
  \BibitemOpen
  \bibfield  {author} {\bibinfo {author} {\bibfnamefont {Assa}\ \bibnamefont
  {Auerbach}}\ and\ \bibinfo {author} {\bibfnamefont {Daniel~P.}\ \bibnamefont
  {Arovas}},\ }\bibfield  {title} {\enquote {\bibinfo {title} {Spin dynamics in
  the square-lattice antiferromagnet},}\ }\href {\doibase
  10.1103/PhysRevLett.61.617} {\bibfield  {journal} {\bibinfo  {journal} {Phys.
  Rev. Lett.}\ }\textbf {\bibinfo {volume} {61}},\ \bibinfo {pages} {617--620}
  (\bibinfo {year} {1988})}\BibitemShut {NoStop}%
\bibitem [{\citenamefont {Arovas}\ and\ \citenamefont
  {Auerbach}(1988)}]{Auerbach1988}%
  \BibitemOpen
  \bibfield  {author} {\bibinfo {author} {\bibfnamefont {Daniel~P.}\
  \bibnamefont {Arovas}}\ and\ \bibinfo {author} {\bibfnamefont {Assa}\
  \bibnamefont {Auerbach}},\ }\bibfield  {title} {\enquote {\bibinfo {title}
  {Functional integral theories of low-dimensional quantum heisenberg
  models},}\ }\href {\doibase 10.1103/PhysRevB.38.316} {\bibfield  {journal}
  {\bibinfo  {journal} {Phys. Rev. B}\ }\textbf {\bibinfo {volume} {38}},\
  \bibinfo {pages} {316--332} (\bibinfo {year} {1988})}\BibitemShut {NoStop}%
\bibitem [{\citenamefont {Liang}\ \emph {et~al.}(1988)\citenamefont {Liang},
  \citenamefont {Doucot},\ and\ \citenamefont {Anderson}}]{Liang1988}%
  \BibitemOpen
  \bibfield  {author} {\bibinfo {author} {\bibfnamefont {S.}~\bibnamefont
  {Liang}}, \bibinfo {author} {\bibfnamefont {B.}~\bibnamefont {Doucot}}, \
  and\ \bibinfo {author} {\bibfnamefont {P.~W.}\ \bibnamefont {Anderson}},\
  }\bibfield  {title} {\enquote {\bibinfo {title} {{Some New Variational
  Resonating-Valence-Bond-Type Wave Functions for the Spin-$\frac{1}{2}$
  Antiferromagnetic Heisenberg Model on a Square Lattice}},}\ }\href {\doibase
  10.1103/PhysRevLett.61.365} {\bibfield  {journal} {\bibinfo  {journal} {Phys.
  Rev. Lett.}\ }\textbf {\bibinfo {volume} {61}},\ \bibinfo {pages} {365--368}
  (\bibinfo {year} {1988})}\BibitemShut {NoStop}%
\bibitem [{\citenamefont {Manousakis}(1991)}]{Manousakis1991}%
  \BibitemOpen
  \bibfield  {author} {\bibinfo {author} {\bibfnamefont {Efstratios}\
  \bibnamefont {Manousakis}},\ }\bibfield  {title} {\enquote {\bibinfo {title}
  {{The spin-$\frac{1}{2}$ Heisenberg Antiferromagnet on a Square Lattice and
  its Application to the Cuprous Oxides}},}\ }\href {\doibase
  10.1103/RevModPhys.63.1} {\bibfield  {journal} {\bibinfo  {journal} {Rev.
  Mod. Phys.}\ }\textbf {\bibinfo {volume} {63}},\ \bibinfo {pages} {1--62}
  (\bibinfo {year} {1991})}\BibitemShut {NoStop}%
\bibitem [{\citenamefont {Anderson}(1973)}]{Anderson1973}%
  \BibitemOpen
  \bibfield  {author} {\bibinfo {author} {\bibfnamefont {P.~W.}\ \bibnamefont
  {Anderson}},\ }\bibfield  {title} {\enquote {\bibinfo {title} {{Resonating
  Valence Bonds: A New Kind of Insulator?}}}\ }\href {\doibase
  10.1016/0025-5408(73)90167-0} {\bibfield  {journal} {\bibinfo  {journal}
  {Mat. Res. Bull.}\ }\textbf {\bibinfo {volume} {8}},\ \bibinfo {pages}
  {153--160} (\bibinfo {year} {1973})}\BibitemShut {NoStop}%
\bibitem [{\citenamefont {Anderson}\ \emph {et~al.}(2004)\citenamefont
  {Anderson}, \citenamefont {Lee}, \citenamefont {Randeria}, \citenamefont
  {Rice}, \citenamefont {Trivedi},\ and\ \citenamefont {Zhang}}]{Anderson2004}%
  \BibitemOpen
  \bibfield  {author} {\bibinfo {author} {\bibfnamefont {P.~W.}\ \bibnamefont
  {Anderson}}, \bibinfo {author} {\bibfnamefont {P.~A.}\ \bibnamefont {Lee}},
  \bibinfo {author} {\bibfnamefont {M.}~\bibnamefont {Randeria}}, \bibinfo
  {author} {\bibfnamefont {T.~M.}\ \bibnamefont {Rice}}, \bibinfo {author}
  {\bibfnamefont {N.}~\bibnamefont {Trivedi}}, \ and\ \bibinfo {author}
  {\bibfnamefont {F.~C.}\ \bibnamefont {Zhang}},\ }\bibfield  {title} {\enquote
  {\bibinfo {title} {{The Physics behind High-Temperature Superconducting
  Cuprates: the Plain Vanilla Version of RVB}},}\ }\href {\doibase
  10.1088/0953-8984/16/24/R02} {\bibfield  {journal} {\bibinfo  {journal} {J.
  Phys.: Condens. Matter}\ }\textbf {\bibinfo {volume} {16}},\ \bibinfo {pages}
  {R755--R769} (\bibinfo {year} {2004})},\ \Eprint
  {http://arxiv.org/abs/cond-mat/0311467} {arXiv:cond-mat/0311467} \BibitemShut
  {NoStop}%
\bibitem [{\citenamefont {Baskaran}\ \emph {et~al.}(1987)\citenamefont
  {Baskaran}, \citenamefont {Zou},\ and\ \citenamefont
  {Anderson}}]{Baskaran1987}%
  \BibitemOpen
  \bibfield  {author} {\bibinfo {author} {\bibfnamefont {G.}~\bibnamefont
  {Baskaran}}, \bibinfo {author} {\bibfnamefont {Z.}~\bibnamefont {Zou}}, \
  and\ \bibinfo {author} {\bibfnamefont {P.~W.}\ \bibnamefont {Anderson}},\
  }\bibfield  {title} {\enquote {\bibinfo {title} {{The Resonating Valence Bond
  State and High-$T_c$ Superconductivity — A Mean Field Theory}},}\ }\href
  {\doibase 10.1016/0038-1098(87)90642-9} {\bibfield  {journal} {\bibinfo
  {journal} {Solid State Commun.}\ }\textbf {\bibinfo {volume} {63}},\ \bibinfo
  {pages} {973--976} (\bibinfo {year} {1987})}\BibitemShut {NoStop}%
\bibitem [{\citenamefont {Kivelson}\ \emph {et~al.}(1987)\citenamefont
  {Kivelson}, \citenamefont {Rokhsar},\ and\ \citenamefont
  {Sethna}}]{Kivelson1987}%
  \BibitemOpen
  \bibfield  {author} {\bibinfo {author} {\bibfnamefont {Steven~A.}\
  \bibnamefont {Kivelson}}, \bibinfo {author} {\bibfnamefont {Daniel~S.}\
  \bibnamefont {Rokhsar}}, \ and\ \bibinfo {author} {\bibfnamefont {James~P.}\
  \bibnamefont {Sethna}},\ }\bibfield  {title} {\enquote {\bibinfo {title}
  {Topology of the resonating valence-bond state: Solitons and high-${T}_{c}$
  superconductivity},}\ }\href {\doibase 10.1103/PhysRevB.35.8865} {\bibfield
  {journal} {\bibinfo  {journal} {Phys. Rev. B}\ }\textbf {\bibinfo {volume}
  {35}},\ \bibinfo {pages} {8865--8868} (\bibinfo {year} {1987})}\BibitemShut
  {NoStop}%
\bibitem [{\citenamefont {Lee}(1989)}]{Lee1989}%
  \BibitemOpen
  \bibfield  {author} {\bibinfo {author} {\bibfnamefont {Patrick~A.}\
  \bibnamefont {Lee}},\ }\bibfield  {title} {\enquote {\bibinfo {title} {Gauge
  field, aharonov-bohm flux, and high-${T}_{c}$ superconductivity},}\ }\href
  {\doibase 10.1103/PhysRevLett.63.680} {\bibfield  {journal} {\bibinfo
  {journal} {Phys. Rev. Lett.}\ }\textbf {\bibinfo {volume} {63}},\ \bibinfo
  {pages} {680--683} (\bibinfo {year} {1989})}\BibitemShut {NoStop}%
\bibitem [{\citenamefont {Fradkin}\ and\ \citenamefont
  {Kivelson}(1990)}]{Fradkin1990}%
  \BibitemOpen
  \bibfield  {author} {\bibinfo {author} {\bibfnamefont {Eduardo}\ \bibnamefont
  {Fradkin}}\ and\ \bibinfo {author} {\bibfnamefont {Steven}\ \bibnamefont
  {Kivelson}},\ }\bibfield  {title} {\enquote {\bibinfo {title} {{Short Range
  Resonating Valence Bond Theories and Superconductivity}},}\ }\href {\doibase
  10.1142/S0217984990000295} {\bibfield  {journal} {\bibinfo  {journal} {Mod.
  Phys. Lett. B}\ }\textbf {\bibinfo {volume} {04}},\ \bibinfo {pages}
  {225--232} (\bibinfo {year} {1990})}\BibitemShut {NoStop}%
\bibitem [{\citenamefont {Lee}\ and\ \citenamefont {Nagaosa}(1992)}]{Lee1992}%
  \BibitemOpen
  \bibfield  {author} {\bibinfo {author} {\bibfnamefont {Patrick~A.}\
  \bibnamefont {Lee}}\ and\ \bibinfo {author} {\bibfnamefont {Naoto}\
  \bibnamefont {Nagaosa}},\ }\bibfield  {title} {\enquote {\bibinfo {title}
  {{Gauge Theory of the Normal State of High-$T_c$ Superconductors}},}\ }\href
  {\doibase 10.1103/PhysRevB.46.5621} {\bibfield  {journal} {\bibinfo
  {journal} {Phys. Rev. B}\ }\textbf {\bibinfo {volume} {46}},\ \bibinfo
  {pages} {5621--5639} (\bibinfo {year} {1992})}\BibitemShut {NoStop}%
\bibitem [{\citenamefont {Wen}\ and\ \citenamefont {Lee}(1996)}]{Wen1996}%
  \BibitemOpen
  \bibfield  {author} {\bibinfo {author} {\bibfnamefont {Xiao-Gang}\
  \bibnamefont {Wen}}\ and\ \bibinfo {author} {\bibfnamefont {Patrick~A.}\
  \bibnamefont {Lee}},\ }\bibfield  {title} {\enquote {\bibinfo {title}
  {{Theory of Underdoped Cuprates}},}\ }\href {\doibase
  10.1103/PhysRevLett.76.503} {\bibfield  {journal} {\bibinfo  {journal} {Phys.
  Rev. Lett.}\ }\textbf {\bibinfo {volume} {76}},\ \bibinfo {pages} {503--506}
  (\bibinfo {year} {1996})},\ \Eprint {http://arxiv.org/abs/cond-mat/9506065}
  {arXiv:cond-mat/9506065} \BibitemShut {NoStop}%
\bibitem [{\citenamefont {Senthil}\ and\ \citenamefont
  {Fisher}(2000)}]{Senthil2000}%
  \BibitemOpen
  \bibfield  {author} {\bibinfo {author} {\bibfnamefont {T.}~\bibnamefont
  {Senthil}}\ and\ \bibinfo {author} {\bibfnamefont {Matthew P.~A.}\
  \bibnamefont {Fisher}},\ }\bibfield  {title} {\enquote {\bibinfo {title}
  {${Z}_{2}$ gauge theory of electron fractionalization in strongly correlated
  systems},}\ }\href {\doibase 10.1103/PhysRevB.62.7850} {\bibfield  {journal}
  {\bibinfo  {journal} {Phys. Rev. B}\ }\textbf {\bibinfo {volume} {62}},\
  \bibinfo {pages} {7850--7881} (\bibinfo {year} {2000})}\BibitemShut {NoStop}%
\bibitem [{\citenamefont {Sorella}\ \emph {et~al.}(2002)\citenamefont
  {Sorella}, \citenamefont {Martins}, \citenamefont {Becca}, \citenamefont
  {Gazza}, \citenamefont {Capriotti}, \citenamefont {Parola},\ and\
  \citenamefont {Dagotto}}]{Sorella2002}%
  \BibitemOpen
  \bibfield  {author} {\bibinfo {author} {\bibfnamefont {S.}~\bibnamefont
  {Sorella}}, \bibinfo {author} {\bibfnamefont {G.~B.}\ \bibnamefont
  {Martins}}, \bibinfo {author} {\bibfnamefont {F.}~\bibnamefont {Becca}},
  \bibinfo {author} {\bibfnamefont {C.}~\bibnamefont {Gazza}}, \bibinfo
  {author} {\bibfnamefont {L.}~\bibnamefont {Capriotti}}, \bibinfo {author}
  {\bibfnamefont {A.}~\bibnamefont {Parola}}, \ and\ \bibinfo {author}
  {\bibfnamefont {E.}~\bibnamefont {Dagotto}},\ }\bibfield  {title} {\enquote
  {\bibinfo {title} {{Superconductivity in the Two-Dimensional $t$-$J$
  Model}},}\ }\href {\doibase 10.1103/PhysRevLett.88.117002} {\bibfield
  {journal} {\bibinfo  {journal} {Phys. Rev. Lett.}\ }\textbf {\bibinfo
  {volume} {88}},\ \bibinfo {pages} {117002} (\bibinfo {year}
  {2002})}\BibitemShut {NoStop}%
\bibitem [{\citenamefont {Edegger}\ \emph {et~al.}(2007)\citenamefont
  {Edegger}, \citenamefont {Muthukumar},\ and\ \citenamefont
  {Gros}}]{Edegger2007}%
  \BibitemOpen
  \bibfield  {author} {\bibinfo {author} {\bibfnamefont {B.}~\bibnamefont
  {Edegger}}, \bibinfo {author} {\bibfnamefont {V.~N.}\ \bibnamefont
  {Muthukumar}}, \ and\ \bibinfo {author} {\bibfnamefont {C.}~\bibnamefont
  {Gros}},\ }\bibfield  {title} {\enquote {\bibinfo {title} {{Gutzwiller–RVB
  Theory of High-Temperature Superconductivity: Results from Renormalized
  Mean-Field Theory and Variational Monte Carlo Calculations}},}\ }\href
  {\doibase 10.1080/00018730701627707} {\bibfield  {journal} {\bibinfo
  {journal} {Adv. Phys.}\ }\textbf {\bibinfo {volume} {56}},\ \bibinfo {pages}
  {927--1033} (\bibinfo {year} {2007})},\ \Eprint
  {http://arxiv.org/abs/0707.1020} {arXiv:0707.1020} \BibitemShut {NoStop}%
\bibitem [{\citenamefont {Anderson}(1990)}]{Anderson1990}%
  \BibitemOpen
  \bibfield  {author} {\bibinfo {author} {\bibfnamefont {P.~W.}\ \bibnamefont
  {Anderson}},\ }\bibfield  {title} {\enquote {\bibinfo {title}
  {{‘‘Luttinger-Liquid'' Behavior of the Normal Metallic State of the 2D
  Hubbard Model}},}\ }\href {\doibase 10.1103/PhysRevLett.64.1839} {\bibfield
  {journal} {\bibinfo  {journal} {Phys. Rev. Lett.}\ }\textbf {\bibinfo
  {volume} {64}},\ \bibinfo {pages} {1839--1841} (\bibinfo {year}
  {1990})}\BibitemShut {NoStop}%
\bibitem [{\citenamefont {Anderson}(1967{\natexlab{a}})}]{Anderson1967a}%
  \BibitemOpen
  \bibfield  {author} {\bibinfo {author} {\bibfnamefont {P.~W.}\ \bibnamefont
  {Anderson}},\ }\bibfield  {title} {\enquote {\bibinfo {title} {{Infrared
  Catastrophe in Fermi Gases with Local Scattering Potentials}},}\ }\href
  {\doibase 10.1103/PhysRevLett.18.1049} {\bibfield  {journal} {\bibinfo
  {journal} {Phys. Rev. Lett.}\ }\textbf {\bibinfo {volume} {18}},\ \bibinfo
  {pages} {1049--1051} (\bibinfo {year} {1967}{\natexlab{a}})}\BibitemShut
  {NoStop}%
\bibitem [{\citenamefont {Anderson}(1967{\natexlab{b}})}]{Anderson1967b}%
  \BibitemOpen
  \bibfield  {author} {\bibinfo {author} {\bibfnamefont {P.~W.}\ \bibnamefont
  {Anderson}},\ }\bibfield  {title} {\enquote {\bibinfo {title} {{Ground State
  of a Magnetic Impurity in a Metal}},}\ }\href {\doibase
  10.1103/PhysRev.164.352} {\bibfield  {journal} {\bibinfo  {journal} {Phys.
  Rev.}\ }\textbf {\bibinfo {volume} {164}},\ \bibinfo {pages} {352--359}
  (\bibinfo {year} {1967}{\natexlab{b}})}\BibitemShut {NoStop}%
\bibitem [{\citenamefont {Sheng}\ \emph {et~al.}(1996)\citenamefont {Sheng},
  \citenamefont {Chen},\ and\ \citenamefont {Weng}}]{Sheng1996}%
  \BibitemOpen
  \bibfield  {author} {\bibinfo {author} {\bibfnamefont {D.~N.}\ \bibnamefont
  {Sheng}}, \bibinfo {author} {\bibfnamefont {Y.~C.}\ \bibnamefont {Chen}}, \
  and\ \bibinfo {author} {\bibfnamefont {Z.~Y.}\ \bibnamefont {Weng}},\
  }\bibfield  {title} {\enquote {\bibinfo {title} {{Phase String Effect in a
  Doped Antiferromagnet}},}\ }\href {\doibase 10.1103/PhysRevLett.77.5102}
  {\bibfield  {journal} {\bibinfo  {journal} {Phys. Rev. Lett.}\ }\textbf
  {\bibinfo {volume} {77}},\ \bibinfo {pages} {5102--5105} (\bibinfo {year}
  {1996})}\BibitemShut {NoStop}%
\bibitem [{\citenamefont {Weng}\ \emph {et~al.}(1997)\citenamefont {Weng},
  \citenamefont {Sheng}, \citenamefont {Chen},\ and\ \citenamefont
  {Ting}}]{Weng1997}%
  \BibitemOpen
  \bibfield  {author} {\bibinfo {author} {\bibfnamefont {Z.~Y.}\ \bibnamefont
  {Weng}}, \bibinfo {author} {\bibfnamefont {D.~N.}\ \bibnamefont {Sheng}},
  \bibinfo {author} {\bibfnamefont {Y.-C.}\ \bibnamefont {Chen}}, \ and\
  \bibinfo {author} {\bibfnamefont {C.~S.}\ \bibnamefont {Ting}},\ }\bibfield
  {title} {\enquote {\bibinfo {title} {{Phase String Effect in the $t$-$J$
  Model: General Theory}},}\ }\href {\doibase 10.1103/PhysRevB.55.3894}
  {\bibfield  {journal} {\bibinfo  {journal} {Phys. Rev. B}\ }\textbf {\bibinfo
  {volume} {55}},\ \bibinfo {pages} {3894--3906} (\bibinfo {year}
  {1997})}\BibitemShut {NoStop}%
\bibitem [{\citenamefont {Wu}\ \emph {et~al.}(2008)\citenamefont {Wu},
  \citenamefont {Weng},\ and\ \citenamefont {Zaanen}}]{Wu2008}%
  \BibitemOpen
  \bibfield  {author} {\bibinfo {author} {\bibfnamefont {K.}~\bibnamefont
  {Wu}}, \bibinfo {author} {\bibfnamefont {Z.~Y.}\ \bibnamefont {Weng}}, \ and\
  \bibinfo {author} {\bibfnamefont {J.}~\bibnamefont {Zaanen}},\ }\bibfield
  {title} {\enquote {\bibinfo {title} {{Sign Structure of the $t$-$J$
  model}},}\ }\href {\doibase 10.1103/PhysRevB.77.155102} {\bibfield  {journal}
  {\bibinfo  {journal} {Phys. Rev. B}\ }\textbf {\bibinfo {volume} {77}},\
  \bibinfo {pages} {155102} (\bibinfo {year} {2008})},\ \Eprint
  {http://arxiv.org/abs/0802.0273} {arXiv:0802.0273} \BibitemShut {NoStop}%
\bibitem [{\citenamefont {Zhang}\ and\ \citenamefont {Weng}(2014)}]{Zhang2014}%
  \BibitemOpen
  \bibfield  {author} {\bibinfo {author} {\bibfnamefont {Long}\ \bibnamefont
  {Zhang}}\ and\ \bibinfo {author} {\bibfnamefont {Zheng~Yu}\ \bibnamefont
  {Weng}},\ }\bibfield  {title} {\enquote {\bibinfo {title} {{Sign structure,
  electron fractionalization, and emergent gauge description of the Hubbard
  model}},}\ }\href {\doibase 10.1103/PhysRevB.90.165120} {\bibfield  {journal}
  {\bibinfo  {journal} {Phys. Rev. B}\ }\textbf {\bibinfo {volume} {90}},\
  \bibinfo {pages} {165120} (\bibinfo {year} {2014})},\ \Eprint
  {http://arxiv.org/abs/1406.6867} {arXiv:1406.6867} \BibitemShut {NoStop}%
\bibitem [{\citenamefont {Zaanen}\ and\ \citenamefont
  {Overbosch}(2011)}]{Zaanen2011}%
  \BibitemOpen
  \bibfield  {author} {\bibinfo {author} {\bibfnamefont {J.}~\bibnamefont
  {Zaanen}}\ and\ \bibinfo {author} {\bibfnamefont {B.~J.}\ \bibnamefont
  {Overbosch}},\ }\bibfield  {title} {\enquote {\bibinfo {title} {{Mottness
  Collapse and Statistical Quantum Criticality}},}\ }\href {\doibase
  10.1098/rsta.2010.0188} {\bibfield  {journal} {\bibinfo  {journal} {Phil.
  Trans. R. Soc. A}\ }\textbf {\bibinfo {volume} {369}},\ \bibinfo {pages}
  {1599--1625} (\bibinfo {year} {2011})}\BibitemShut {NoStop}%
\bibitem [{\citenamefont {{Zheng}}\ and\ \citenamefont
  {{Weng}}(2018)}]{Zheng2018a}%
  \BibitemOpen
  \bibfield  {author} {\bibinfo {author} {\bibfnamefont {Wayne}\ \bibnamefont
  {{Zheng}}}\ and\ \bibinfo {author} {\bibfnamefont {Zheng-Yu}\ \bibnamefont
  {{Weng}}},\ }\bibfield  {title} {\enquote {\bibinfo {title} {{Charge-Spin
  Mutual Entanglement: A Case Study by Exact Diagonalization of the One Hole
  Doped $t$-$J$ loop}},}\ }\href {\doibase 10.1038/s41598-018-21775-2}
  {\bibfield  {journal} {\bibinfo  {journal} {Sci. Rep.}\ }\textbf {\bibinfo
  {volume} {8}},\ \bibinfo {eid} {3612} (\bibinfo {year} {2018})},\ \Eprint
  {http://arxiv.org/abs/1703.04255} {arXiv:1703.04255} \BibitemShut {NoStop}%
\bibitem [{\citenamefont {Weng}(2011{\natexlab{a}})}]{Weng2011b}%
  \BibitemOpen
  \bibfield  {author} {\bibinfo {author} {\bibfnamefont {Zheng-Yu}\
  \bibnamefont {Weng}},\ }\bibfield  {title} {\enquote {\bibinfo {title} {{Mott
  Physics, Sign Structure, Ground State Wavefunction, and High-$T_c$
  Superconductivity}},}\ }\href {\doibase 10.1007/s11467-011-0220-1} {\bibfield
   {journal} {\bibinfo  {journal} {Front. Phys.}\ }\textbf {\bibinfo {volume}
  {6}},\ \bibinfo {pages} {370--378} (\bibinfo {year} {2011}{\natexlab{a}})},\
  \Eprint {http://arxiv.org/abs/1110.0546} {arXiv:1110.0546} \BibitemShut
  {NoStop}%
\bibitem [{\citenamefont {Weng}(2011{\natexlab{b}})}]{Weng2011a}%
  \BibitemOpen
  \bibfield  {author} {\bibinfo {author} {\bibfnamefont {Zheng-Yu}\
  \bibnamefont {Weng}},\ }\bibfield  {title} {\enquote {\bibinfo {title}
  {{Superconducting Ground State of a Doped Mott Insulator}},}\ }\href
  {\doibase 10.1088/1367-2630/13/10/103039} {\bibfield  {journal} {\bibinfo
  {journal} {New J. Phys.}\ }\textbf {\bibinfo {volume} {13}},\ \bibinfo
  {pages} {103039} (\bibinfo {year} {2011}{\natexlab{b}})},\ \Eprint
  {http://arxiv.org/abs/1105.3027} {arXiv:1105.3027} \BibitemShut {NoStop}%
\bibitem [{\citenamefont {Weng}\ \emph {et~al.}(1998)\citenamefont {Weng},
  \citenamefont {Sheng},\ and\ \citenamefont {Ting}}]{Weng1998}%
  \BibitemOpen
  \bibfield  {author} {\bibinfo {author} {\bibfnamefont {Z.~Y.}\ \bibnamefont
  {Weng}}, \bibinfo {author} {\bibfnamefont {D.~N.}\ \bibnamefont {Sheng}}, \
  and\ \bibinfo {author} {\bibfnamefont {C.~S.}\ \bibnamefont {Ting}},\
  }\bibfield  {title} {\enquote {\bibinfo {title} {{Bosonic
  Resonating-Valence-Bond Description of a Doped Antiferromagnet}},}\ }\href
  {\doibase 10.1103/PhysRevLett.80.5401} {\bibfield  {journal} {\bibinfo
  {journal} {Phys. Rev. Lett.}\ }\textbf {\bibinfo {volume} {80}},\ \bibinfo
  {pages} {5401--5404} (\bibinfo {year} {1998})}\BibitemShut {NoStop}%
\bibitem [{\citenamefont {Weng}\ \emph {et~al.}(1999)\citenamefont {Weng},
  \citenamefont {Sheng},\ and\ \citenamefont {Ting}}]{Weng1999}%
  \BibitemOpen
  \bibfield  {author} {\bibinfo {author} {\bibfnamefont {Z.~Y.}\ \bibnamefont
  {Weng}}, \bibinfo {author} {\bibfnamefont {D.~N.}\ \bibnamefont {Sheng}}, \
  and\ \bibinfo {author} {\bibfnamefont {C.~S.}\ \bibnamefont {Ting}},\
  }\bibfield  {title} {\enquote {\bibinfo {title} {{Mean-Field Description of
  the Phase String Effect in the $t$-$J$ Model}},}\ }\href {\doibase
  10.1103/PhysRevB.59.8943} {\bibfield  {journal} {\bibinfo  {journal} {Phys.
  Rev. B}\ }\textbf {\bibinfo {volume} {59}},\ \bibinfo {pages} {8943--8955}
  (\bibinfo {year} {1999})}\BibitemShut {NoStop}%
\bibitem [{\citenamefont {Ma}\ \emph {et~al.}(2014)\citenamefont {Ma},
  \citenamefont {Ye},\ and\ \citenamefont {Weng}}]{Ma2014}%
  \BibitemOpen
  \bibfield  {author} {\bibinfo {author} {\bibfnamefont {Yao}\ \bibnamefont
  {Ma}}, \bibinfo {author} {\bibfnamefont {Peng}\ \bibnamefont {Ye}}, \ and\
  \bibinfo {author} {\bibfnamefont {Zheng-Yu}\ \bibnamefont {Weng}},\
  }\bibfield  {title} {\enquote {\bibinfo {title} {{Low-Temperature Pseudogap
  Phenomenon: Precursor of High-$T_c$ Superconductivity}},}\ }\href {\doibase
  10.1088/1367-2630/16/8/083039} {\bibfield  {journal} {\bibinfo  {journal}
  {New J. Phys.}\ }\textbf {\bibinfo {volume} {16}},\ \bibinfo {pages} {083039}
  (\bibinfo {year} {2014})},\ \Eprint {http://arxiv.org/abs/1311.3395}
  {arXiv:1311.3395} \BibitemShut {NoStop}%
\bibitem [{\citenamefont {Wang}\ \emph {et~al.}(2015)\citenamefont {Wang},
  \citenamefont {Zhu}, \citenamefont {Qi},\ and\ \citenamefont
  {Weng}}]{Wang2015}%
  \BibitemOpen
  \bibfield  {author} {\bibinfo {author} {\bibfnamefont {Qing-Rui}\
  \bibnamefont {Wang}}, \bibinfo {author} {\bibfnamefont {Zheng}\ \bibnamefont
  {Zhu}}, \bibinfo {author} {\bibfnamefont {Yang}\ \bibnamefont {Qi}}, \ and\
  \bibinfo {author} {\bibfnamefont {Zheng-Yu}\ \bibnamefont {Weng}},\
  }\bibfield  {title} {\enquote {\bibinfo {title} {{Variational Wave Function
  for an Anisotropic Single-hole-doped $t$-$J$ Ladder}},}\ }\href@noop {} {\
  (\bibinfo {year} {2015})},\ \Eprint {http://arxiv.org/abs/1509.01260}
  {arXiv:1509.01260} \BibitemShut {NoStop}%
\bibitem [{\citenamefont {Zhu}\ \emph {et~al.}(2016)\citenamefont {Zhu},
  \citenamefont {Wang}, \citenamefont {Sheng},\ and\ \citenamefont
  {Weng}}]{Zhu2016}%
  \BibitemOpen
  \bibfield  {author} {\bibinfo {author} {\bibfnamefont {Zheng}\ \bibnamefont
  {Zhu}}, \bibinfo {author} {\bibfnamefont {Qing-Rui}\ \bibnamefont {Wang}},
  \bibinfo {author} {\bibfnamefont {D.~N.}\ \bibnamefont {Sheng}}, \ and\
  \bibinfo {author} {\bibfnamefont {Zheng-Yu}\ \bibnamefont {Weng}},\
  }\bibfield  {title} {\enquote {\bibinfo {title} {{Exact Sign Structure of the
  $t$-$J$ Chain and the Single Hole Ground State}},}\ }\href {\doibase
  10.1016/j.nuclphysb.2015.12.004} {\bibfield  {journal} {\bibinfo  {journal}
  {Nucl. Phys. B}\ }\textbf {\bibinfo {volume} {903}},\ \bibinfo {pages}
  {51--77} (\bibinfo {year} {2016})},\ \Eprint
  {http://arxiv.org/abs/1510.07634} {arXiv:1510.07634} \BibitemShut {NoStop}%
\bibitem [{\citenamefont {Chen}\ \emph {et~al.}(2019)\citenamefont {Chen},
  \citenamefont {Wang}, \citenamefont {Qi}, \citenamefont {Sheng},\ and\
  \citenamefont {Weng}}]{Chen2019}%
  \BibitemOpen
  \bibfield  {author} {\bibinfo {author} {\bibfnamefont {Shuai}\ \bibnamefont
  {Chen}}, \bibinfo {author} {\bibfnamefont {Qing-Rui}\ \bibnamefont {Wang}},
  \bibinfo {author} {\bibfnamefont {Yang}\ \bibnamefont {Qi}}, \bibinfo
  {author} {\bibfnamefont {D.~N.}\ \bibnamefont {Sheng}}, \ and\ \bibinfo
  {author} {\bibfnamefont {Zheng-Yu}\ \bibnamefont {Weng}},\ }\bibfield
  {title} {\enquote {\bibinfo {title} {{Single-Hole Wave Function in Two
  Dimensions: A Case Study of the Doped Mott Insulator}},}\ }\href {\doibase
  10.1103/PhysRevB.99.205128} {\bibfield  {journal} {\bibinfo  {journal} {Phys.
  Rev. B}\ }\textbf {\bibinfo {volume} {99}},\ \bibinfo {pages} {205128}
  (\bibinfo {year} {2019})},\ \Eprint {http://arxiv.org/abs/1812.05627}
  {arXiv:1812.05627} \BibitemShut {NoStop}%
\bibitem [{\citenamefont {Zheng}\ \emph {et~al.}(2018)\citenamefont {Zheng},
  \citenamefont {Zhu}, \citenamefont {Sheng},\ and\ \citenamefont
  {Weng}}]{Zheng2018b}%
  \BibitemOpen
  \bibfield  {author} {\bibinfo {author} {\bibfnamefont {Wayne}\ \bibnamefont
  {Zheng}}, \bibinfo {author} {\bibfnamefont {Zheng}\ \bibnamefont {Zhu}},
  \bibinfo {author} {\bibfnamefont {D.~N.}\ \bibnamefont {Sheng}}, \ and\
  \bibinfo {author} {\bibfnamefont {Zheng-Yu}\ \bibnamefont {Weng}},\
  }\bibfield  {title} {\enquote {\bibinfo {title} {{Hidden Spin Current in
  Doped Mott Antiferromagnets}},}\ }\href {\doibase 10.1103/PhysRevB.98.165102}
  {\bibfield  {journal} {\bibinfo  {journal} {Phys. Rev. B}\ }\textbf {\bibinfo
  {volume} {98}},\ \bibinfo {pages} {165102} (\bibinfo {year} {2018})},\
  \Eprint {http://arxiv.org/abs/1802.05977} {arXiv:1802.05977} \BibitemShut
  {NoStop}%
\bibitem [{\citenamefont {Chen}\ \emph {et~al.}(2018)\citenamefont {Chen},
  \citenamefont {Zhu},\ and\ \citenamefont {Weng}}]{Chen2018}%
  \BibitemOpen
  \bibfield  {author} {\bibinfo {author} {\bibfnamefont {Shuai}\ \bibnamefont
  {Chen}}, \bibinfo {author} {\bibfnamefont {Zheng}\ \bibnamefont {Zhu}}, \
  and\ \bibinfo {author} {\bibfnamefont {Zheng-Yu}\ \bibnamefont {Weng}},\
  }\bibfield  {title} {\enquote {\bibinfo {title} {{Two-hole Ground State
  Wavefunction: Non-BCS Pairing in a $t$-$J$ Two-leg Ladder}},}\ }\href
  {\doibase 10.1103/PhysRevB.98.245138} {\bibfield  {journal} {\bibinfo
  {journal} {Phys. Rev. B}\ }\textbf {\bibinfo {volume} {98}},\ \bibinfo
  {pages} {245138} (\bibinfo {year} {2018})},\ \Eprint
  {http://arxiv.org/abs/1808.06173} {arXiv:1808.06173} \BibitemShut {NoStop}%
\bibitem [{\citenamefont {{Zhu}}\ \emph {et~al.}(2018)\citenamefont {{Zhu}},
  \citenamefont {{Sheng}},\ and\ \citenamefont {{Weng}}}]{Zhu2018}%
  \BibitemOpen
  \bibfield  {author} {\bibinfo {author} {\bibfnamefont {Zheng}\ \bibnamefont
  {{Zhu}}}, \bibinfo {author} {\bibfnamefont {D.~N.}\ \bibnamefont {{Sheng}}},
  \ and\ \bibinfo {author} {\bibfnamefont {Zheng-Yu}\ \bibnamefont {{Weng}}},\
  }\bibfield  {title} {\enquote {\bibinfo {title} {{Pairing Versus Phase
  Coherence of Doped Holes in Distinct Quantum Spin Backgrounds}},}\ }\href
  {\doibase 10.1103/PhysRevB.97.115144} {\bibfield  {journal} {\bibinfo
  {journal} {Phys. Rev. B}\ }\textbf {\bibinfo {volume} {97}},\ \bibinfo {eid}
  {115144} (\bibinfo {year} {2018})},\ \Eprint
  {http://arxiv.org/abs/1706.02305} {arXiv:1706.02305} \BibitemShut {NoStop}%
\bibitem [{\citenamefont {Zhu}\ \emph {et~al.}(2014)\citenamefont {Zhu},
  \citenamefont {Jiang}, \citenamefont {Sheng},\ and\ \citenamefont
  {Weng}}]{Zhu2014}%
  \BibitemOpen
  \bibfield  {author} {\bibinfo {author} {\bibfnamefont {Zheng}\ \bibnamefont
  {Zhu}}, \bibinfo {author} {\bibfnamefont {Hong~Chen}\ \bibnamefont {Jiang}},
  \bibinfo {author} {\bibfnamefont {D.~N.}\ \bibnamefont {Sheng}}, \ and\
  \bibinfo {author} {\bibfnamefont {Zheng~Yu}\ \bibnamefont {Weng}},\
  }\bibfield  {title} {\enquote {\bibinfo {title} {{Nature of Strong Hole
  Pairing in Doped Mott Antiferromagnets}},}\ }\href {\doibase
  10.1038/srep05419} {\bibfield  {journal} {\bibinfo  {journal} {Sci. Rep.}\
  }\textbf {\bibinfo {volume} {4}},\ \bibinfo {pages} {5419} (\bibinfo {year}
  {2014})},\ \Eprint {http://arxiv.org/abs/1312.6893} {arXiv:1312.6893}
  \BibitemShut {NoStop}%
\bibitem [{\citenamefont {Brinkman}\ and\ \citenamefont
  {Rice}(1970)}]{Brinkman1970}%
  \BibitemOpen
  \bibfield  {author} {\bibinfo {author} {\bibfnamefont {W.~F.}\ \bibnamefont
  {Brinkman}}\ and\ \bibinfo {author} {\bibfnamefont {T.~M.}\ \bibnamefont
  {Rice}},\ }\bibfield  {title} {\enquote {\bibinfo {title} {{Single-Particle
  Excitations in Magnetic Insulators}},}\ }\href {\doibase
  10.1103/PhysRevB.2.1324} {\bibfield  {journal} {\bibinfo  {journal} {Phys.
  Rev. B}\ }\textbf {\bibinfo {volume} {2}},\ \bibinfo {pages} {1324--1338}
  (\bibinfo {year} {1970})}\BibitemShut {NoStop}%
\bibitem [{\citenamefont {Schmitt-Rink}\ \emph {et~al.}(1988)\citenamefont
  {Schmitt-Rink}, \citenamefont {Varma},\ and\ \citenamefont
  {Ruckenstein}}]{Schmitt-Rink1988}%
  \BibitemOpen
  \bibfield  {author} {\bibinfo {author} {\bibfnamefont {S.}~\bibnamefont
  {Schmitt-Rink}}, \bibinfo {author} {\bibfnamefont {C.~M.}\ \bibnamefont
  {Varma}}, \ and\ \bibinfo {author} {\bibfnamefont {A.~E.}\ \bibnamefont
  {Ruckenstein}},\ }\bibfield  {title} {\enquote {\bibinfo {title} {{Spectral
  Function of Holes in a Quantum Antiferromagnet}},}\ }\href {\doibase
  10.1103/PhysRevLett.60.2793} {\bibfield  {journal} {\bibinfo  {journal}
  {Phys. Rev. Lett.}\ }\textbf {\bibinfo {volume} {60}},\ \bibinfo {pages}
  {2793--2796} (\bibinfo {year} {1988})}\BibitemShut {NoStop}%
\bibitem [{\citenamefont {Kane}\ \emph {et~al.}(1989)\citenamefont {Kane},
  \citenamefont {Lee},\ and\ \citenamefont {Read}}]{Kane1989}%
  \BibitemOpen
  \bibfield  {author} {\bibinfo {author} {\bibfnamefont {C.~L.}\ \bibnamefont
  {Kane}}, \bibinfo {author} {\bibfnamefont {P.~A.}\ \bibnamefont {Lee}}, \
  and\ \bibinfo {author} {\bibfnamefont {N.}~\bibnamefont {Read}},\ }\bibfield
  {title} {\enquote {\bibinfo {title} {{Motion of a Single Hole in a Quantum
  Antiferromagnet}},}\ }\href {\doibase 10.1103/PhysRevB.39.6880} {\bibfield
  {journal} {\bibinfo  {journal} {Phys. Rev. B}\ }\textbf {\bibinfo {volume}
  {39}},\ \bibinfo {pages} {6880--6897} (\bibinfo {year} {1989})}\BibitemShut
  {NoStop}%
\bibitem [{\citenamefont {Martinez}\ and\ \citenamefont
  {Horsch}(1991)}]{Martinez1991}%
  \BibitemOpen
  \bibfield  {author} {\bibinfo {author} {\bibfnamefont {Gerardo}\ \bibnamefont
  {Martinez}}\ and\ \bibinfo {author} {\bibfnamefont {Peter}\ \bibnamefont
  {Horsch}},\ }\bibfield  {title} {\enquote {\bibinfo {title} {{Spin Polarons
  in the $t$-$J$ Model}},}\ }\href {\doibase 10.1103/PhysRevB.44.317}
  {\bibfield  {journal} {\bibinfo  {journal} {Phys. Rev. B}\ }\textbf {\bibinfo
  {volume} {44}},\ \bibinfo {pages} {317--331} (\bibinfo {year}
  {1991})}\BibitemShut {NoStop}%
\bibitem [{\citenamefont {Ba{\l}a}\ \emph {et~al.}(1995)\citenamefont
  {Ba{\l}a}, \citenamefont {Ole{\'{s}}},\ and\ \citenamefont
  {Zaanen}}]{Bala1995}%
  \BibitemOpen
  \bibfield  {author} {\bibinfo {author} {\bibfnamefont {Jan}\ \bibnamefont
  {Ba{\l}a}}, \bibinfo {author} {\bibfnamefont {Andrzej~M.}\ \bibnamefont
  {Ole{\'{s}}}}, \ and\ \bibinfo {author} {\bibfnamefont {Jan}\ \bibnamefont
  {Zaanen}},\ }\bibfield  {title} {\enquote {\bibinfo {title} {{Spin Polarons
  in the $t$-$t'$-$J$ Model}},}\ }\href {\doibase 10.1103/PhysRevB.52.4597}
  {\bibfield  {journal} {\bibinfo  {journal} {Phys. Rev. B}\ }\textbf {\bibinfo
  {volume} {52}},\ \bibinfo {pages} {4597--4606} (\bibinfo {year}
  {1995})}\BibitemShut {NoStop}%
\bibitem [{\citenamefont {Sorella}(1992)}]{Sorella1992}%
  \BibitemOpen
  \bibfield  {author} {\bibinfo {author} {\bibfnamefont {S.}~\bibnamefont
  {Sorella}},\ }\bibfield  {title} {\enquote {\bibinfo {title} {{Quantum Monte
  Carlo Study of a Single Hole in a Quantum Antiferromagnet}},}\ }\href
  {\doibase 10.1103/PhysRevB.46.11670} {\bibfield  {journal} {\bibinfo
  {journal} {Phys. Rev. B}\ }\textbf {\bibinfo {volume} {46}},\ \bibinfo
  {pages} {11670--11680} (\bibinfo {year} {1992})}\BibitemShut {NoStop}%
\bibitem [{\citenamefont {Brunner}\ \emph {et~al.}(2000)\citenamefont
  {Brunner}, \citenamefont {Assaad},\ and\ \citenamefont
  {Muramatsu}}]{Brunner2000}%
  \BibitemOpen
  \bibfield  {author} {\bibinfo {author} {\bibfnamefont {Michael}\ \bibnamefont
  {Brunner}}, \bibinfo {author} {\bibfnamefont {Fakher~F.}\ \bibnamefont
  {Assaad}}, \ and\ \bibinfo {author} {\bibfnamefont {Alejandro}\ \bibnamefont
  {Muramatsu}},\ }\bibfield  {title} {\enquote {\bibinfo {title} {{Single-Hole
  Dynamics in the $t$-$J$ Model on a Square Lattice}},}\ }\href {\doibase
  10.1103/PhysRevB.62.15480} {\bibfield  {journal} {\bibinfo  {journal} {Phys.
  Rev. B}\ }\textbf {\bibinfo {volume} {62}},\ \bibinfo {pages} {15480--15492}
  (\bibinfo {year} {2000})}\BibitemShut {NoStop}%
\bibitem [{\citenamefont {Charlebois}\ and\ \citenamefont
  {Imada}(2020)}]{Imada2020}%
  \BibitemOpen
  \bibfield  {author} {\bibinfo {author} {\bibfnamefont {Maxime}\ \bibnamefont
  {Charlebois}}\ and\ \bibinfo {author} {\bibfnamefont {Masatoshi}\
  \bibnamefont {Imada}},\ }\bibfield  {title} {\enquote {\bibinfo {title}
  {Single-particle spectral function formulated and calculated by variational
  monte carlo method with application to $d$-wave superconducting state},}\
  }\href {\doibase 10.1103/PhysRevX.10.041023} {\bibfield  {journal} {\bibinfo
  {journal} {Phys. Rev. X}\ }\textbf {\bibinfo {volume} {10}},\ \bibinfo
  {pages} {041023} (\bibinfo {year} {2020})}\BibitemShut {NoStop}%
\bibitem [{\citenamefont {Shraiman}\ and\ \citenamefont
  {Siggia}(1988)}]{Shraiman1988a}%
  \BibitemOpen
  \bibfield  {author} {\bibinfo {author} {\bibfnamefont {Boris~I.}\
  \bibnamefont {Shraiman}}\ and\ \bibinfo {author} {\bibfnamefont {Eric~D.}\
  \bibnamefont {Siggia}},\ }\bibfield  {title} {\enquote {\bibinfo {title}
  {{Mobile Vacancies in a Quantum Heisenberg Antiferromagnet}},}\ }\href
  {\doibase 10.1103/PhysRevLett.61.467} {\bibfield  {journal} {\bibinfo
  {journal} {Phys. Rev. Lett.}\ }\textbf {\bibinfo {volume} {61}},\ \bibinfo
  {pages} {467--470} (\bibinfo {year} {1988})}\BibitemShut {NoStop}%
\bibitem [{\citenamefont {Shraiman}\ and\ \citenamefont
  {Siggia}(1989{\natexlab{a}})}]{Shraiman1989}%
  \BibitemOpen
  \bibfield  {author} {\bibinfo {author} {\bibfnamefont {Boris~I.}\
  \bibnamefont {Shraiman}}\ and\ \bibinfo {author} {\bibfnamefont {Eric~D.}\
  \bibnamefont {Siggia}},\ }\bibfield  {title} {\enquote {\bibinfo {title}
  {{Spiral Phase of a Doped Quantum Antiferromagnet}},}\ }\href {\doibase
  10.1103/PhysRevLett.62.1564} {\bibfield  {journal} {\bibinfo  {journal}
  {Phys. Rev. Lett.}\ }\textbf {\bibinfo {volume} {62}},\ \bibinfo {pages}
  {1564--1567} (\bibinfo {year} {1989}{\natexlab{a}})}\BibitemShut {NoStop}%
\bibitem [{\citenamefont {White}\ and\ \citenamefont
  {Scalapino}(1997)}]{White1997}%
  \BibitemOpen
  \bibfield  {author} {\bibinfo {author} {\bibfnamefont {Steven~R.}\
  \bibnamefont {White}}\ and\ \bibinfo {author} {\bibfnamefont
  {D.}~\bibnamefont {Scalapino}},\ }\bibfield  {title} {\enquote {\bibinfo
  {title} {{Hole and Pair Structures in the $t$-$J$ Model}},}\ }\href {\doibase
  10.1103/PhysRevB.55.6504} {\bibfield  {journal} {\bibinfo  {journal} {Phys.
  Rev. B}\ }\textbf {\bibinfo {volume} {55}},\ \bibinfo {pages} {6504--6517}
  (\bibinfo {year} {1997})},\ \Eprint {http://arxiv.org/abs/cond-mat/9605143}
  {arXiv:cond-mat/9605143} \BibitemShut {NoStop}%
\bibitem [{\citenamefont {Chernyshev}\ \emph {et~al.}(1998)\citenamefont
  {Chernyshev}, \citenamefont {Leung},\ and\ \citenamefont
  {Gooding}}]{Chernyshev1998}%
  \BibitemOpen
  \bibfield  {author} {\bibinfo {author} {\bibfnamefont {A.~L.}\ \bibnamefont
  {Chernyshev}}, \bibinfo {author} {\bibfnamefont {P.~W.}\ \bibnamefont
  {Leung}}, \ and\ \bibinfo {author} {\bibfnamefont {R.~J.}\ \bibnamefont
  {Gooding}},\ }\bibfield  {title} {\enquote {\bibinfo {title} {{Comprehensive
  Numerical and Analytical Study of Two Holes Doped into the Two-dimensional
  $t$-$J$ Model}},}\ }\href {\doibase 10.1103/PhysRevB.58.13594} {\bibfield
  {journal} {\bibinfo  {journal} {Phys. Rev. B}\ }\textbf {\bibinfo {volume}
  {58}},\ \bibinfo {pages} {13594--13613} (\bibinfo {year} {1998})}\BibitemShut
  {NoStop}%
\bibitem [{\citenamefont {Poilblanc}(1994)}]{Poilblanc1994}%
  \BibitemOpen
  \bibfield  {author} {\bibinfo {author} {\bibfnamefont {Didier}\ \bibnamefont
  {Poilblanc}},\ }\bibfield  {title} {\enquote {\bibinfo {title} {{Internal
  Structure of the Singlet $d_{x^2-y^2}$ Hole Pair in an Antiferromagnet}},}\
  }\href {\doibase 10.1103/PhysRevB.49.1477} {\bibfield  {journal} {\bibinfo
  {journal} {Phys. Rev. B}\ }\textbf {\bibinfo {volume} {49}},\ \bibinfo
  {pages} {1477--1479} (\bibinfo {year} {1994})}\BibitemShut {NoStop}%
\bibitem [{\citenamefont {Mezzacapo}\ \emph {et~al.}(2016)\citenamefont
  {Mezzacapo}, \citenamefont {Angelone},\ and\ \citenamefont
  {Pupillo}}]{Mezzacapo2016}%
  \BibitemOpen
  \bibfield  {author} {\bibinfo {author} {\bibfnamefont {Fabio}\ \bibnamefont
  {Mezzacapo}}, \bibinfo {author} {\bibfnamefont {Adriano}\ \bibnamefont
  {Angelone}}, \ and\ \bibinfo {author} {\bibfnamefont {Guido}\ \bibnamefont
  {Pupillo}},\ }\bibfield  {title} {\enquote {\bibinfo {title} {{Two Holes in a
  Two-Dimensional Quantum Antiferromagnet: A Variational Study Based on
  Entangled-Plaquette States}},}\ }\href {\doibase 10.1103/PhysRevB.94.155120}
  {\bibfield  {journal} {\bibinfo  {journal} {Phys. Rev. B}\ }\textbf {\bibinfo
  {volume} {94}},\ \bibinfo {pages} {155120} (\bibinfo {year} {2016})},\
  \Eprint {http://arxiv.org/abs/1610.04205} {arXiv:1610.04205} \BibitemShut
  {NoStop}%
\bibitem [{\citenamefont {Feynman}\ and\ \citenamefont
  {Cohen}(1956)}]{Feynman1956}%
  \BibitemOpen
  \bibfield  {author} {\bibinfo {author} {\bibfnamefont {R.~P.}\ \bibnamefont
  {Feynman}}\ and\ \bibinfo {author} {\bibfnamefont {Michael}\ \bibnamefont
  {Cohen}},\ }\bibfield  {title} {\enquote {\bibinfo {title} {{Energy Spectrum
  of the Excitations in Liquid Helium}},}\ }\href {\doibase
  10.1103/PhysRev.102.1189} {\bibfield  {journal} {\bibinfo  {journal} {Phys.
  Rev.}\ }\textbf {\bibinfo {volume} {102}},\ \bibinfo {pages} {1189--1204}
  (\bibinfo {year} {1956})}\BibitemShut {NoStop}%
\bibitem [{\citenamefont {Kou}\ and\ \citenamefont
  {Weng}(2003{\natexlab{a}})}]{Kou2003a}%
  \BibitemOpen
  \bibfield  {author} {\bibinfo {author} {\bibfnamefont {Su-Peng}\ \bibnamefont
  {Kou}}\ and\ \bibinfo {author} {\bibfnamefont {Zheng-Yu}\ \bibnamefont
  {Weng}},\ }\bibfield  {title} {\enquote {\bibinfo {title} {{Topological Gauge
  Structure and Phase Diagram for Weakly Doped Antiferromagnets}},}\ }\href
  {\doibase 10.1103/PhysRevLett.90.157003} {\bibfield  {journal} {\bibinfo
  {journal} {Phys. Rev. Lett.}\ }\textbf {\bibinfo {volume} {90}},\ \bibinfo
  {pages} {157003} (\bibinfo {year} {2003}{\natexlab{a}})}\BibitemShut
  {NoStop}%
\bibitem [{\citenamefont {Ye}\ \emph {et~al.}(2011)\citenamefont {Ye},
  \citenamefont {Tian}, \citenamefont {Qi},\ and\ \citenamefont
  {Weng}}]{Ye2011}%
  \BibitemOpen
  \bibfield  {author} {\bibinfo {author} {\bibfnamefont {Peng}\ \bibnamefont
  {Ye}}, \bibinfo {author} {\bibfnamefont {Chu-Shun}\ \bibnamefont {Tian}},
  \bibinfo {author} {\bibfnamefont {Xiao-Liang}\ \bibnamefont {Qi}}, \ and\
  \bibinfo {author} {\bibfnamefont {Zheng-Yu}\ \bibnamefont {Weng}},\
  }\bibfield  {title} {\enquote {\bibinfo {title} {{Confinement-Deconfinement
  Interplay in Quantum Phases of Doped Mott Insulators}},}\ }\href {\doibase
  10.1103/PhysRevLett.106.147002} {\bibfield  {journal} {\bibinfo  {journal}
  {Phys. Rev. Lett.}\ }\textbf {\bibinfo {volume} {106}},\ \bibinfo {pages}
  {147002} (\bibinfo {year} {2011})},\ \Eprint {http://arxiv.org/abs/1310.6496}
  {arXiv:1310.6496} \BibitemShut {NoStop}%
\bibitem [{\citenamefont {Kou}\ and\ \citenamefont
  {Weng}(2003{\natexlab{b}})}]{Kou2003b}%
  \BibitemOpen
  \bibfield  {author} {\bibinfo {author} {\bibfnamefont {Su-Peng}\ \bibnamefont
  {Kou}}\ and\ \bibinfo {author} {\bibfnamefont {Zheng-Yu}\ \bibnamefont
  {Weng}},\ }\bibfield  {title} {\enquote {\bibinfo {title} {{Holes as Dipoles
  in a Doped Antiferromagnet and Stripe Instabilities}},}\ }\href {\doibase
  10.1103/PhysRevB.67.115103} {\bibfield  {journal} {\bibinfo  {journal} {Phys.
  Rev. B}\ }\textbf {\bibinfo {volume} {67}},\ \bibinfo {pages} {115103}
  (\bibinfo {year} {2003}{\natexlab{b}})}\BibitemShut {NoStop}%
\bibitem [{\citenamefont {Capati}\ \emph {et~al.}(2015)\citenamefont {Capati},
  \citenamefont {Caprara}, \citenamefont {{Di Castro}}, \citenamefont {Grilli},
  \citenamefont {Seibold},\ and\ \citenamefont {Lorenzana}}]{Capati2015}%
  \BibitemOpen
  \bibfield  {author} {\bibinfo {author} {\bibfnamefont {M.}~\bibnamefont
  {Capati}}, \bibinfo {author} {\bibfnamefont {S.}~\bibnamefont {Caprara}},
  \bibinfo {author} {\bibfnamefont {C.}~\bibnamefont {{Di Castro}}}, \bibinfo
  {author} {\bibfnamefont {M.}~\bibnamefont {Grilli}}, \bibinfo {author}
  {\bibfnamefont {G.}~\bibnamefont {Seibold}}, \ and\ \bibinfo {author}
  {\bibfnamefont {J.}~\bibnamefont {Lorenzana}},\ }\bibfield  {title} {\enquote
  {\bibinfo {title} {{Electronic Polymers and Soft-Matter-Like Broken
  Symmetries in Underdoped Cuprates}},}\ }\href {\doibase 10.1038/ncomms8691}
  {\bibfield  {journal} {\bibinfo  {journal} {Nat. Commun.}\ }\textbf {\bibinfo
  {volume} {6}},\ \bibinfo {pages} {7691} (\bibinfo {year} {2015})},\ \Eprint
  {http://arxiv.org/abs/1505.01847} {arXiv:1505.01847} \BibitemShut {NoStop}%
\bibitem [{\citenamefont {Dagotto}\ \emph {et~al.}(1990)\citenamefont
  {Dagotto}, \citenamefont {Riera},\ and\ \citenamefont {Young}}]{Dagotto1990}%
  \BibitemOpen
  \bibfield  {author} {\bibinfo {author} {\bibfnamefont {Elbio}\ \bibnamefont
  {Dagotto}}, \bibinfo {author} {\bibfnamefont {Jose}\ \bibnamefont {Riera}}, \
  and\ \bibinfo {author} {\bibfnamefont {A.~P.}\ \bibnamefont {Young}},\
  }\bibfield  {title} {\enquote {\bibinfo {title} {{Dynamical Pair
  Susceptibilities in the $t$-$J$ and Hubbard Models}},}\ }\href {\doibase
  10.1103/PhysRevB.42.2347} {\bibfield  {journal} {\bibinfo  {journal} {Phys.
  Rev. B}\ }\textbf {\bibinfo {volume} {42}},\ \bibinfo {pages} {2347--2352}
  (\bibinfo {year} {1990})}\BibitemShut {NoStop}%
\bibitem [{\citenamefont {Poilblanc}(1993)}]{Poilblanc1993}%
  \BibitemOpen
  \bibfield  {author} {\bibinfo {author} {\bibfnamefont {Didier}\ \bibnamefont
  {Poilblanc}},\ }\bibfield  {title} {\enquote {\bibinfo {title} {{Binding of
  Holes and the Pair Spectral Function in the $t$-$J$ model}},}\ }\href
  {\doibase 10.1103/PhysRevB.48.3368} {\bibfield  {journal} {\bibinfo
  {journal} {Phys. Rev. B}\ }\textbf {\bibinfo {volume} {48}},\ \bibinfo
  {pages} {3368--3374} (\bibinfo {year} {1993})}\BibitemShut {NoStop}%
\bibitem [{\citenamefont {Jiang}\ \emph {et~al.}(2018)\citenamefont {Jiang},
  \citenamefont {Weng},\ and\ \citenamefont {Kivelson}}]{Jiang2018}%
  \BibitemOpen
  \bibfield  {author} {\bibinfo {author} {\bibfnamefont {Hong-Chen}\
  \bibnamefont {Jiang}}, \bibinfo {author} {\bibfnamefont {Zheng-Yu}\
  \bibnamefont {Weng}}, \ and\ \bibinfo {author} {\bibfnamefont {Steven~A.}\
  \bibnamefont {Kivelson}},\ }\bibfield  {title} {\enquote {\bibinfo {title}
  {{Superconductivity in the Doped $t$-$J$ Model: Results for Four-leg
  Cylinders}},}\ }\href {\doibase 10.1103/PhysRevB.98.140505} {\bibfield
  {journal} {\bibinfo  {journal} {Phys. Rev. B}\ }\textbf {\bibinfo {volume}
  {98}},\ \bibinfo {pages} {140505(R)} (\bibinfo {year} {2018})},\ \Eprint
  {http://arxiv.org/abs/1805.11163} {arXiv:1805.11163} \BibitemShut {NoStop}%
\bibitem [{\citenamefont {Jiang}\ and\ \citenamefont
  {Devereaux}(2019)}]{Jiang2019}%
  \BibitemOpen
  \bibfield  {author} {\bibinfo {author} {\bibfnamefont {Hong~Chen}\
  \bibnamefont {Jiang}}\ and\ \bibinfo {author} {\bibfnamefont {Thomas~P.}\
  \bibnamefont {Devereaux}},\ }\bibfield  {title} {\enquote {\bibinfo {title}
  {{Superconductivity in the Doped Hubbard model and Its Interplay with
  Next-nearest Hopping $t'$}},}\ }\href {\doibase 10.1126/science.aal5304}
  {\bibfield  {journal} {\bibinfo  {journal} {Science}\ }\textbf {\bibinfo
  {volume} {365}},\ \bibinfo {pages} {1424--1428} (\bibinfo {year} {2019})},\
  \Eprint {http://arxiv.org/abs/arXiv:1806.01465v2} {arXiv:arXiv:1806.01465v2}
  \BibitemShut {NoStop}%
\bibitem [{\citenamefont {Shih}\ \emph {et~al.}(1998)\citenamefont {Shih},
  \citenamefont {Chen}, \citenamefont {Lin},\ and\ \citenamefont
  {Lee}}]{Shih1998}%
  \BibitemOpen
  \bibfield  {author} {\bibinfo {author} {\bibfnamefont {C.~T.}\ \bibnamefont
  {Shih}}, \bibinfo {author} {\bibfnamefont {Y.~C.}\ \bibnamefont {Chen}},
  \bibinfo {author} {\bibfnamefont {H.~Q.}\ \bibnamefont {Lin}}, \ and\
  \bibinfo {author} {\bibfnamefont {T.~K.}\ \bibnamefont {Lee}},\ }\bibfield
  {title} {\enquote {\bibinfo {title} {{$d$-Wave Pairing Correlation in the
  Two-Dimensional $t$-$J$ Model}},}\ }\href {\doibase
  10.1103/PhysRevLett.81.1294} {\bibfield  {journal} {\bibinfo  {journal}
  {Phys. Rev. Lett.}\ }\textbf {\bibinfo {volume} {81}},\ \bibinfo {pages}
  {1294--1297} (\bibinfo {year} {1998})}\BibitemShut {NoStop}%
\bibitem [{\citenamefont {Misawa}\ and\ \citenamefont
  {Imada}(2014)}]{Misawa2014}%
  \BibitemOpen
  \bibfield  {author} {\bibinfo {author} {\bibfnamefont {Takahiro}\
  \bibnamefont {Misawa}}\ and\ \bibinfo {author} {\bibfnamefont {Masatoshi}\
  \bibnamefont {Imada}},\ }\bibfield  {title} {\enquote {\bibinfo {title}
  {{Origin of high-$T_c$ Superconductivity in Doped Hubbard Models and their
  Extensions: Roles of Uniform Charge Fluctuations}},}\ }\href {\doibase
  10.1103/PhysRevB.90.115137} {\bibfield  {journal} {\bibinfo  {journal} {Phys.
  Rev. B}\ }\textbf {\bibinfo {volume} {90}},\ \bibinfo {pages} {115137}
  (\bibinfo {year} {2014})},\ \Eprint {http://arxiv.org/abs/1306.1434}
  {arXiv:1306.1434} \BibitemShut {NoStop}%
\bibitem [{\citenamefont {Shraiman}\ and\ \citenamefont
  {Siggia}(1989{\natexlab{b}})}]{Shraiman1989b}%
  \BibitemOpen
  \bibfield  {author} {\bibinfo {author} {\bibfnamefont {Boris~I.}\
  \bibnamefont {Shraiman}}\ and\ \bibinfo {author} {\bibfnamefont {Eric~D}\
  \bibnamefont {Siggia}},\ }\bibfield  {title} {\enquote {\bibinfo {title}
  {{Mean-Field Theory for Vacancies in a Quantum Antiferromagnet}},}\ }\href
  {\doibase 10.1103/PhysRevB.40.9162} {\bibfield  {journal} {\bibinfo
  {journal} {Phys. Rev. B}\ }\textbf {\bibinfo {volume} {40}},\ \bibinfo
  {pages} {9162--9166} (\bibinfo {year} {1989}{\natexlab{b}})}\BibitemShut
  {NoStop}%
\bibitem [{\citenamefont {Shraiman}\ and\ \citenamefont
  {Siggia}(1990)}]{Shraiman1990}%
  \BibitemOpen
  \bibfield  {author} {\bibinfo {author} {\bibfnamefont {Boris~I.}\
  \bibnamefont {Shraiman}}\ and\ \bibinfo {author} {\bibfnamefont {Eric~D.}\
  \bibnamefont {Siggia}},\ }\bibfield  {title} {\enquote {\bibinfo {title}
  {{Mobile Vacancy in a Quantum Antiferromagnet: Effective Hamiltonian}},}\
  }\href {\doibase 10.1103/PhysRevB.42.2485} {\bibfield  {journal} {\bibinfo
  {journal} {Phys. Rev. B}\ }\textbf {\bibinfo {volume} {42}},\ \bibinfo
  {pages} {2485--2500} (\bibinfo {year} {1990})}\BibitemShut {NoStop}%
\bibitem [{\citenamefont {Schrieffer}\ \emph {et~al.}(1988)\citenamefont
  {Schrieffer}, \citenamefont {Wen},\ and\ \citenamefont
  {Zhang}}]{Schrieffer1988}%
  \BibitemOpen
  \bibfield  {author} {\bibinfo {author} {\bibfnamefont {J.~R.}\ \bibnamefont
  {Schrieffer}}, \bibinfo {author} {\bibfnamefont {X.-G.}\ \bibnamefont {Wen}},
  \ and\ \bibinfo {author} {\bibfnamefont {S.-C.}\ \bibnamefont {Zhang}},\
  }\bibfield  {title} {\enquote {\bibinfo {title} {{Spin-Bag Mechanism of
  High-Temperature Superconductivity}},}\ }\href {\doibase
  10.1103/PhysRevLett.60.944} {\bibfield  {journal} {\bibinfo  {journal} {Phys.
  Rev. Lett.}\ }\textbf {\bibinfo {volume} {60}},\ \bibinfo {pages} {944--947}
  (\bibinfo {year} {1988})}\BibitemShut {NoStop}%
\bibitem [{\citenamefont {Weng}\ \emph {et~al.}(1990)\citenamefont {Weng},
  \citenamefont {Ting},\ and\ \citenamefont {Lee}}]{Weng1990}%
  \BibitemOpen
  \bibfield  {author} {\bibinfo {author} {\bibfnamefont {Z.~Y.}\ \bibnamefont
  {Weng}}, \bibinfo {author} {\bibfnamefont {C.~S.}\ \bibnamefont {Ting}}, \
  and\ \bibinfo {author} {\bibfnamefont {T.~K.}\ \bibnamefont {Lee}},\
  }\bibfield  {title} {\enquote {\bibinfo {title} {{Mobile Spin Bags and Their
  Interaction in the Spin-Density-Wave Background}},}\ }\href {\doibase
  10.1103/PhysRevB.41.1990} {\bibfield  {journal} {\bibinfo  {journal} {Phys.
  Rev. B}\ }\textbf {\bibinfo {volume} {41}},\ \bibinfo {pages} {1990--2002}
  (\bibinfo {year} {1990})}\BibitemShut {NoStop}%
\bibitem [{\citenamefont {Zhong}\ \emph {et~al.}(2016)\citenamefont {Zhong},
  \citenamefont {Wang}, \citenamefont {Han}, \citenamefont {Lv}, \citenamefont
  {Wang}, \citenamefont {Zhang}, \citenamefont {Ding}, \citenamefont {Zhang},
  \citenamefont {Wang}, \citenamefont {He}, \citenamefont {Zhong},
  \citenamefont {Schneeloch}, \citenamefont {Gu}, \citenamefont {Song},
  \citenamefont {Ma},\ and\ \citenamefont {Xue}}]{Zhong2016}%
  \BibitemOpen
  \bibfield  {author} {\bibinfo {author} {\bibfnamefont {Yong}\ \bibnamefont
  {Zhong}}, \bibinfo {author} {\bibfnamefont {Yang}\ \bibnamefont {Wang}},
  \bibinfo {author} {\bibfnamefont {Sha}\ \bibnamefont {Han}}, \bibinfo
  {author} {\bibfnamefont {Yan~Feng}\ \bibnamefont {Lv}}, \bibinfo {author}
  {\bibfnamefont {Wen~Lin}\ \bibnamefont {Wang}}, \bibinfo {author}
  {\bibfnamefont {Ding}\ \bibnamefont {Zhang}}, \bibinfo {author}
  {\bibfnamefont {Hao}\ \bibnamefont {Ding}}, \bibinfo {author} {\bibfnamefont
  {Yi~Min}\ \bibnamefont {Zhang}}, \bibinfo {author} {\bibfnamefont {Lili}\
  \bibnamefont {Wang}}, \bibinfo {author} {\bibfnamefont {Ke}~\bibnamefont
  {He}}, \bibinfo {author} {\bibfnamefont {Ruidan}\ \bibnamefont {Zhong}},
  \bibinfo {author} {\bibfnamefont {John~A.}\ \bibnamefont {Schneeloch}},
  \bibinfo {author} {\bibfnamefont {Gen~Da}\ \bibnamefont {Gu}}, \bibinfo
  {author} {\bibfnamefont {Can~Li}\ \bibnamefont {Song}}, \bibinfo {author}
  {\bibfnamefont {Xu~Cun}\ \bibnamefont {Ma}}, \ and\ \bibinfo {author}
  {\bibfnamefont {Qi~Kun}\ \bibnamefont {Xue}},\ }\bibfield  {title} {\enquote
  {\bibinfo {title} {{Nodeless Pairing in Superconducting Copper-Oxide
  Monolayer Films on $\mathrm{Bi_2Sr_2CaCu_2O}_{8+\delta}$}},}\ }\href
  {\doibase 10.1007/s11434-016-1145-4} {\bibfield  {journal} {\bibinfo
  {journal} {Sci. Bull.}\ }\textbf {\bibinfo {volume} {61}},\ \bibinfo {pages}
  {1239--1247} (\bibinfo {year} {2016})}\BibitemShut {NoStop}%
\bibitem [{\citenamefont {Ren}\ \emph {et~al.}(2016)\citenamefont {Ren},
  \citenamefont {Yan}, \citenamefont {Zhang},\ and\ \citenamefont
  {Feng}}]{Ren2016}%
  \BibitemOpen
  \bibfield  {author} {\bibinfo {author} {\bibfnamefont {Ming-Qiang}\
  \bibnamefont {Ren}}, \bibinfo {author} {\bibfnamefont {Ya-Jun}\ \bibnamefont
  {Yan}}, \bibinfo {author} {\bibfnamefont {Tong}\ \bibnamefont {Zhang}}, \
  and\ \bibinfo {author} {\bibfnamefont {Dong-Lai}\ \bibnamefont {Feng}},\
  }\bibfield  {title} {\enquote {\bibinfo {title} {{Possible Nodeless
  Superconducting Gaps in $\mathrm{Bi_2Sr_2CaCu_2O_{8+\delta}}$ and
  $\mathrm{YBa_2Cu_3O_{7-x}}$ Revealed by Cross-Sectional Scanning Tunneling
  Spectroscopy}},}\ }\href {\doibase 10.1088/0256-307X/33/12/127402} {\bibfield
   {journal} {\bibinfo  {journal} {Chin. Phys. Lett.}\ }\textbf {\bibinfo
  {volume} {33}},\ \bibinfo {pages} {127402} (\bibinfo {year} {2016})},\
  \Eprint {http://arxiv.org/abs/1611.04220} {arXiv:1611.04220} \BibitemShut
  {NoStop}%
\bibitem [{\citenamefont {Zhu}\ \emph {et~al.}(2021)\citenamefont {Zhu},
  \citenamefont {Liao}, \citenamefont {Zhang}, \citenamefont {Xie},
  \citenamefont {Meng}, \citenamefont {Liu}, \citenamefont {Bai}, \citenamefont
  {Ji}, \citenamefont {Zhang}, \citenamefont {Jiang}, \citenamefont {Zhong},
  \citenamefont {Schneeloch}, \citenamefont {Gu}, \citenamefont {Gu},
  \citenamefont {Ma}, \citenamefont {Zhang},\ and\ \citenamefont
  {Xue}}]{Zhu2021}%
  \BibitemOpen
  \bibfield  {author} {\bibinfo {author} {\bibfnamefont {Yuying}\ \bibnamefont
  {Zhu}}, \bibinfo {author} {\bibfnamefont {Menghan}\ \bibnamefont {Liao}},
  \bibinfo {author} {\bibfnamefont {Qinghua}\ \bibnamefont {Zhang}}, \bibinfo
  {author} {\bibfnamefont {Hong-yi}\ \bibnamefont {Xie}}, \bibinfo {author}
  {\bibfnamefont {Fanqi}\ \bibnamefont {Meng}}, \bibinfo {author}
  {\bibfnamefont {Yaowu}\ \bibnamefont {Liu}}, \bibinfo {author} {\bibfnamefont
  {Zhonghua}\ \bibnamefont {Bai}}, \bibinfo {author} {\bibfnamefont {Shuaihua}\
  \bibnamefont {Ji}}, \bibinfo {author} {\bibfnamefont {Jin}\ \bibnamefont
  {Zhang}}, \bibinfo {author} {\bibfnamefont {Kaili}\ \bibnamefont {Jiang}},
  \bibinfo {author} {\bibfnamefont {Ruidan}\ \bibnamefont {Zhong}}, \bibinfo
  {author} {\bibfnamefont {John}\ \bibnamefont {Schneeloch}}, \bibinfo {author}
  {\bibfnamefont {Genda}\ \bibnamefont {Gu}}, \bibinfo {author} {\bibfnamefont
  {Lin}\ \bibnamefont {Gu}}, \bibinfo {author} {\bibfnamefont {Xucun}\
  \bibnamefont {Ma}}, \bibinfo {author} {\bibfnamefont {Ding}\ \bibnamefont
  {Zhang}}, \ and\ \bibinfo {author} {\bibfnamefont {Qi-Kun}\ \bibnamefont
  {Xue}},\ }\bibfield  {title} {\enquote {\bibinfo {title} {{Presence of
  $s$-Wave Pairing in Josephson Junctions Made of Twisted Ultrathin
  $\mathrm{Bi_2Sr_2CaCu_2O}_{8+x}$ Flakes}},}\ }\href {\doibase
  10.1103/PhysRevX.11.031011} {\bibfield  {journal} {\bibinfo  {journal} {Phys.
  Rev. X}\ }\textbf {\bibinfo {volume} {11}},\ \bibinfo {pages} {031011}
  (\bibinfo {year} {2021})}\BibitemShut {NoStop}%
\bibitem [{\citenamefont {Sun}\ \emph {et~al.}(2019)\citenamefont {Sun},
  \citenamefont {Zhu},\ and\ \citenamefont {Weng}}]{Sun2019}%
  \BibitemOpen
  \bibfield  {author} {\bibinfo {author} {\bibfnamefont {Rong-Yang}\
  \bibnamefont {Sun}}, \bibinfo {author} {\bibfnamefont {Zheng}\ \bibnamefont
  {Zhu}}, \ and\ \bibinfo {author} {\bibfnamefont {Zheng-Yu}\ \bibnamefont
  {Weng}},\ }\bibfield  {title} {\enquote {\bibinfo {title} {{Localization in a
  $t$-$J$-Type Model with Translational Symmetry}},}\ }\href {\doibase
  10.1103/PhysRevLett.123.016601} {\bibfield  {journal} {\bibinfo  {journal}
  {Phys. Rev. Lett.}\ }\textbf {\bibinfo {volume} {123}},\ \bibinfo {pages}
  {016601} (\bibinfo {year} {2019})},\ \Eprint
  {http://arxiv.org/abs/1811.06120} {arXiv:1811.06120} \BibitemShut {NoStop}%
\bibitem [{\citenamefont {Sandvik}\ and\ \citenamefont
  {Evertz}(2010)}]{Sandvik2010}%
  \BibitemOpen
  \bibfield  {author} {\bibinfo {author} {\bibfnamefont {Anders~W.}\
  \bibnamefont {Sandvik}}\ and\ \bibinfo {author} {\bibfnamefont {Hans~Gerd}\
  \bibnamefont {Evertz}},\ }\bibfield  {title} {\enquote {\bibinfo {title}
  {{Loop Updates for Variational and Projector Quantum Monte Carlo Simulations
  in the Valence-Bond Basis}},}\ }\href {\doibase 10.1103/PhysRevB.82.024407}
  {\bibfield  {journal} {\bibinfo  {journal} {Phys. Rev. B}\ }\textbf {\bibinfo
  {volume} {82}},\ \bibinfo {pages} {024407} (\bibinfo {year} {2010})},\
  \Eprint {http://arxiv.org/abs/0807.0682} {arXiv:0807.0682} \BibitemShut
  {NoStop}%
\end{thebibliography}%

\end{document}